\documentclass[fleqn,usenatbib]{mnras}

\usepackage{amsmath}
\usepackage{array}
\usepackage{xfrac}
\usepackage{multirow}
\newcommand{\lagr}{\mathcal{L}}

\usepackage[T1]{fontenc}
\usepackage{ae,aecompl}
\hypersetup{draft}

\title[Milky Way Analogues in MaNGA]{Milky Way Analogues in MaNGA: Multi-Parameter Homogeneity and Comparison to the Milky Way}

\author[Boardman et al.]{
N.~Boardman$^{1}$\thanks{E-mail: nick.boardman@astro.utah.edu},
G.~Zasowski$^{1}$,
A.~Seth$^{1}$,
J.~Newman$^{2}$,
B.~Andrews$^{2}$, \newauthor
M.~Bershady$^{3,4}$,
J.~Bird$^{5}$,
C.~Chiappini$^{6}$,
C.~Fielder$^{2}$,
A.~Fraser-McKelvie$^{7}$, \newauthor
A.~Jones$^{8}$,
T.~Licquia$^{2,9}$,
K.~L.~Masters$^{10,11}$,
I. Minchev$^{6}$,
R.~P.~Schiavon$^{12}$, \newauthor
J.~R.~Brownstein$^{1}$,
N.~Drory$^{13}$,
R.~R.~Lane$^{14,15}$
\\
$^{1}$Department of Physics \& Astronomy, University of Utah, Salt Lake City, UT, 84112, USA\\
$^{2}$Department of Physics \& Astronomy and PITT PACC, University of Pittsburgh, Pittsburgh, PA 15260, USA\\
$^{3}$Department of Astronomy, University of Wisconsin-Madison, 475N. Charter St., Madison WI 53703, USA\\
$^{4}$South African Astronomical Observatory, P.O. Box 9, Observatory 7935, Cape Town, South Africa\\
$^{5}$Department of Physics and Astronomy, Vanderbilt University, VU Station 1807, Nashville, TN 37235, USA\\
$^{6}$Leibniz-Institut f\" ur Astrophysik Potsdam (AIP), An der Sternwarte 16, 14482 Potsdam, Germany\\
$^{7}$School of Physics \& Astronomy, University of Nottingham, Nottingham, NG7 2RD, UK\\
$^{8}$Department of Physics \& Astronomy, University of Alabama, Tuscaloosa, AL, 35487-0324, USA\\
$^{9}$Advanced Analytics, The Dow Chemical Company, 2200 W. Salzburg Road, PO Box 994, Auburn, MI 48611\\
$^{10}$Haverford College, Department of Physics and Astronomy, 370 Lancaster Ave, Haverford, PA 19041\\
$^{11}$Institute of Cosmology \& Gravitation, University of Portsmouth, Dennis Sciama Building, Portsmouth, PO1 3FX, UK\\
$^{12}$Astrophysics Research Institute, Liverpool John Moores University, 146 Brownlow Hill, Liverpool L3 5RF, UK\\
$^{13}$McDonald Observatory, The University of Texas at Austin, 1 University Station, Austin, TX 78712, USA\\
$^{14}$Pontificia Universidad Cat\'  olica de Chile, Instituto de Astrofısica, Av. Vicuna Mackenna 4860, 782-0436 Macul, Santiago, Chile\\
$^{15}$Millennium Institute of Astrophysics, Av. Vicu\~ na Mackenna 4860, 782-0436 Macul, Santiago, Chile}

\date{Accepted XXX. Received YYY; in original form ZZZ}
\pubyear{2019}
\begin{document} 
\label{firstpage}
\pagerange{\pageref{firstpage}--\pageref{lastpage}}
\maketitle

\begin{abstract}

The Milky Way provides an ideal laboratory to test our understanding of galaxy evolution, owing to our ability to observe our Galaxy over fine scales. However, connecting the Galaxy to the wider galaxy population remains difficult, due to the challenges posed by our internal perspective and to the different observational techniques employed. Here, we present a sample of galaxies identified as Milky Way Analogs (MWAs) on the basis of their stellar masses and bulge-to-total ratios, observed as part of the Mapping Nearby Galaxies at Apache Point Observatory (MaNGA) survey. We analyse the galaxies in terms of their stellar kinematics and populations as well as their ionised gas contents.  We find our sample to contain generally young stellar populations in their outskirts. However, we find a wide range of stellar ages in their central regions, and we detect central AGN-like or composite-like activity in roughly half of the sample galaxies, with the other half consisting of galaxies with central star-forming emission or emission consistent with old stars. We measure gradients in gas metallicity and stellar metallicity that are generally flatter in physical units than those measured for the Milky Way; however, we find far better agreement with the Milky Way when scaling gradients by galaxies' disc scale lengths. From this, we argue much of the discrepancy in metallicity gradients to be due to the relative compactness of the Milky Way, with differences in observing perspective also likely to be a factor.

\end{abstract}


\section{Introduction}

The formation and evolution of galaxies remains a key unsolved problem in astronomy and astrophysics, in spite of the substantial progress made over the last two decades. Data from the Sloan Digital Sky Survey \citep[SDSS;][]{york2000} has conclusively demonstrated the existence of a bimodality in the optical colours of nearby galaxies \citep{strateva2001,baldry2004}. Nearby galaxies can mostly be classified as ``blue cloud'' or ``red sequence'' on the basis of their optical colors, with a minority of galaxies appearing to populate a ``green valley'' transition regime between the two. It is now widely accepted that galaxies transition from the star-forming blue cloud to the quiescent red sequence over the course of their lives \citep[e.g.][]{bell2004,faber2007}. However, the precise mechanisms behind such transitions remain only partially understood.

Our understanding of resolved galaxy properties has significantly improved in more recent years in large part due to the advent of large sample, integral field unit (IFU) spectroscopic surveys, such as ATLAS3D \citep{cappellari2011}, CALIFA \citep{sanchez2012}, SAMI \citep{croom2012}, MAD \citep{errozferror2019}, and MaNGA \citep{bundy2015}. Quiescent galaxies show a range of kinematic behaviours, with some being dispersion-dominated and others retaining a significant level of rotational support \citep[e.g.][]{emsellem2007,emsellem2011}. Galaxies appear to grow their bulge fraction over time \citep[e.g.][]{cappellari2013a,li2018}, suggesting a link between bulge-growth and the quenching of star formation, and are typically younger and more metal-poor in their outer parts than in their inner regions \citep[e.g.][]{gonzalezdelgado2015,zheng2017}. A number of possible physical reasons exist for this connection, including the stabilisation of gas disks against collapse \citep{martig2009,ceverino2010} as well as feedback effects \citep[e.g.][]{croton2006,hopkins2006}. Galaxy quenching appears to involve both a reduction in galaxies' inner star formation efficiency as well as a reduced molecular gas content \citep{sanchez2018}.

The existence of a connection between bulge growth and quenching is consistent with an inside-out view of galaxy evolution, in which galaxies have more efficient star-formation histories in their inner parts than the regions further out \citep{perez2013, ibarramedel2016, garciabenito2017,lopezfernandez2018}. Such a view is also consistent with colour gradients observed in disk galaxies \citep[e.g.][]{bell2000, macarthur2004, pohlen2006}, as well as with galaxies' negative oxygen abundance gradients \citep[e.g.][]{searle1971,diaz1989,villascostas1992,sanchez2012b}. The availability of large IFU datasets has allowed the latter finding to be explored in far greater detail. \citet{sanchez2012} analyse IFU data of 38 face-on spiral galaxies and report a characteristic oxygen abundance gradient of $\sim -0.1$ dex/$r_e$, and also show the oxygen abundance around the solar region to be in good consistency with the measured profiles for their galaxies. \citet{sanchez2014} analyze 306 disk galaxies from the CALIFA survey and likewise report a characteristic negative abundance gradient of $\sim -0.1$ dex/$r_e$ for all undisturbed galaxies, with little scatter. CALIFA oxygen abundance gradients have been further explored in a nymber of subsequent studies \citep{ho2015,sm2016,pm2016}. Negative oxygen abundance gradients have also been detected in MaNGA \citep{belfiore2017} along with several other recent IFU datasets \citep[e.g.][]{ho2015,kaplan2016, sm2018,poetrodjojo2018}.

A number of evolutionary pathways appear to be required to explain the range of galaxies' observed properties. Red disk galaxies were likely quenched slowly over many Gyr, following a cessation in gas inflow; \citet{schawinski2014} find from UV-optical colors that such galaxies quench on time-scales of more than 1 Gyr, while early-type galaxies (ETGs) appear from the same diagnostic to quench in less than 250 Myr. The elemental abundance ratios of ETGs likewise support a fast quenching timescale \citep[e.g.][]{thomas1999}. The present-day green-valley population, meanwhile, appears to be dominated by disk galaxies quanching on intermediate-to-long time-scales \citep{smethurst2015}.  The quenching of star formation in ETGs appears to coincide with the growth of a central bulge \citep{cappellari2013a,cappellari2016}, based on the correlation between ETGs' bulge-mass fractions and other properties, such as galaxy age and optical color below a characteristic mass of $M \simeq 2\times10^{11} M_\odot$.  Regardless of their precise evolutionary tracks, modern-day quiescent galaxies formed stars at much higher rates in the past, implying such galaxies to have experienced a significant drop in star formation at redshifts $z \simeq 0.3$ \citep{sanchez2019}.

Much progress has also been made during this same time period towards understanding the structure of our own Milky Way (MW) Galaxy. Several surveys have been undertaken in recent years with the aim of understanding the Milky Way's stellar kinematics and stellar populations, including RAVE \citep{steinmetz2006}, LEGUE \citep{deng2012}, LAMOST \citep{Cui_2012_lamost}, the Gaia-ESO survey \citep{gilmore2012}, GALAH \citep{desilva2015}, {\it Gaia} \citep{gaia}, and APOGEE \citep{majewski2017}. The MW's stellar metallicities decline with Galactocentric radius along the disk plane \citep[e.g.][]{genovali2014}, with metallicity gradients significantly flatter for older stars than for younger stars or for the interstellar medium \citep[ISM; e.g.][]{anders2017,hasselquist2019}. ISM abundance gradients were first noted for the MW by \citet{peimbert1978}, based on a sample of eight HII regions; \citet{shaver1983} subsequently used a larger sample to firmly demonstrate the presence of a large-scale abundance gradient in the MW's ISM, using both radio and optical spectroscopy. A bimodality in the abundance ratios of the $\alpha$ elements is apparent at intermediate stellar metallicities for the MW, indicating the existence of at least two distinct stellar populations or evolutionary sequences \citep[e.g.][]{fuhrmann1998,fuhrmann2011,anders2014,nidever2014,mikolaitis2014,reciobianco2014,hayden2015}. This pattern is generally understood in terms of the existence of a chemical ``thick disk'' \citep{yoshii1982,gilmore1983} that comprises largely old stars enhanced in the $\alpha$ elements, along with a younger chemical ``thin disk'' of roughly solar abundance \citep[e.g.][]{chiappini1997,bensby2003,haywood2013,xiang2017,wu2019}. 

The position of the MW in the context of the general galaxy population remains of particular interest. However, the MW's global properties have  remained relatively poorly constrained, due to the relative difficulty in studying such properties from our position within the MW's disk (for reviews, see \citet{bh2016} and \citet{barbuy2018}). For example, interstellar dust preferentially reddens the colours of distant stars \citep[while also obscuring much of the UV/optical MW from view; e.g.][]{schlegel1998,schlafly2011}, and large fractional distance uncertainties complicate the interpretation of observed kinematical and chemical patterns. In addition, the observables and methods used for studying the MW are different than for external galaxies - utilizing properties of individual stars as opposed to integrated quantities - and so suffer from different systematics and biases. As a result, the MW's position relative to the wider galaxy population remains only partially understood. This is an unfortunate situation, as MW observations allow us to study galaxy evolution on far smaller scales than is possible with extragalactic objects (see, for instance, the reviews of \citet{ivezic2013} and \citet{hayer2015}). 

A further complication is that the MW may not be a typical star-forming spiral galaxy as has traditionally been assumed. After converting \citet{vdk1986}'s Johnson $(B-V)$ color measurement of the MW to an SDSS AB $(u-r)$ color, \citet{mutch2011} find the MW to be consistent with the green valley on the color--mass diagram; however, the uncertainty in the MW's color was too large to draw a strong conclusion at the time. The MW also appears to be unusually compact for a spiral galaxy of its stellar mass \citep[e.g.][]{licquia2016a}, as well as having satellite galaxies that are fewer in number \citep[e.g.][]{bullock2010} and more compactly distributed \citep[e.g.][]{yniguez2014} than around other galaxies of similar stellar mass. The MW has also previously been claimed to be an outlier from the Tully-Fisher relation \citep{tully1977}, appearing deficient in stellar luminosity with respect to its rotational velocity \citep[e.g.][]{hammer2007}; however, \citet{licquia2016a} argue such findings to be due to measurement systematics, and find the MW to be consistent with the Tully-Fisher relation from more recent MW measurements.

Milky Way Analogs (MWAs) offer a chance to reconcile Galactic and extragalactic observations by providing observers with an external, global view of MW-like objects. MWAs allow one to estimate integrated properties of the MW that cannot otherwise be calculated robustly or directly, by assuming the MWA properties to be representative of what would be calculated for the MW by an external observer or, more conservatively, that the properties of the Milky Way should lie within the range spanned by an MWA sample. In addition, observations of MWAs may be compared directly with observations of the MW in certain cases. As such, MWAs allow us to place the MW within the context of the wider galaxy population. 

There is no single definition for what makes a galaxy a MWA; rather, the definition may (or must) be tailored depending on the goals of a particular study. Galaxies may be identified as being MW-like on the basis of having similar qualitative characteristics to the MW, such as the presence of a boxy/peanut bulge \citep[e.g.][]{georgiev2019,kormendy2019}, or on the basis of their position relative to the MW in a given parameter space. \citet[][hereafter L15]{L15}, select a large photometric MWA sample on the basis of total stellar mass and current star formation rate. L15 obtain MW values of these parameters from \citet{licquia2015}, who employ a hierarchical Bayesian analysis to combine the results of numerous previously-reported measurements. L15 use their sample to estimate the integrated color and optical magnitude of the MW via the Copernican assumption that the position of the MW in color-magnitude space should not be extraordinary compared to a sample of galaxies spanning the range of its possible stellar mass and star formation rate. L15 find their results to be consistent with the MW occupying the green valley region of the color--magnitude diagram, in agreement with \citet{mutch2011} but with significantly less uncertainty.

Recently, a sample of MWAs was observed as part of the Mapping Galaxies at Apache Point Observatory survey \citep[MaNGA;][]{bundy2015}, to expand upon the MWA technique and take advantage of the rich IFU data offered by the MaNGA dataset. In this paper, we explore the properties of these MWAs as revealed through MaNGA. We investigate the range of properties displayed by the MWAs in terms of stellar kinematics, stellar populations, and ionised gas. We also compare the metallicity gradients we measure for the MWAs with gradients measured for the MW. In this way, we aim to test the self-similarity of the MaNGA MWA samples, as well as to assess how truly representative they are of the MW itself.

The structure of this paper is as follows: we introduce our MWA sample in \autoref{sample}, describe the methods used for analysing our sample in \autoref{specmod}, and present our results in \autoref{results}. We discuss our findings in \autoref{disc}, before summarizing and concluding in \autoref{conclusion}.

\section{Sample and Data}\label{sample}

We employ a sample of MWAs observed by MaNGA \citep[][]{bundy2015}, which is part of the Sloan Digital Sky Survey~IV programme \citep[SDSS-IV;][]{blanton2017}. The final MaNGA sample will consist of approximately 10,000 galaxies with redshifts $0.01<z<0.18$, spanning a wide range in galaxy morphologies and selected to be have an approximately flat distribution in terms of log(mass) \citep{yan2016b,wake2017}. Observations are taken using the BOSS spectrograph \citep{smee2013} on the 2.5~m Sloan telescope at Apache Point Observatory \citep{gunn2006}. The spectra have a wavelength range of 3600--10000~\AA\ and a spectral resolution of $R \simeq 2000$ \citep{drory2015}; such a wide wavelength range is useful for breaking the age-metallicity degeneracy, which is easier to constrain with wider spectral ranges \citep[e.g.][]{pforr2012}. MaNGA's individual IFUs consist of hexagonal fibre bundles containing 19--127 optical fibres (each with diameter 2$^{\prime\prime}$), with a three-point dithering pattern employed during observations in order to fully sample the targeted field of view \citep{drory2015,law2015}. Observations are reduced through the MaNGA Data Reduction Pipeline \citep[DRP;][]{law2016,yan2016a}, which flux-calibrates and sky-subtracts spectra before drizzling observations onto spaxels of width $0.5^{\prime\prime} \times 0.5^{\prime\prime}$. Flux calibration is performed using standard stars observed with fibre bundles of 7 fibres apiece. The final reconstructed datacubes have a point-spread function (PSF) full-width at half-maximum (FWHM) of approximately 2.5$^{\prime\prime}$ \citep{law2015}.

In the remainder of this section, we describe the sample of MWAs used in our analysis. We will refer to individual galaxies from this point onwards by their SDSS plate number and MaNGA IFU designation; for instance, the galaxy 8137-12703 has a plate number of 8137 and an IFU designation of 12703.  

\subsection{Sample selection}

Our galaxy sample is drawn from a ancillary  MaNGA observation program aimed at observing MWA galaxies. This program set out to observe ``structural'' MW analogs, selected based on their stellar masses ($M_*$) and bulge-to-total (B/T) ratios. Stellar masses were obtained from the MPA-JHU catalog\footnote{\url{https://www.sdss.org/dr14/spectro/galaxy_mpajhu/}}, while $r$-band B/T values were obtained from the PSF-convolved bulge-disk decompositions of \citet{simard2011}. 
Galaxies were then selected as analogs in a manner similar to that described in L15: pairs of values in $M_*$--B/T space were randomly drawn based on the MW's $M_*$--B/T joint probability distribution function (PDF) as derived in  \citet{licquia2016}, who use a heirarchical Bayesian analysis to combine the results of many individual MW studies in an almost-identical manner to \citet{licquia2015}. For each drawn pair of values, a nearby galaxy was then randomly selected for inclusion in the sample. A statistical MWA sample can then be constructed by repeating this process thousands of times. 

In principle, one could attempt to select MWAs using a larger number of simultaneous selection parameters, as well as more stringent constraints on a given parameter, in order to create a sample of galaxies with properties that may be closer to the true properties of the Milky Way. In practice, however, such an approach has a number of issues. One problem is that our understanding of the MW is still evolving, meaning that MWA samples with highly-specific criteria may well become obsolete as our understanding of the MW grows. At the same time, sample sizes for MWAs - even drawing from large datasets such as SDSS - quickly become vanishingly small as an increasing number of selection criteria are considered.

We illustrate the latter problem with the following exercise. First, we select from the catalog of \citet{simard2011} all galaxies with disk scale radii between 2.51 and 2.93~kpc and with $r$-band B/T values between 0.13 and 0.19, using the results from the $n=4$ bulge-plus-disc Sersic fits described in that paper. Our parameter ranges are based on the 1$\sigma$ confidence ranges reported in \citet{licquia2016}. This yields a sample of 6648 galaxies. Next, we cross-reference with the MPA-JHU catalogue, finding 4120 out of the 6648 galaxies selected from \citet{simard2011}, and we extract global stellar masses and star formation rates for all found galaxies. Of these 4120 galaxies, we then select those with stellar masses between $4.6 \times 10^{10} ~M_\odot$ and $7.2 \times 10^{10}~M_\odot$, and with current star formation rates between $1.46~M_\odot~{\rm yr}^{-1}$ and $1.84~M_\odot~{\rm yr}^{-1}$, based on the 1$\sigma$ confidence ranges of \citet{licquia2016} and \citet{licquia2015} respectively. We find only 15 galaxies that satisfy these additional cuts, and none of these galaxies are found when cross-matching with version 2.5.3 of the MaNGA \textit{drpall} file, which catalogs all galaxies in MaNGS Produce Launch 8 (MPL-8). Even if we drop our star formation rate criterion, we find just one galaxy in \textit{drpall} that satisfies the remaining criteria. A similar example of this problem can be found in \citet{frasermckelvie2019b}, who select a sample of MWAs based on M* and B/T along with the presence of bar and spiral features; out of the entire SDSS galaxy sample, \citet{frasermckelvie2019b} find just 176 ($\sim$0.01\%) galaxies that satisfy their criteria.

By exploring a sample based on stellar mass and B/T ratio, we are selecting MW analogs using only a fraction of the full range of possible selection criteria; as such, the selected MWAs can be expected to span a range of values in alternative parameter spaces. We do not assert that these galaxies represent the bulk population of sampled galaxies in the covered mass range, nor do we assert that each individual MWA in our sample is a perfect match to the MW. Rather, we simply make the Copernican assumption that that the MW should not be extraordinary among a set of galaxies selected to span its possible properties. As such, by studying MWAs in MaNGA, we can begin to visualise how the MW's parameter PDFs may be sampled through IFU observations, and how similarity in certain types of galaxy parameter is or is not predictive of similarity in other types of parameters.

\subsection{MWA sample overview}

Our MWA sample comprises 62 structural analogs included within the 6507 MaNGA galaxies in the MPL-8 data release. The data from MPL-8 will be be released in SDSS Data Release 17 (DR17). 

As part of our analysis, we have collated a number of parameters for each galaxy that have been previously published in the literature. We use position angles, ellipticities, NASA-Sloan Atlas (NSA\footnote{ http://www.nsatlas.com}) redshifts and half-light radii from version 2.5.3 of the MaNGA \textit{drpall} file; in all cases, these are obtained from the SDSS elliptical Petrosian apertures. We obtain radial disk scale-lengths and $r$-band B/T ratios from \citet{simard2011} and global SFRs from the MPA-JHU catalogue, in line with the sample selection performed in L15. The properties of the galaxy sample are summarized in \autoref{table1}. We note a variety of IFU sizes, as indicated by galaxies' IFU designation numbers. It is tempting, at first glance, to narrow our focus on galaxies with the largest IFU sizes when considering galaxy properties as functions of position; however, restricting samples to large IFU sizes is expected to bias samples towards blue galaxies of lower surface brightness \citep{wake2017}, and so we made no such restriction when selecting this sample. Thus, as a consequence of the MaNGA sample selection, we include galaxies that are observed with smaller IFU sizes - and so not fully resolved in their centres - in order to avoid this bias.

\begin{table*}
\begin{center}
\begin{tabular}{c|c|c|c|c|c|c|c|c|c}
MaNGA ID & Plate & IFU Dsgn. & RA (deg.) & DEC (deg.) & $\log_{10}(M_*/M_\odot)$  & $(B/T)_r$  & $\log_{10}$SFR ($M_\odot/yr$)&  & z \\[0.3 pt]
\hline
\hline
1-603793 & 8079 & 12705 & 42.811312 & -0.32925214 & 11.1$\pm$0.1 & 0.11 & 1.2$\pm$0.1 & 5.04 & 0.067\\[0.3 pt]
1-604085 & 8085 & 12704 & 52.619896 & 0.81149714 & 10.5$\pm$0.09 & 0.10 & 0.034$\pm$0.3 & 4.60 & 0.031\\[0.3 pt]
1-338796 & 8137 & 12703 & 116.16926 & 44.144409 & 10.7$\pm$0.09 & 0.16 & 0.32$\pm$0.3 & 2.10 & 0.038\\[0.3 pt]
1-338771 & 8141 & 12701 & 116.73264 & 44.265699 & 10.6$\pm$0.09 & 0.20 & 0.39$\pm$0.3 & 5.61 & 0.031\\[0.3 pt]
1-152596 & 8146 & 12702 & 116.83569 & 29.412934 & 11.0$\pm$0.1 & 0.19 & 0.13$\pm$0.6 & 8.14 & 0.063\\[0.3 pt]
1-460912 & 8241 & 3704 & 126.56891 & 17.362452 & 11.0$\pm$0.09 & 0.10 & 1.2$\pm$0.2 & 1.77 & 0.066\\[0.3 pt]
1-47120 & 8244 & 6101 & 130.45491 & 50.786425 & 11.1$\pm$0.09 & 0.18 & 0.88$\pm$0.3 & 8.16 & 0.054\\[0.3 pt]
1-47890 & 8244 & 9101 & 133.78758 & 52.470442 & 11.1$\pm$0.09 & 0.14 & -0.31$\pm$0.6 & 5.09 & 0.058\\[0.3 pt]
1-277552 & 8257 & 12705 & 167.03456 & 45.984624 & 10.7$\pm$0.09 & 0.14 & -0.11$\pm$0.5 & 6.17 & 0.036\\[0.3 pt]
1-282762 & 8263 & 6104 & 187.28568 & 46.881492 & 10.8$\pm$0.10 & 0.08 & 0.17$\pm$0.4 & 2.49 & 0.040\\[0.3 pt]
1-322787 & 8315 & 12705 & 235.92049 & 39.540357 & 10.6$\pm$0.1 & 0.24 & 0.85$\pm$0.2 & 7.22 & 0.063\\[0.3 pt]
1-422040 & 8444 & 12703 & 201.41111 & 33.680433 & 11.0$\pm$0.09 & 0.19 & 0.88$\pm$0.5 & 4.24 & 0.039\\[0.3 pt]
1-134239 & 8549 & 3703 & 241.41644 & 46.846561 & 10.8$\pm$0.10 & 0.23 & -0.66$\pm$0.7 & 6.06 & 0.057\\[0.3 pt]
1-351566 & 8567 & 3701 & 118.10651 & 47.777979 & 10.5$\pm$0.10 & 0.16 & -0.01$\pm$0.3 & 5.00 & 0.031\\[0.3 pt]
1-274323 & 8568 & 6103 & 155.37679 & 38.308947 & 10.9$\pm$0.10 & 0.04 & 0.074$\pm$0.3 & 5.99 & 0.055\\[0.3 pt]
1-197702 & 8595 & 3702 & 219.04276 & 50.518822 & 11.0$\pm$0.10 & 0.19 & -0.20$\pm$0.7 & 5.36 & 0.069\\[0.3 pt]
1-135545 & 8601 & 6103 & 247.53037 & 40.880157 & 10.4$\pm$0.09 & 0.14 & 0.12$\pm$0.4 & 3.26 & 0.030\\[0.3 pt]
1-136125 & 8613 & 12702 & 254.04414 & 34.836520 & 10.5$\pm$0.09 & 0.18 & -0.24$\pm$0.5 & 4.28 & 0.032\\[0.3 pt]
1-177061 & 8614 & 6101 & 256.75708 & 34.389276 & 10.8$\pm$0.08 & 0.06 & 0.88$\pm$0.2 & 5.38 & 0.071\\[0.3 pt]
1-180039 & 8615 & 9102 & 321.47404 & 0.41605877 & 10.5$\pm$0.09 & 0.09 & 0.34$\pm$0.3 & 2.63 & 0.032\\[0.3 pt]
1-403703 & 8934 & 12701 & 196.16279 & 28.628904 & 10.6$\pm$0.09 & 0.19 & 0.49$\pm$0.3 & 4.79 & 0.057\\[0.3 pt]
1-174618 & 8947 & 12703 & 172.88669 & 49.857504 & 11.1$\pm$0.09 & 0.12 & 0.069$\pm$0.6 & 9.64 & 0.069\\[0.3 pt]
1-135626 & 8978 & 9102 & 248.51518 & 41.240258 & 10.8$\pm$0.09 & 0.15 & -0.75$\pm$0.7 & 4.38 & 0.061\\[0.3 pt]
1-248420 & 8979 & 6102 & 241.82339 & 41.403604 & 10.8$\pm$0.09 & 0.10 & 0.024$\pm$0.4 & 5.48 & 0.035\\[0.3 pt]
1-248405 & 8979 & 9101 & 241.27859 & 42.037358 & 11.0$\pm$0.09 & 0.16 & 0.084$\pm$0.6 & 5.57 & 0.077\\[0.3 pt]
1-416039 & 8985 & 9102 & 205.87340 & 31.003218 & 11.0$\pm$0.10 & 0.23 & 0.32$\pm$0.5 & 5.59 & 0.065\\[0.3 pt]
1-296824 & 9028 & 9101 & 243.02559 & 28.498866 & 10.9$\pm$0.09 & 0.15 & 0.86$\pm$0.2 & 8.18 & 0.053\\[0.3 pt]
1-134620 & 9031 & 12703 & 242.81922 & 45.121318 & 10.6$\pm$0.10 & 0.22 & -0.78$\pm$0.6 & 7.22 & 0.055\\[0.3 pt]
1-296659 & 9040 & 6102 & 243.41436 & 27.592856 & 10.8$\pm$0.10 & 0.05 & 0.21$\pm$0.4 & 4.84 & 0.066\\[0.3 pt]
1-296537 & 9040 & 9102 & 245.47816 & 26.600483 & 10.9$\pm$0.09 & 0.14 & -0.09$\pm$0.5 & 1.86 & 0.064\\[0.3 pt]
1-318334 & 9095 & 9102 & 243.08489 & 23.002020 & 10.6$\pm$0.09 & 0.10 & 0.43$\pm$0.2 & 4.53 & 0.032\\[0.3 pt]
1-51296 & 9189 & 9101 & 50.416587 & -6.6356147 & 10.6$\pm$0.09 & 0.19 & 0.33$\pm$0.3 & 3.17 & 0.035\\[0.3 pt]
1-37601 & 9192 & 12704 & 46.285743 & 0.34344304 & 10.8$\pm$0.09 & 0.20 & 0.65$\pm$0.3 & 4.11 & 0.044\\[0.3 pt]
1-24822 & 9196 & 6104 & 262.31772 & 54.205924 & 10.9$\pm$0.09 & 0.21 & 0.99$\pm$0.2 & 7.45 & 0.079\\[0.3 pt]
1-24814 & 9196 & 9101 & 262.65063 & 54.101229 & 10.8$\pm$0.1 & 0.07 & -0.08$\pm$0.5 & 4.07 & 0.062\\[0.3 pt]
1-122002 & 9485 & 1901 & 120.77803 & 37.023557 & 10.7$\pm$0.09 & 0.24 & 1.2$\pm$0.09 & 3.42 & 0.071\\[0.3 pt]
1-72207 & 9486 & 12702 & 121.12960 & 40.206045 & 10.8$\pm$0.09 & 0.10 & 0.83$\pm$0.3 & 4.94 & 0.040\\[0.3 pt]
1-44618 & 9487 & 12701 & 122.52287 & 46.193025 & 10.7$\pm$0.09 & 0.16 & 0.12$\pm$0.3 & 6.25 & 0.032\\[0.3 pt]
1-382809 & 9491 & 12704 & 119.53574 & 19.550146 & 10.6$\pm$0.09 & 0.13 & 0.80$\pm$0.2 & 5.75 & 0.062\\[0.3 pt]
1-298298 & 9496 & 9102 & 120.53836 & 21.252763 & 10.6$\pm$0.08 & 0.17 & -0.91$\pm$0.6 & 2.81 & 0.030\\[0.3 pt]
1-145679 & 9499 & 12703 & 118.42323 & 26.492699 & 10.8$\pm$0.09 & 0.13 & 0.00$\pm$0.5 & 3.24 & 0.037\\[0.3 pt]
1-605531 & 9506 & 3701 & 133.56990 & 27.266526 & 10.9$\pm$0.08 & 0.19 & 1.2$\pm$0.2 & 3.82 & 0.064\\[0.3 pt]
1-41746 & 9514 & 9101 & 31.682736 & 13.371629 & 10.7$\pm$0.10 & 0.25 & 0.24$\pm$0.4 & 6.64 & 0.061\\[0.3 pt]
1-271967 & 9866 & 12701 & 242.35067 & 33.000246 & 10.9$\pm$0.1 & 0.17 & 0.26$\pm$0.4 & 3.37 & 0.053\\[0.3 pt]
1-321271 & 9868 & 12705 & 218.87457 & 46.202391 & 10.9$\pm$0.09 & 0.17 & 0.69$\pm$0.3 & 6.43 & 0.073\\[0.3 pt]
1-294374 & 9894 & 6104 & 250.90783 & 21.691287 & 10.6$\pm$0.09 & 0.13 & 0.52$\pm$0.2 & 3.20 & 0.035\\[0.3 pt]
1-122611 & 10213 & 1902 & 123.45312 & 41.649479 & 10.9$\pm$0.1 & 0.17 & 0.39$\pm$0.5 & 9.34 & 0.049\\[0.3 pt]
1-153613 & 10215 & 9102 & 122.39068 & 36.985267 & 10.4$\pm$0.09 & 0.13 & -0.05$\pm$0.5 & 5.28 & 0.041\\[0.3 pt]
1-153693 & 10215 & 12705 & 122.98448 & 36.424268 & 10.7$\pm$0.10 & 0.07 & 0.27$\pm$0.4 & 3.28 & 0.074\\[0.3 pt]
1-297871 & 10216 & 6102 & 116.73020 & 18.482376 & 10.8$\pm$0.09 & 0.09 & 0.26$\pm$0.3 & 9.52 & 0.044\\[0.3 pt]
1-297854 & 10216 & 6104 & 117.43277 & 18.918596 & 10.9$\pm$0.1 & 0.18 & 0.39$\pm$0.3 & 3.96 & 0.048\\[0.3 pt]
1-458727 & 10217 & 3703 & 119.29321 & 14.658764 & 10.8$\pm$0.09 & 0.21 & 0.72$\pm$0.3 & 3.62 & 0.048\\[0.3 pt]
1-412119 & 10218 & 12701 & 117.85779 & 16.601841 & 10.7$\pm$0.09 & 0.12 & 0.33$\pm$0.3 & 3.32 & 0.054\\[0.3 pt]
1-202008 & 10220 & 6104 & 121.71814 & 32.250329 & 10.7$\pm$0.10 & 0.18 & 0.27$\pm$0.2 & 2.87 & 0.040\\[0.3 pt]
1-153142 & 10220 & 12702 & 120.48960 & 33.164591 & 10.9$\pm$0.09 & 0.07 & -0.24$\pm$0.7 & 3.60 & 0.038\\[0.3 pt]
1-201934 & 10220 & 12705 & 120.55032 & 32.315051 & 10.8$\pm$0.09 & 0.10 & 0.47$\pm$0.3 & 3.14 & 0.037\\[0.3 pt]
1-352102 & 10493 & 3701 & 122.85428 & 52.530381 & 10.8$\pm$0.09 & 0.21 & 0.66$\pm$0.3 & 5.08 & 0.071\\[0.3 pt]
1-352513 & 10494 & 6101 & 123.25528 & 54.521178 & 10.7$\pm$0.09 & 0.19 & 0.29$\pm$0.3 & 5.29 & 0.041\\[0.3 pt]
1-383135 & 10497 & 6101 & 119.99119 & 17.847843 & 11.0$\pm$0.09 & 0.12 & 0.85$\pm$0.3 & 4.07 & 0.062\\[0.3 pt]
1-385544 & 10498 & 3701 & 130.71740 & 25.642697 & 10.8$\pm$0.09 & 0.14 & 0.32$\pm$0.3 & 3.34 & 0.054\\[0.3 pt]
1-278191 & 10509 & 3704 & 167.56430 & 46.441525 & 10.7$\pm$0.09 & 0.13 & 0.78$\pm$0.1 & 9.39 & 0.036\\[0.3 pt]
1-78856 & 10517 & 12703 & 151.39912 & 4.2794104 & 11.0$\pm$0.09 & 0.11 & 0.72$\pm$0.4 & 3.94 & 0.063\\[0.3 pt]
\end{tabular}
\end{center}
\caption{Summary of MWA sample parameters. Columns from left to right: MaNGA ID number, SDSS plate number, IFU designation, MPA-JHU stellar mass, MPA-JHU star formation rate, bulge-to-total ratio in the r-band, disk scale-length, NSA heliocentric redshift. Bulge-to-total ratios and disk scale lengths are from \citet{simard2011}. SFRs are in units of $M_\odot$/yr.}
\label{table1}
\end{table*}

\begin{figure}
\begin{center}
	\includegraphics[trim = 9cm 2cm 10cm 12cm,scale=0.5]{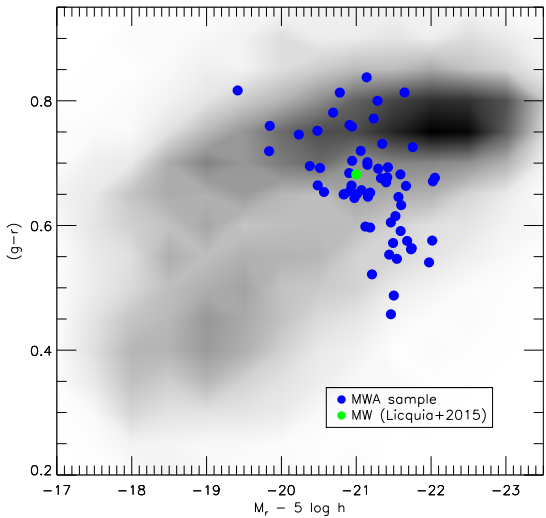}
	\caption{Color--magnitude diagram of the MWA sample, with a density map for all galaxies in MaNGA MPL-8 with available color--magnitude data as the background greyscale. Blue circles indicate MWAs, and the green point shows the best-fitting MW position from L15.
	} 
	\label{sample1}
	\end{center}
\end{figure}

We plot the color--magnitude distribution of the MWA sample in \autoref{sample1}, along with that of the full MaNGA MPL-8 dataset and the L15 MW values. Our MWA sample comprises objects of intermediate color and magnitude, with significant scatter. This distribution is similar to what was shown for the star formation analogs of L15. We obtain both colors and magnitudes for the MaNGA galaxies from version 2.5.3 of the MaNGA \textit{drpall} file, using values from Sersic fits to SDSS imaging.

\begin{figure}
\begin{center}
	\includegraphics[trim = 5cm 9cm 4cm 3cm,scale=0.5]{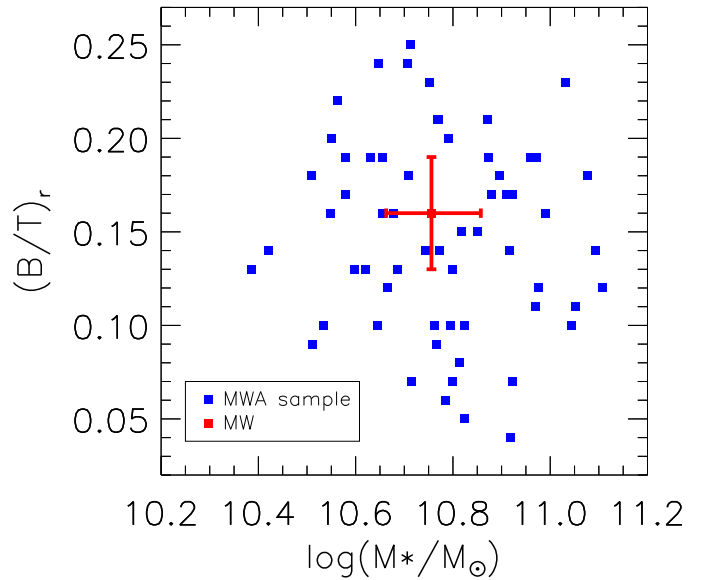}
	\caption{$R$-band bulge-to-total (B/T) ratio plotted against galaxy stellar mass. The Milky Way mass and B/T values from \citet{licquia2015} are also shown in red.}

	\label{sample2}
	\end{center}
\end{figure}

In \autoref{sample2}, we present our MWA sample's distribution of stellar mass and B/T, with the MW values also shown for comparison \citep{licquia2015}. We see that the MWA sample centres around the MW values as expected, though with non-negligible levels of scatter. This scatter is due in part to the analog sample being designed to sample the PDF of the MW, rather than solely being selected around the current best-fitting MW values. In addition we note that many of the currently available MWAs are serendipitously-observed galaxies from the main MaNGA sample rather than being high-priority MWA targets; such galaxies were independently selected for inclusion in the target list for the MWA MaNGA ancillary program, but were subsequently observed instead as members of the main MaNGA galaxy sample. We will further explore the MWAs' similarity to the MW in M-B/T space and also in other parameter spaces in \autoref{paramcomparison}

\subsection{MaNGA Data Analysis pipeline}

We employ the unbinned per-spaxel emission line measurements from the MaNGA data analysis pipeline \citep[DAP;][]{westfall2019, belfiore2019}. We retrieved DAP cubes using the Marvin interface \citep{cherinka2018} to obtain equivalent width (EW) and flux measurements for the following emission features, calculated via Gaussian emission line fits: H$\alpha$, H$\beta$, [O~\textsc{III}]$_{4959,5007}$,  [N~\textsc{II}]$_{6548,6583}$ and [S~\textsc{II}]$_{6716,6731}$. We also extract model emission line spectra for each individual spaxel.

\begin{figure}
\begin{center}
	\includegraphics[trim = 9cm 11cm 10cm 3cm,scale=0.5]{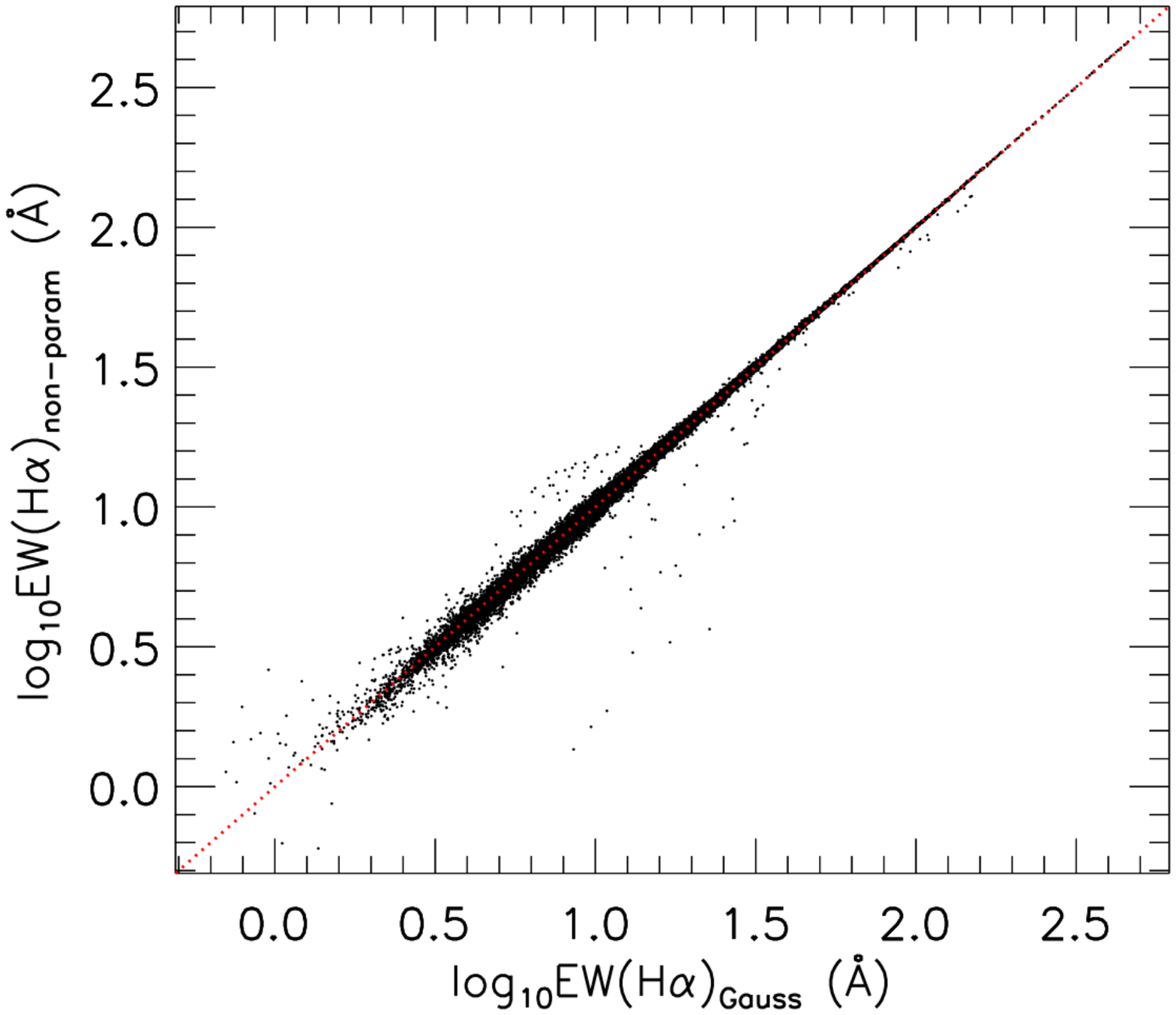}
	\caption{Comparison of H$\alpha$ EWs measured via Gaussian fits to emission line features vs those obtained non-parametrically. We show all spaxels for which the EWs are greater than 0.5~\AA\ and the S/N greater than 4;
	these show a 1--1 relation with small scatter.}
	\label{gewsew}
	\end{center}
\end{figure}

To assess the impact of using Gaussian fit measurements, we also extracted H$\alpha$ EWs from the DAP as obtained from non-parametric emission line flux measurements. The non-parametric flux measurements and subsequent EW measurements are described in Section 9.1 of \citet{westfall2019}. In brief, spaxel spectra are continuum-subtracted and flux measurements are then performed by summing the resulting emission spectra over all wavelength ranges of interest, after subtracting a linear baseline from each emission feature. We selected all spaxels with S/N values greater than 4, and then further selected all of those spaxels for which both H$\alpha$ EW values were greater than 0.5~Angstrom. We compare the resulting two sets of values in \autoref{gewsew}. Outside of a small number of outliers, we find the values to follow a 1--1 relation with little scatter. From this, we conclude that the choice of Gaussian over direct flux measurements does not impact the analysis.

\section{Spectral modelling}\label{specmod}
In this section we describe our analysis of the MaNGA spectra, carried out using the Penalised Pixel Fitting (pPXF) method of \citet{cappellari2004}, including the upgrade described in \citet{cappellari2017}. The pPXF routine fits an observed galaxy spectrum $G(x)$ with an optimised template $G_{mod}(x)$. This fit is carried out directly in pixel-space (as opposed to using Fourier-based techniques), with both $G(x)$ and $G_{mod}(x)$ logarithmically binned in wavelength. 

For the remainder of our analysis, we restrict our galaxy spectra to a rest-frame wavelength range of 3900--6800 Angstroms, based on galaxies' NSA heliocentric redshifts. This wavelength range is chosen to best match the range of the ELODIE spectral library of observed stars \citep{prugniel2001}, which we employ in our measurement of stellar kinematics (\autoref{kin}).

Before applying pPXF, we subtracted the DAP emission line models from the MaNGA datacubes, and then spatially binned spectra in the galaxies' outer regions in order to improve the signal-to-noise ratio (S/N). We carried out this binning using the Voronoi binning method of \citet{cappellari2003}. We used a target S/N of 20 when possible, though for one galaxy (9196-9101) we found it necessary to lower this target to 15 to prevent excessive binning. We obtained S/N values for each spaxel from the MaNGA DAP, and we approximated the covariance between neighbouring MaNGA spaxels using Equation 9 of \citet{law2016}. We restrict our binning (and subsequent analysis of binned properties) to spaxels with S/N values of 4 or greater, of which 11\% have S/N values greater than 20 before binning.  Overall, our binning target is aimed at improving the S/N for measuring stellar populations while preserving spatial information as much as possible. For the remainder of this paper, we refer to all spectra in the Voronoi-binned datacubes as ``binned'' spectra, including spectra from unbinned spaxels in these cubes. Likewise, we will refer to all Voronoi regions in a given datacube as ``bins'', including those regions which consist of a single (unbinned) spaxel.

We obtain an average of 215 Voronoi bins for each galaxy, with a standard deviation of 154. In turn, we achieve a mean S/N of 19.4 with a standard deviation of 5.8. We note a minority of bins to fall significantly below our desired S/N target, with 13.7\%\ of bins yielding S/N values less than 15. Such bins are typically located in the outer regions of our sample galaxies.

\subsection{Stellar Kinematics}\label{kin}

We extracted stellar kinematics by using pPXF to fit ELODIE medium-resolution stellar spectra \citep{prugniel2001} to our galaxy spectra. We chose to use ELODIE for stellar kinematics due to its significantly higher spectral resolution ($R = 10,000$) when compared to the MaNGA data. We broadened the ELODIE spectra to match the median MaNGA resolution FWHM for each galaxy separately over the wavelength range of interest. We excluded from this process any ELODIE templates identified as emission-line stars. These templates were identified by measuring the flux over a 16~\AA\ region centred around H$\alpha$ and then measuring the flux of a region defined by drawing a straight line between the pixels at the edges of the 16~\AA\ region; for stars with an H$\alpha$ absorption feature, the former flux will be lower, while for stars with an emission feature it will be higher. We exluded templates for which the former flux was more than 1\% higher than the latter flux, resulting in a library of 1876 empirical ELODIE templates.

We fitted each binned spectrum for the stellar line-of-sight velocity and velocity dispersion. For our implementation, we also allow for a 6th-order multiplicative Legendre polynomial. As such, the model spectrum $G_{mod}(x)$ takes the form 

\begin{equation}\label{ppxf2}
G_{mod}(x) = \sum_{k=1}^K w_k[\lagr (cx)*T_k(x)] \times \sum_{l=1}^L b_l\mathcal{P}_l(x),
\end{equation}  

\noindent
where spectra have been rebinned such that $x = \ln{\lambda}$. The broadening function is $\lagr(cx)$, $T$ is a set of distinct stellar templates, and $w_k$ is the array of optimal weights of those templates, with $*$ describing convolution. $\mathcal{P}_l(x)$ are Legendre polynomials of order $l$, with $b_l$ the corresponding weights, that allow pPXF to correct any low-frequency differences between model and data.

We begin our implementation by performing an initial pPXF fit to the center-most spectrum of each galaxy, for a systemic velocity estimate to use as a starting guess for all subsequent pPXF calls. We mask the 5577~\AA\ and 5899~\AA\ sky emission lines from this fit, along with the NaID region. We perform this masking for all subsequent fits for stellar kinematics. We have chosen to mask the NaID feature due to its sensitivity to dust extinction \citep[e.g.][]{richmond1994,poznanski2012}, which we expected to be significant for the more edge-on galaxies in our sample. Our pPXF implementation cannot account for such an effect, leading to high residuals around this region in cases where the dust extinction is high. 

We pre-select ELODIE spectra for each galaxy by performing an additional pair of pPXF fits --- to the center-most spectrum for that galaxy and to the total integrated galaxy spectrum --- and then selecting the ELODIE templates given non-zero weight in either of the two fits. An alternative approach would be to use the full ELODIE library to fit each binned galaxy spectrum individually; we elect not to do this due to such a method being computationally prohibitive. 

While running pPXF for all binned galaxy spectra,
we determine random measurement errors by adding Gaussian noise to the spectra and rerunning the fits over 100 Monte-Carlo resimulations with zero bias; such a technique is commonly used for underdstanding uncertainty in spectral fitting, in the case of pPXF along with other spectral fitting codes \citep[e.g.][]{cidfernandes2014,garciabenito2015,sanchez2016b}. In each case, we recorded measurement errors as the standard deviation of the parameters obtained from the 100 resimulations; we obtain the standard deviation in this case and in all subsequent cases using the \textit{robust\_sigma} IDL routine. In each case, we set the noise level for these resimulations as follows: We calculate the standard deviation of the residuals for the initial pPXF to that spectrum, and we divide the median flux of the spectrum though by this value to estimate a ``fit S/N'' \citep[similarly to, e.g.][]{krajnovic2015}, hereafter $\rm (S/N)_{fit}$. For each simulation, the pixels in the spectrum are then perturbed by a number drawn from a random normal distribution with width equal to the flux of the pixel divided by $\rm (S/N)_{fit}$.    

We find median overall error values of 3.7~km~s$^{-1}$ and 7.9~km~s$^{-1}$ for velocity and velocity dispersion, respectively. By design, the calculated errors depend significantly on the achieved $\rm (S/N)_{fit}$. For binned spectra with $\rm (S/N)_{fit} > 20$, we find median error values of 3.4~km~s$^{-1}$ and 5.3~km~s$^{-1}$ for velocity and velocity dispersion, respectively, while for spectra with $\rm (S/N)_{fit} < 20$ we find median velocity and dispersion error values of 5.5~km~s$^{-1}$ and 10.6~km~s$^{-1}$. We note here that the calculated velocity dispersions in our galaxies' outskirts are frequently well below the MaNGA instrumental resolution of $\sim$77~km~s$^{-1}$, which will affect the accuracy of derived dispersions; this problem is discussed in the context of MaNGA data in \citet{penny2016} and \citet{westfall2019}.

In a future paper, we will employ dynamical models to further study the kinematics of the MWA sample and their dynamical mass contents.

\subsection{Stellar populations}\label{pop}

We next employ pPXF to measure the ages, metallicities ([Z/H]), and $r$-band mass-to-light ratios ($M_*/L_r$) of the MWAs' stellar populations. We note that our chosen wavelength range includes a number of absorption features that are individually sensitive to stellar metallicity and stellar age, and so we expect the impact of the age-metallicity degeneracy on our fitting to be limited so long as a good fit is achieved to a given spectrum. Our pPXF fits allow for a \citet{calzetti2000} gas reddening law, with the amount of reddening fitted along with the stellar templates. We do not include regularisation in these fits.

We use for templates here the E-MILES library of simple stellar population (SSP) models \citep{vazdekis2016}. We use base-Fe models with BaSTI isochrones \citep{hidalgo2018} and a revised \citet{kroupa2001} initial mass function (IMF), which takes into account the effects of unresolved binaries. We select SSP models to span a wide grid in age and metallicity, with approximately logarithmic spacing in terms of both of these parameters. We use model ages of 0.03, 0.05, 0.1, 0.2, 0.3, 0.5, 0.7, 1, 1.5, 2.25, 3.25, 4.5, 6.5, 9.5, and 14 Gyr. We use model metallicity [Z/H] values of $-1.49$, $-1.26$, $-0.96$, $-0.66$, $-0.35$, $+0.06$, and $+0.26$; this metallicity range is chosen to reside entirely within the MILES SAFE ranges. We assume that $\rm [Z/H] = [Fe/H]$ for these models; however, given that input stars for these models follow the MW abundance distribution, this assumption is only strictly true at high metallicities \citep[][Table 6]{schiavon2007}\footnote{\url{http://www.iac.es/proyecto/miles/pages/ssp-models/name-convention.php}}.

For these pPXF fits, we again masked the 5577 and 5899 Angstrom sky emission regions, as well as the NaID absorption region. We broadened the model spectra to match the MaNGA resolution in the same manner as the ELODIE spectra (\autoref{kin}). To assess the impact of fitted stellar kinematics on derived population parameters, and in turn to assess the importance of the metallicity-dispersion degeneracy \citep[e.g.][]{koleva2008,sb2011} on our results, we performed two sets of pPXF fits: in one set, we fix the stellar velocity dispersions to those derived by ELODIE, while in the other set we leave the kinematics as free parameters.

We calculated light-weighted values for the log(age), metallicity, and mass-to-light ratio using

\begin{equation}\label{popeq2}
X_{\rm light} = \frac{\sum_{k=1}^K w_k\times x_k \times F_k}{\sum_{k=1}^K w_k \times F_k},
\end{equation}  
\noindent 
where $x$ represents the parameter value of a given template, $X$ the final weighted value of that parameter, $w_k$ the template weights and $F_k$ the mean flux of a given template. Likewise, for mass-weighted ages and metallicities we use 

\begin{equation}\label{popeq3}
X_{\rm mass} = \frac{\sum_{k=1}^K w_k\times x_k}{\sum_{k=1}^K w_k},
\end{equation}  
\noindent
where all symbols are as in \autoref{popeq2}. We derive errors on our stellar population values by performing 100 Monte-Carlo resimulations with added Gaussian noise, using zero regularisation. We note that our $M_*/L_r$ values are pure population estimates, and so do not take the effects of dust-reddening into account. We have elected to use logarithmic ages over linear ages due to the approximately logarithmic spacing of the SSP model grid in age.

\begin{figure*}
\begin{center}
	\includegraphics[trim = 1cm 1cm 0cm 15cm,scale=0.95,clip = true]{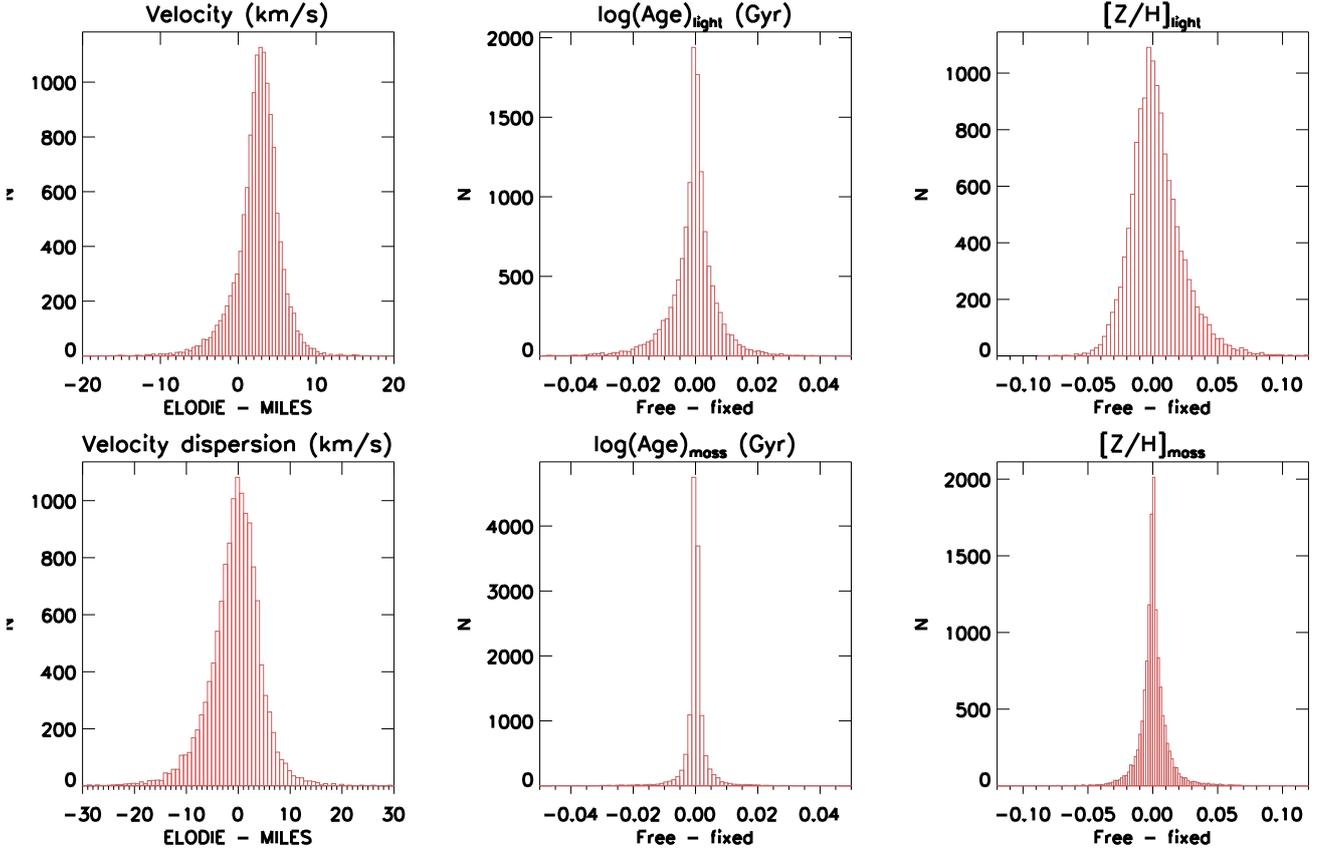}
	\caption{Comparison of ELODIE-derived kinematics with those derived from E-MILES fits (left two windows), along with comparisons of light-weighted and mass weighted ages (middle two windows) and light-weights and mass-weighted metallicities (right two windows), for the cases with E-MILES kinematics freely fit and with E-MILES stellar velocity dispersions fixed to ELODIE values. We find that the choice of fitting method makes has only a small effect on output stellar population parameters, though we note the velocity dispersion from E-MILES to be higher on average.}
	\label{fixnofix}
	\end{center}
\end{figure*}

In \autoref{fixnofix}, we compare the population parameters extracted from the two sets of pPXF population fits. We also compare the SSP-derived kinematics with those calculated from fits with ELODIE, in the case where the former is fitted freely. We find that population parameters from the two methods to have very similar values overall, though with a slight tendancy towards higher velocity dispersions when the E-MILES kinematics are fit freely along with a small velocity offset of approximately 3 km/s. For the remainder of our analysis, we will refer to our ELODIE calculations for kinematics, and we will refer to E-MILES fits with kinematics fixed to the ELODIE stellar velocity dispersions for the purpose of calculating stellar population properties.

\subsection{Methods summary}

\begin{figure*}
\begin{center}
	\includegraphics[trim = 1cm 4cm 0cm 10cm,scale=1]{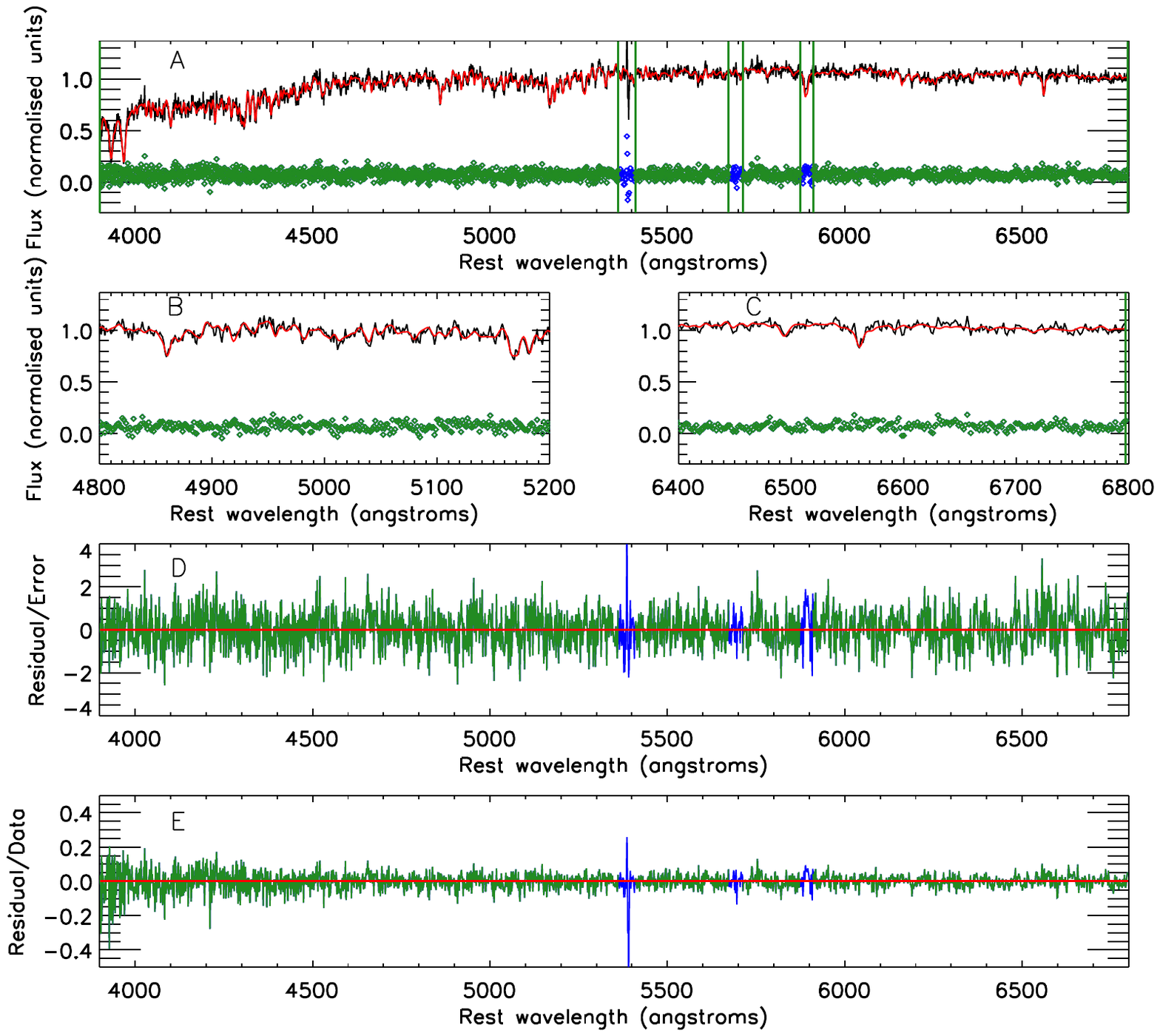}
	\caption{Panel A: example pPXF fit to a gas-cleaned spectrum from galaxy 8979-6102. The black line shows the observed data, and the red line the fitted ppxf model. Vertical green lines indicate masked regions, with the green and blue points indicating the fit and masked residuals. Panel B: zoom-in on the region around the H$\beta$ absorption feature. Panel C: zoom-in around the H$\alpha$ absorption feature. Panel D: residuals of the pPXF fit, plotted relative to the size of the corresponding error spectrum. Panel E: residuals of the pPXF fit, divided by the input galaxy spectrum.
	}
	\label{ppxfexample}
\end{center}
\end{figure*}

In summary, we employ the pPXF fitting procedure in order to measure the following parameters of interest over a 2D FOV for each galaxy: stellar velocities, stellar velocity dispersions, stellar population ages, stellar population metallicities, and $r$-band mass-to-light ratios. For stellar ages and metallicities, we obtain both mass-weighted and light-weighted quantities. In \autoref{ppxfexample}, we show an example pPXF fit to a gas-cleaned spectrum from galaxy 8979-6102.

\begin{figure*}
\begin{center}
	\includegraphics[trim = 2cm 10cm 0cm 0cm,scale=0.75]{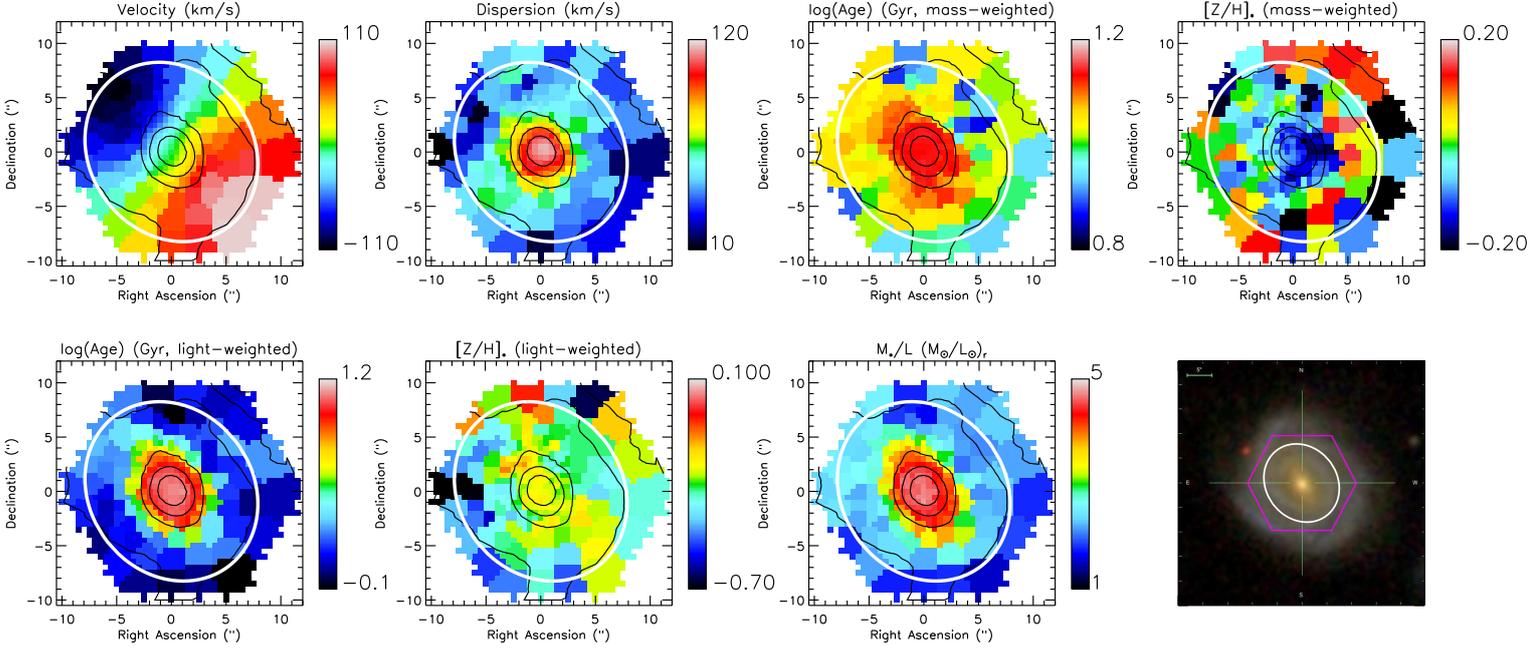}
	\caption{pPXF-derived maps for galaxy 8979-6102, along with an SDSS thumbnail with the MaNGA FOV overlaid (magenta hexagon). The black contours show mean r-band flux, with the white ellipses indicating one half-light radius.}
	\label{ppxfmaps}
	\end{center}
\end{figure*}

In \autoref{ppxfmaps}, we summarize the measurements described thus far with parameter maps for this galaxy, along with an optical SDSS image for reference.

\section{Results}\label{results}

In this section, we present results for our MWA sample in terms of stellar kinematics (\autoref{results_kin}), stellar populations (\autoref{results_pops}), and ionised gas (\autoref{results_iongas}). 

\subsection{Stellar kinematics results} \label{results_kin}

We use our pPXF-derived stellar kinematics to construct 1-dimensional spin profiles for each galaxy in our sample, using the spin parameter $\lambda_R$ \citep{emsellem2007}. We note that $\lambda_R$ could also be derived from DAP data products or else from the MaNGA Pipe3D VAC \citep{sanchez2018}, but for consistency with our stellar population results we report it here as derived from our  spectral fitting procedure. In the case of two-dimensional spectroscopy, $\lambda_R$ can be written as

\begin{equation}\label{lit1}
\lambda_R = \frac{\langle \sum_{i=1}^{N_P} F_i R_i |V_i|\rangle}{\langle \sum_{i=1}^{N_P} F_i R_i \sqrt{V_i^2 + \sigma_i^2}\rangle}
\end{equation}
\noindent
where $R_i$ represents the circular radius of a data point, $F_i$ the flux, $V_i$ the stellar line-of-sight velocity, and $\sigma_i$ the velocity dispersion. $N_p$ represents total number of datapoints within a given ellipse.

Before applying \autoref{lit1}, we correct our derived velocity maps for their systemic velocity components. We perform this correction by first subtracting the mean stellar velocity from each map and then applying the FIT\_KINEMATIC\_PA IDL routine\footnote{\url{https://www-astro.physics.ox.ac.uk/~mxc/software/}}, which implements the method described in Appendix C of \citet{kraj2006} and calculates a systemic velocity correction as part of fitting for the kinematic position angle. We then project onto each spaxel the kinematics values of its corresponding Voronoi bin.

We apply \autoref{lit1} over a series of elliptical apertures, setting the aperture position angles equal to the kinematic position angle for each galaxy, with the aperture axis ratio $b/a$ set equal to the SDSS elliptical Petrosian aperture axis ratio for a given galaxy. We calculate $\lambda_R$ over successively larger apertures, with the maximum aperture size chosen to be the largest value for which the aperture is at least 85\%\ filled with spectra.

\begin{figure}
\begin{center}
	\includegraphics[trim = 1cm 8cm 0cm 5cm,scale=0.5]{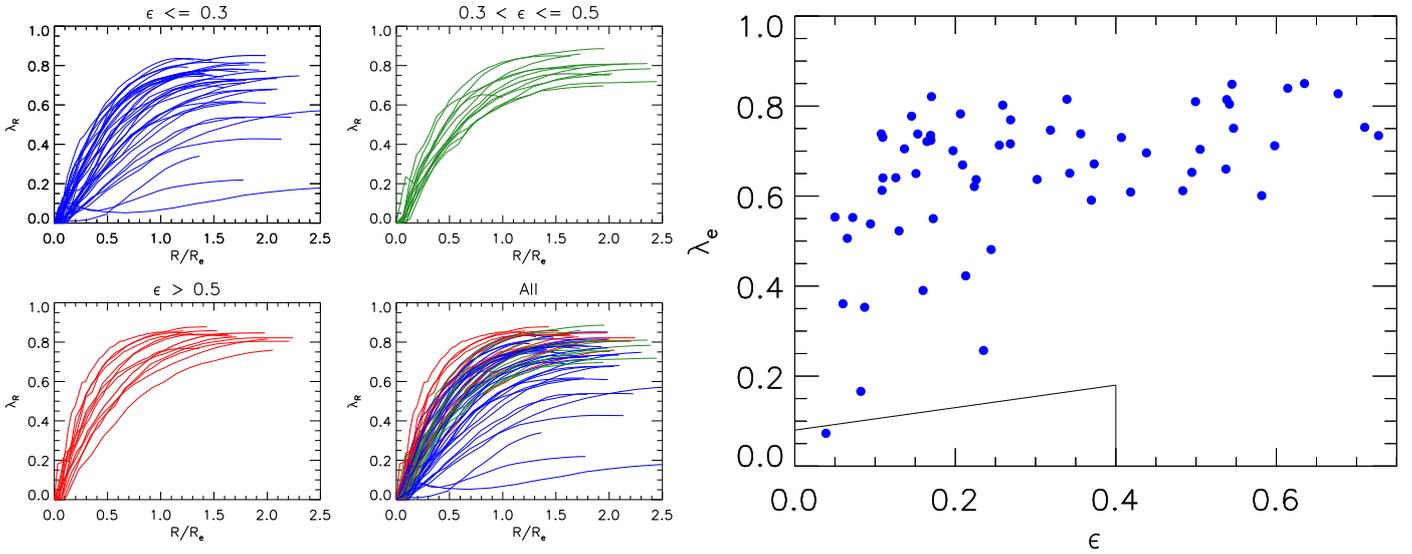}
	\caption{Profiles of spin parameter $\lambda_R$ for the MWA sample, colored according to axis ratio $\epsilon= 1- b/a$. Higher values of $\lambda_R$ indicate a greater degree of rotational support. We find the MWA profiles generally increase with radius over the tested FOV. The lowest blue line corresponds to galaxy 9485-1901; this line extends to 3.4~$R_e$ and rises throughout.}
	\label{lamrprofs}
\end{center}
\end{figure}

We present $\lambda_R$ profiles for our sample in \autoref{lamrprofs}, wherein we separate the MWAs in terms of ellipticity $\epsilon= 1-b/a$. Although the maximum values of the profiles show the expected dependence with ellipticity, the qualitative behaviour of the profiles is similar: for the vast majority of the MWAs, we find the $\lambda_R$ profiles to rise throughout the studied FOV. We note that we have not corrected our spin parameter values for seeing in this case, which was done for MaNGA galaxies in \citet{graham2018}. Such a correction would increase the measured spin parameter values, though with significant corresponding uncertainties.

\begin{figure}
\begin{center}
	\includegraphics[trim = 2cm 1cm 0cm 14cm,scale=0.55]{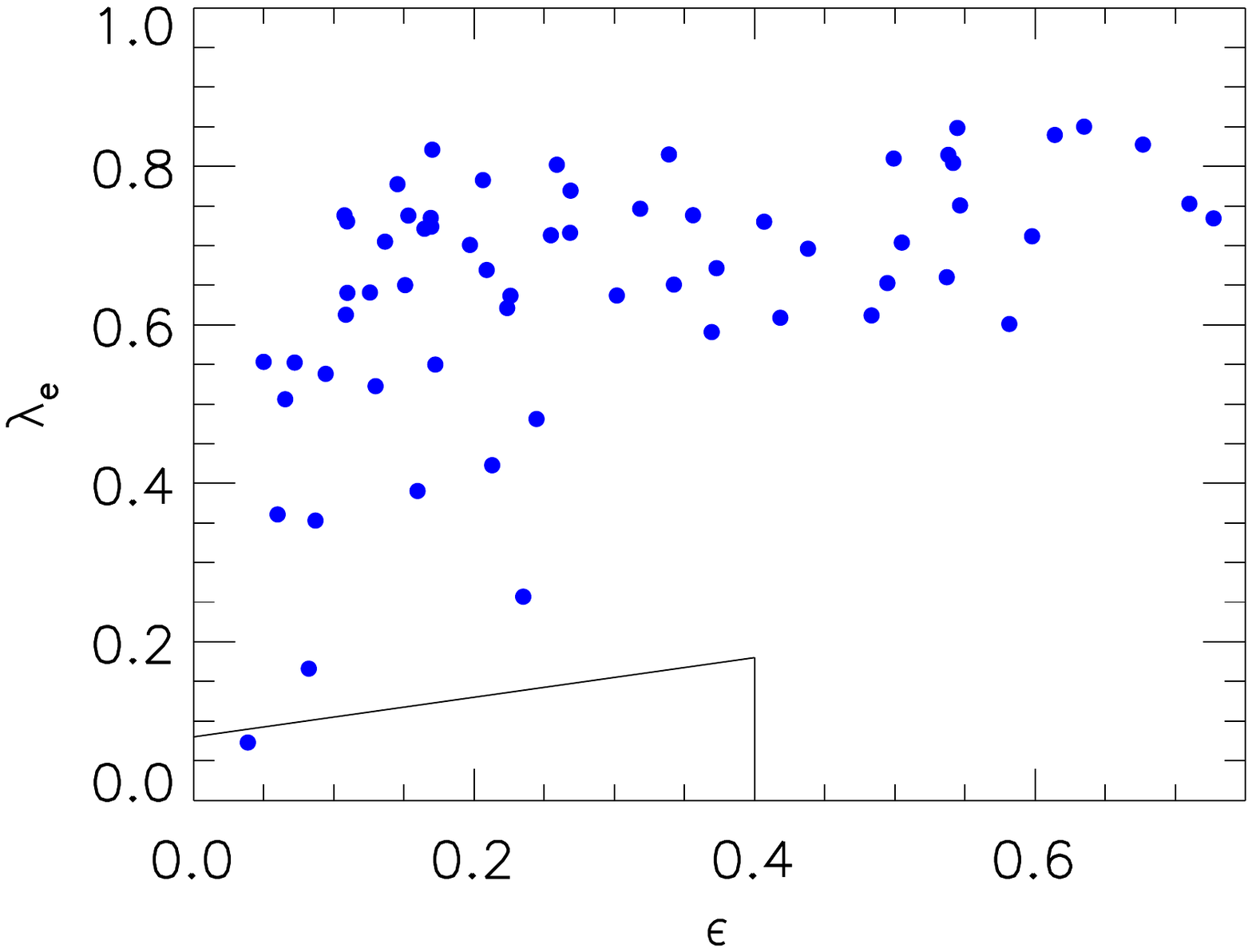}
	\caption{Spin parameter $\lambda_R$ values at $1R_e$, compared to galaxy ellipticity.  The black line shows the fast/slow rotator dividing line of \citet{cappellari2016} Our sample is dominated by galaxies with significant rotation, as expected given the sample selection criteria.}
	\label{lamres}
\end{center}
\end{figure}

We compare the spin at 1~$R_e$, $\lambda_R(R_e)$ (hereafter $\lambda_e$), with galaxy ellipticity in \autoref{lamres}. We overplot the fast/slow rotator dividing line suggested in \citet{cappellari2016}. We find our MWA sample is overwhelmingly comprised of  rotation-supported systems, with only one galaxy (9485-1901) falling within the slow rotator region. Such a finding is a natural consequence of the selection criteria behind this MWA sample, which prioritises galaxies with comparitively low bulge-to-total ratios and so in turn favours the inclusion of rotating disk galaxies. 

\subsection{Stellar populations results} \label{results_pops}

We use our pPXF stellar population maps to construct 1-dimensional stellar populations profiles. 

We replicate all Voronoi bin parameter values onto individual unbinned spaxels. We place a series of elliptical annuli over our MaNGA maps, with each annulus boundary spaced by $0.1 R_e$ along the major axis. We set position angles and ellipticities equal to the SDSS Petrosian aperture position angles and axis ratios. The outermost annulus for each galaxy is chosen such that the annuli remain within the maximum radius calculated for the $\lambda_R$ profiles; if this results in any annuli falling wholly outside the data region, then these annuli are discounted. We then take for each annulus the encompassed spaxels' flux-weighted parameter values and flux-weighted elliptical radii.

We estimate the errors on all stellar population profile parameters using

\begin{equation}
    \sigma_{\rm ann} = \sqrt{\left(\frac{\sum{F_i\sigma_i^2}}{\sum{F_i}}\right)}
\end{equation}
\noindent
where $\sigma_{\rm ann}$ denotes the error in an annulus value, $\sigma_i$ the error of each Voronoi bin, and $F_i$ the flux of each Voronoi bin contained within the annulus. 

We then calculated the radial gradients in the stellar population parameters by performing maximum-likelihood fits of linear functions to profiles over a range of $0.5-1.5R_e$. The minimum radius value is chosen to minimise the impact of the MaNGA PSF on derived slopes as well as to minimise the influence of the galaxies' bulge regions \citep[e.g.][]{sanchez2012b,sanchez2014,gonzalezdelgado2015,belfiore2017}, while we have selected the maximum value to ensure good coverage by the FOVs of the majority of our galaxies. We chose to restrict our analysis to galaxies for which full coverage of the this radial range was available, thus obtaining gradients for 44 MWAs overall. Taking the means and standard deviations of the gradients, we find a light-weighted age gradient of $-0.22 \pm 0.22$~dex~$R_e^{-1}$ and a mass-weighted age gradient of  $-0.10 \pm 0.05$~dex~$R_e^{-1}$. For light-weighted metallicity, we find a gradient of $-0.13 \pm 0.15$~dex~$R_e^{-1}$, while for the mass-weighted case we find a metallicity gradient of $-0.16 \pm 0.13$~dex~$R_e^{-1}$.

\begin{figure*}
\begin{center}
	\includegraphics[trim = 1cm 1cm 0cm 18cm,scale=1]{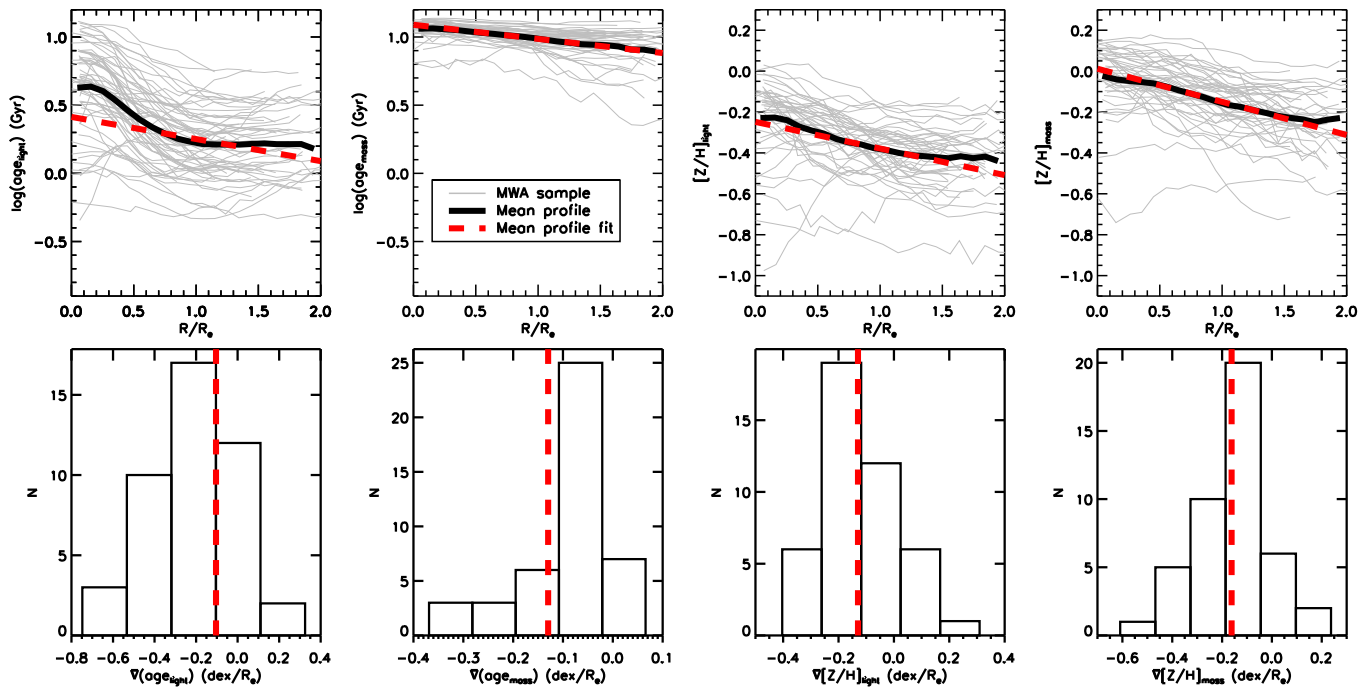}
	\vspace{0.2cm}
	\caption{Stellar population profiles (top) and histograms of linear gradients fitted between $0.5 R_e$ to $1.5 R_e$ for galaxies with sufficient radial coverage. The black solid lines in the top panels show the mean profiles calculated over $0.1R_e$ radial bins, with radius given as the mean of all radii in a given bin, with the red dashed line showing the results of a linear fit between $0.5 R_e$ to $1.5 R_e$.}
	\label{mangapops}
\end{center}
\end{figure*}

In \autoref{mangapops}, we present light-weighted and mass-weighted profiles of stellar population ages and metallicities calculated for the 62 MWAs. We also present histograms of gradients for the 44 MWAs with profiles that fully span the range $0.5-1.5R_e$. We note an offset of $\sim$0.2~dex between the light-weighted and mass-weighted metallicities; this is due to our pPXF fits favouring old and metal-rich SSP model components, which influence the mass-weighted values but have a smaller impact on the light-weighted values. We also note that the light-weighted age  profiles are not ideally fit by a single line, with a break in the mean profiles apparent between 0.5 and 1$R_e$. However, we continue to use linear fits for more straightforward comparisons with existing literature, including that based on analysis of MaNGA data.

We obtain a wide range of central (light-weighted) ages for the galaxies, though the mass-weighted age fits are predominantly old. This range of light-weighted ages can be understood by the scatter in color-magnitude space among the MWAs. Given the Milky Way's apparent status as a green valley galaxy, we interpret the MWAs as being in different stages of a transition from star-forming to quiescent, with small changes in color corresponding to large changes in transition stage and hence large changes in the inferred central galaxy stellar age.

In \autoref{table2}, we present our derived gradients for the MWAs along with results from other, larger galaxy samples in the literature. We compare our results with those of \citet{zheng2017} for MaNGA disc galaxies, along with CALIFA results for disc galaxies from \citet{sb2014}. We also compare with \citet{gonzalezdelgado2015}, who study a sample of 300 CALIFA galaxies spanning a range of morphologies.  We find our mean mass-weighted gradients to be similar to those of Zheng et al. 2017. We note our populations gradients to be somewhat steeper than those calculated \citet{sb2014} on average, possibly due to \citet{sb2014} specifically calculating gradients over regions of disk domination. We also find the age gradients of \citet{gonzalezdelgado2015} to be significantly steeper on average than what we find in this present study. This difference can be understood in light of the sensitivity of age calculations to different weighting schemes, as well as the different fitting methods used to obtain stellar population parameters between studies. The different radial ranges employed for calculating gradients in different studies are another relevant factor.

\begin{table*}
\begin{center}
\begin{tabular}{c|c|c|c|c}

$\rm \nabla log(age)$ & $\rm \nabla [Z/H]$ & Weighting & Range & Ref.\\[2pt]
\hline
\hline
$-0.22 \pm 0.22$ dex$/R_e$ & $-0.13 \pm 0.15$ dex$/R_e$ & Light & 0.5-1.5$R_e$ & This paper\\[2pt]

$-0.10 \pm 0.05$ dex$/R_e$ & $-0.16 \pm 0.13$ dex$/R_e$ & Mass & 0.5-1.5$R_e$ & This paper\\[2pt]

$\approx -0.4$ dex$/R_e$& $\approx -0.1$ dex$/R_e$ & Mass & 0-1$R_e$ & \citet{gonzalezdelgado2015}\\[2pt]
$-0.08 \pm 0.02^a$ dex$/R_e$ & $-0.14 \pm 0.02^a$ dex$/R_e$ & Mass & 0.5-1.5$R_e$ & \citet{zheng2017}\\[2pt]
$-0.116 \pm 0.200$ dex$/r_e$ & $-0.089 \pm 0.151$ dex$/r_e$ & Light & $r_{d0}$-1.5$r_e$\textsuperscript{b} & \citet{sb2014}\\[2pt]
$-0.014 \pm 0.135$ dex$/r_e$ & $-0.051 \pm 0.26$ dex$/r_e$ & Mass & $r_{d0}$-1.5$r_e$\textsuperscript{b} & \citet{sb2014}\\[2pt]

\hline
\end{tabular}
\caption{Stellar population gradients reported in this work, along with results from other recent IFU studies. Gradients from our work and those from \citet{sb2014} are given as means and standard deviations, while the \citet{zheng2017} gradients (marked with $^a$) are given as means and errors of means. \textsuperscript{b}: $r_{d0}$ marks the minimum radius at which the disk dominates, while $r_e$ signifies disk effective radii.}
\label{table2}
\end{center}
\end{table*}

When comparing our calculated metallicity gradients to those calculated for the Milky Way, we assume for the MILES models that $\rm [Z/H] = [Fe/H]$. Taking the galaxies for which we obtain sufficient radial coverage, we obtain light-weighted metallicity gradients with mean and standard deviation $-0.022 \pm 0.024$ dex~kpc$^{-1}$, while for the mass-weighted case we obtain $-0.026 \pm 0.023$ dex~kpc$^{-1}$. We compare this to the MW radial gradients in metallicity obtained by \citet{hasselquist2019} along the Galactic midplane using APOGEE observations of red giant branch stars. \citet{hasselquist2019} calculate profiles of [Fe/H] for stars of various ages and find young stars to display steeper metallicity gradients than older stars, in line with previous findings \citep[e.g.,][]{anders2017}, as well as with the results of MW chemodynamic models \citep[e.g.,][]{minchev2013,minchev2014}. They report a gradient of $-0.06$~dex~kpc$^{-1}$ for stars younger than 2.5~Gyr and a gradient of $-0.016$~dex~kpc$^{-1}$ for stars older than 9~Gyr, both measured over Galactocentric radii between $6-12$~kpc. 

As such, our gradients appear in good consistency with the MW measurement for old ($>9$ Gyr) stars, while being significantly flatter than measured for younger ($<2.5$ Gyr) MW stars. Given the young light-weighted stellar ages we observe for most MWAs beyond $1R_e$ (see the top left window of \autoref{mangapops}), and given that the light-weighted metallicity gradients would be expected to be dominated by the influence of young stars, this is not what would naively be expected.

When scaling by disk scale length, however, we find our gradients to be in closer agreement with that of young MW stars. Taking our previous-reported gradient in dex~kpc$^{-1}$ and scaling by disk scale length, we find a light-weighted and mass-weighted metallicity gradient of $-0.09 \pm 0.11$ and $-0.11 \pm 0.09$~dex~$R_d^{-1}$, respectively, where the stated errors are again the robust standard deviations. If we scale the \citet{hasselquist2019} gradients by a MW disk scale length of 2.71~kpc \citep[calculated by][from optical data]{licquia2016}, meanwhile, we find metallicity gradients of  $-0.16$~dex~$R_d^{-1}$ and $-0.043$~dex~$R_d^{-1}$ from the fits to young and old MW stars, respectively. It is worthwhile to consider the disk scale length along with the (total) effective radius in this case, since the former measure specifically quantifies the extent of galaxies' disk components; disk effective radii \citep[e.g.][]{sanchez2014} would also be viable in this regard.

\begin{figure}
\begin{center}
	\includegraphics[trim = 1.0cm 2cm 0cm 13cm,scale=0.65]{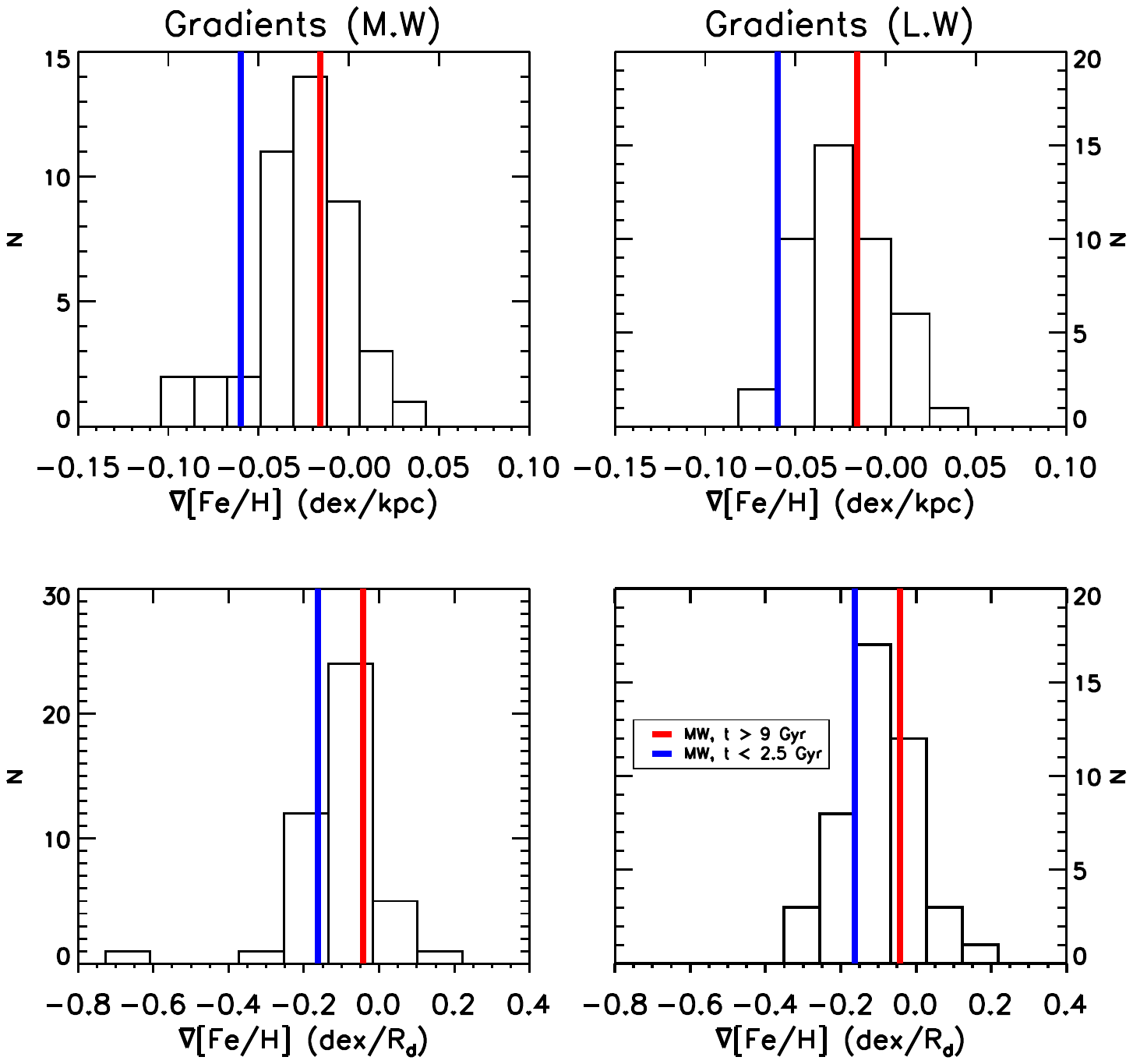}
	\caption{Histograms of mass-weighted (left) and light-weighted (right) linear gradients fitted between $0.5 R_e$ and $1.5 R_e$ for galaxies with sufficient coverage in that range, scaled in units of dex~kpc$^{-1}$ (top) and dex~$R_d^{-1}$ (bottom). The blue line shows the \citet{hasselquist2019} gradient for MW stars younger than 2.5 Gyr, and the red line shows the gradient for stars older than 9 Gyr. Our metallicity gradients are largely inconsistent in physical units with the blue line, but closer achievement is achieved when gradients are scaled by $R_d$.	}
	\label{mwpophist}
\end{center}
\end{figure}

\autoref{mwpophist} compares our MWA metallicity gradients to gradients of the Milky Way, in units of both dex~kpc$^{-1}$ and dex~$R_d^{-1}$. We find that the MWA gradients are largely inconsistent with the MW young star gradient in physical units, but find greater agreement (particularly for our mass-weighted metallicity gradients) when scaled with $R_d$. As such, we argue that the MW's relatively small disc scale radius can at least partially explain offsets between the MWA gradients and the MW gradient for young stars.

\subsection{Ionised gas results} \label{results_iongas}

We now consider what we can learn from the emission line measurements described in the previous section. In \autoref{iongas_classification}, we classify galaxies in terms of their ionised gas emission, and in \autoref{gas_metallicity}, we study the gas phase metallicities of the MWA sample.

\subsubsection{Emission source classification} \label{iongas_classification}

We classify the sources of gas ionisation in our galaxies using three complementary metrics. First, we construct BPT diagrams \citep{bpt,veilleux1987} using the BPT-NII and BPT-SII metrics as presented in \citet{kewley2006}. These two metrics can be used to identify gas emission consistent with ionisation due to a central active galactic nucleus (AGN) by comparing the ratio of [OIII] and H$\beta$ emission with that of [NII] and H$\alpha$ (for BPT-NII) or [SII] and H$\alpha$ (for BPT-SII). We use the maximum starburst line of \citet{kewley2001}, along with the line proposed by \citet{kauffmann2003}, to separate pure star-forming emission from composite emission.

We note at this point that the term ``composite'' can in fact be misleading. While ``composite'' emission can indeed be produced by a combination of star-forming and AGN emission \citep[e.g.][]{davies2017}, similar emission line ratios can also be produced from single ionising sources such as nitrogen-enhanced HII regions \citep[e.g.][]{ho1997} and shock ionisation \citep[e.g.][]{allen2008}.

It is also the case that numerous other ionising sources can produce emission consistent with AGN on a BPT diagram, including diffuse ionised gas (DIG) and post-asymptotic giant branch (post-AGB) stars. As such, it is possible for galaxies in AGN regions of BPT diagrams to be simply galaxies that have stopped forming new stars \citep{binette1994,stasinska2008,sarzi2010,singh2013,belfiore2016}. Such galaxies display low H$\alpha$ equivalent widths (hereafter $EW(\textrm{H}\alpha)$) in their ionised gas emission \citep{papaderos2013,sanchez2014,gomes2016,belfiore2017} and may be detected and ruled out as AGN candidates on that basis. Thus, we make use of the complementary WHAN diagnostic introduced by \citet{cidfernandes2010,cidfernandes2011}, in which $EW(\textrm{H}\alpha)$ is plotted against the ratio of [NII] and H$\alpha$. Such a combination of BPT and WHAN diagnostics has often been used for analysing galaxies' gas contents in IFU studies \citep[e.g.][]{belfiore2015,rembold2017}, along with other combinations of BPT diagnostics with the H$\alpha$ equivalent width \citep[e.g.][]{sanchez2014,sm2016,lacerda2018}. These diagnostics provide a way to distinguish AGN activity from diffuse ionised gas emission, the latter of which is powered by old stars and has weaker H$\alpha$ emission. 

We use WHAN diagnostic lines selected to match the BPT-NII diagnostic lines; as in \citet{belfiore2015}, we place a boundary at $\rm [NII]/H\alpha = -0.1$ beyond which a galaxy's emission is deemed AGN-like. We also place a line at $\rm [NII]/H\alpha = -0.32$, as in \citet{cidfernandes2010}, to separate star-forming emission from composite-like emission; this value is based on the \citet{kauffmann2003} line described previously. For AGN and composite regions, we assume spaxels with $EW(H\alpha) < 3$\AA\ to represent a retired galaxy regions regardless of their $\rm [NII]/H\alpha$ value. For spaxels $\rm [NII]/H\alpha$ in the star-forming region, we instead employ a cut of 6\AA\ to separate star-forming and non-star-forming emission, and we likewise describe the latter regions as "retired" for the remainder of this article.

For all three metrics, we have corrected our emission line fluxes by assuming an intrinsic Balmer decrement of $H\alpha / H\beta = 2.86$ and then fitting a \citet{calzetti2000} dust reddening law. This correction is only reliable if both $H\alpha$ and $H\beta$ have been clearly detected in a spectrum. As such, we do not perform the correction for spaxels in which the amplitude-over-noise ratio ($A/N$) of the $H\beta$ feature is less than three. For affected spaxels in the retired region of the WHAN diagram, we assume this to be due to low levels of emission, and we continue to use these spaxels in our analysis. We deem affected spaxels in other regions of the WHAN diagram to be unreliable, and we mask them from subsequent analysis. Such spaxels are almost entirely found at the edges of a given galaxy's FOV, and so masking them has no effect on the results that we present.

\begin{figure}
\begin{center}
	\includegraphics[trim = 2cm 3cm 0cm 7.5cm,scale=0.7]{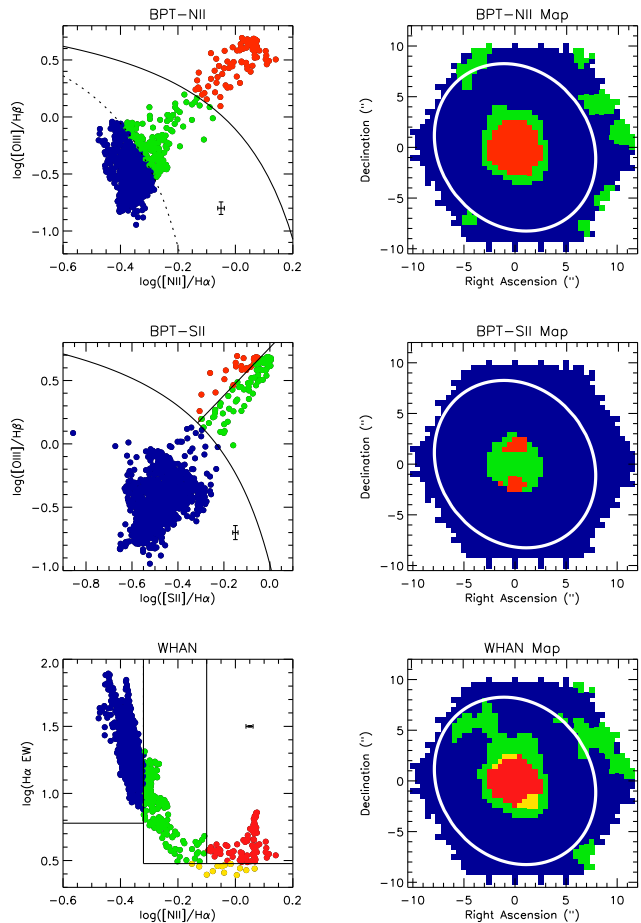}
	\caption{BPT and WHAN diagrams for galaxy 8979-6102. Top: BPT-NII diagram (left) and map (right) of diagram classifications. A mixture of star-forming (blue) and AGN-like (red) emission is detected, along with a region of composite emission (green). Middle: BPT-SII diagram and map. We find LINER-like (green) and Seyfert-like (red) emission in a handful of spaxels near the centre, along with substantial star-forming emission (blue). Bottom: WHAN diagram and map. We find a region of high $H\alpha$ emission near the centre, corresponding to central AGN-like emission (red, orange); this is surrounded by a composite region (blue, green) which also contains three bins consistent with ionisation by old stars. Thus, the WHAN diagram supports and complements the picture implied by the BPT diagrams. The black error bars show the median error for each parameter.
	}

	\label{bptmaps1}
\end{center}
\end{figure}

We show maps of these diagnostics in \autoref{bptmaps1}, using galaxy 8979-6102 as an example. In this particular case, the BPT-NII and BPT-SII distributions are both consistent with AGN-like gas emission in the galaxy's centre; the WHAN diagnostic, meanwhile rules out the BPT features being the result of old stellar populations. In this particular case, we see that the WHAN diagnostic alone is consistent with an AGN-like central region.

Beyond approximately 0.5$R_e$, the majority of our galaxies are dominated by star-forming emission, with the exceptions being galaxies that are heavily inclined. Within their central regions, however, our galaxies display a wider range of behaviour. We therefore classify the MWAs according to the emission sources inferred around the galaxy centres, using four classifications:

\begin{itemize}
\item Centrally star-forming (``SF'') galaxies are those for which the inner effective radius is dominated by star-forming emission; both the BPT-NII and BPT-SII diagnostics show the majority of inner spaxels lie in the star-forming region, with values of EW(H$\alpha$) greater than 6\AA. Given that the galaxies in our sample largely contain star-forming emission away from their centres, galaxies classified as centrally star-forming are thus dominated by star-formation emission over the whole of the studied FOV. 

\item Centrally Intermediate (``CI'') galaxies have a central region consistent with composite-like emission on the basis of the BPT-NII diagnostic. The EW(H$\alpha$) of central spaxels in these galaxies is 3\AA\ or greater.

\item ``AGN'' galaxies are those for which we detect ionisation consistent with central AGN activity via both the BPT-NII and BPT-SII diagnostics; in addition, the spaxels corresonding to this activity display EW(H$\alpha$) values of greater than 3.

\item Centrally Retired (``CR'') galaxies have central values of $EW(\textrm{H}\alpha)$ too low to be consistent with any other classification. Such galaxies may have central emission line ratios that appear AGN-like when viewed in the two BPT diagrams. We classify a galaxy as CR if its central region is consistent with star-forming emission from the two BPT diagnostics and $EW(\textrm{H}\alpha) \leq$6~\AA, or if it has an AGN-like or composite-like central region from the BPT diagnostics and $EW(\textrm{H}\alpha)\leq$3~\AA.
\end{itemize}

\begin{figure*}
\begin{center}
	\includegraphics[trim = 2cm 2cm 1cm 4cm,scale=0.9]{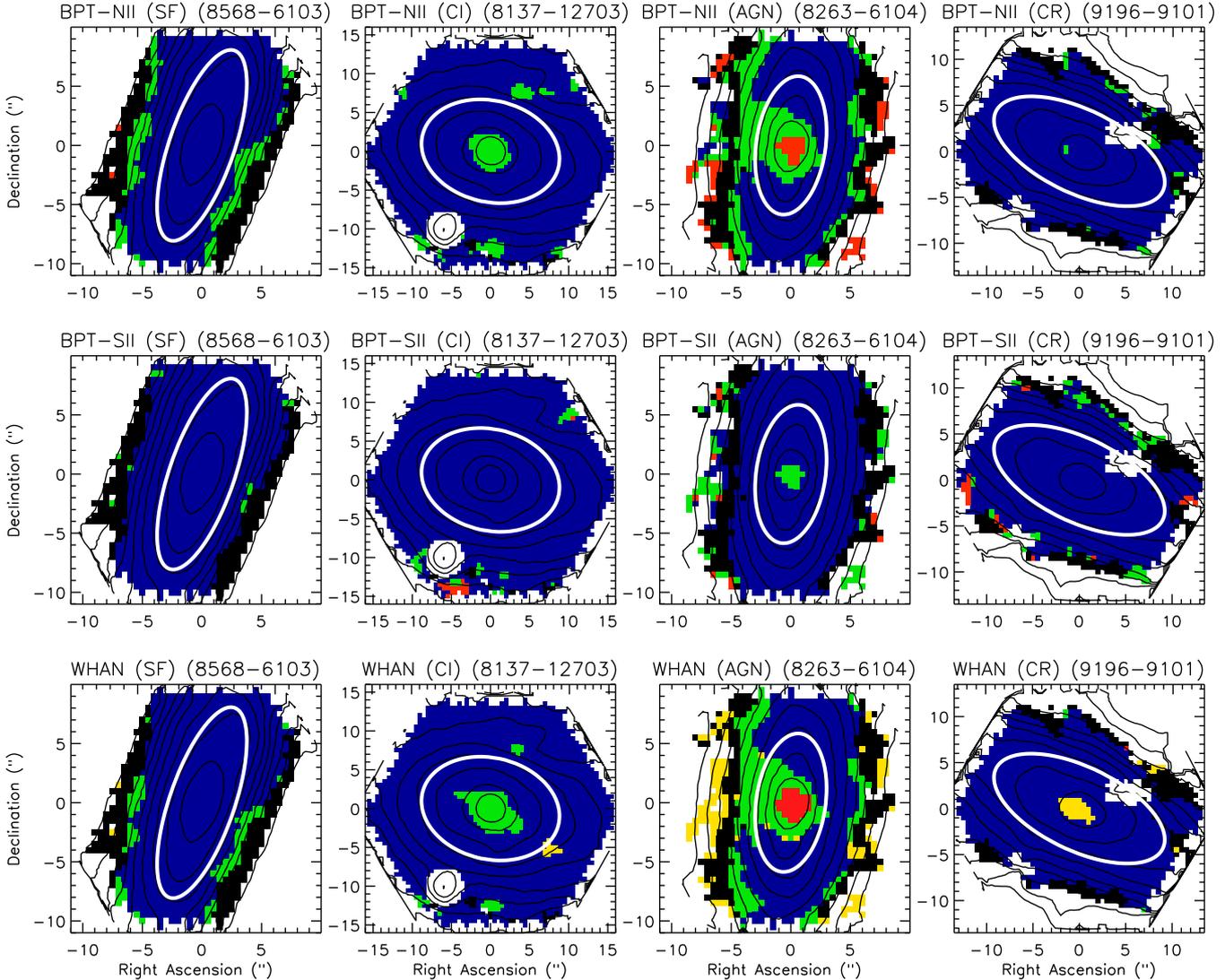}
	\caption{Example BPT and WHAN maps for our four ionisation source classifications: centrally star-forming (SF), centrally intermediate (CI), AGN-like (AGN) and centrally retired (CR). Colors are as follows. For BPT-NII: blue regions are star-forming, green regions composite and red regions AGN. For BPT-SII: blue regions are star-forming, green regions are LINER-like and red regions Seyfert-like. For WHAN: blue regions are star-forming, green  regions are composite regions, red regions are AGN-like regions, and yellow regions are retired regions. Black regions are those regions with unreliable dust corrections, as explained in the text.}
	\label{bptexamples}
	\end{center}
\end{figure*}

We present examples of each classification in \autoref{bptexamples}. We note that we do not directly use the WHAN diagnostic in our classifications, though it remains useful for visualisation purposes. Of the 62 MWAs in the sample, we identify 12 (19\%) as being centrally retired and 22 (35\%) as centrally star-forming. We find 7 galaxies (11\%) to have central AGN-like activity (8979-6102, 8146-12702, 8263-6104, 8257-12705, 9040-9102, 9496-9102, 9499-12703) in terms of BPT diagnostics and EW(H$\alpha$) values, though we note that none of these display broad-line spectra when inspected by eye. We have verified that the ``AGN'' galaxies are inconsistent with any other classifications within the uncertainties of their emission line flux and $EW(\textrm{H}\alpha)$ values. We identify a further 21 galaxies (33\%) as having centrally intermediate emission. That so many of the MWAs display hints of AGN activity is interesting in light of suggestions that the MW itself hosted an AGN in its past \citep[e.g.][]{nicastro2016}, and we suggest that MWAs could be used to investigate this further. 

Candidate AGN galaxy catalogs have been previously published in MaNGA, both in \citet{rembold2017} and also in \citet{sanchez2018}. Both catalogs used MPL-5 MaNGA data, for which four of our AGN galaxies (8979-6102, 8146-12702, 8263-6104, 8257-12705) have data available. Of these four, galaxy 8979-6102 is the only one listed in \citet{sanchez2018}, while both 8979-6102 and 8257-12705 are listed in \citet{rembold2017}. When considering these differences, it is important to note that we employ different selection critia to the quoted studies, which in turn use different selection criteria to each other. \citet{rembold2017} performs selections using the BPT-NII and WHAN diagnostics, based on SDSS aperture spectra. \citet{sanchez2018}, meanwhile, performs selection using 3''-3'' MaNGA spaxel regions; they use the BPT-OI diagnostic along with the BPT-NII and BPT-SII diagnostics and EW(H$\alpha$).

The presence of a non-neglibile number of CR objects objects agrees well with the idea of the Milky Way being in the green valley region of the color--magnitude diagram, which is consistent with our stellar population results (\autoref{results_pops}). Most likely, these centrally quiescent galaxies are in the process of a gradual cessation of star formation and transition onto the red sequence. Such a gradual transformation is proposed for disk galaxies in \citet{schawinski2014}, and is supported by the results of \citet{sanchez2019}.

\begin{figure}
\begin{center}
	\includegraphics[trim = 2cm 2cm 0cm 10cm,scale=0.5]{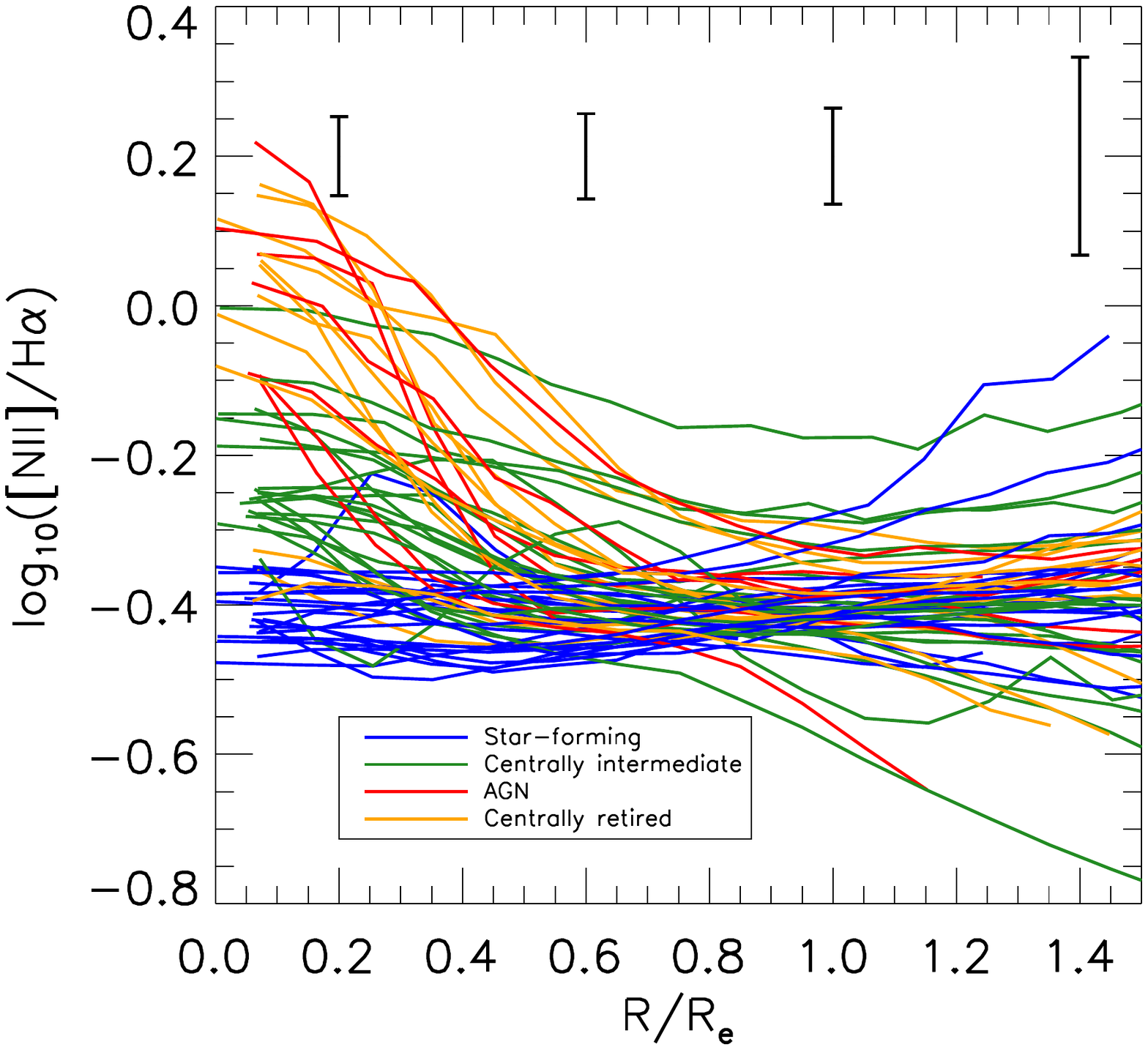}
	\caption{Profiles of $\rm \log([NII]/H\alpha )$, colored according to gas-ionisation classification. Representative error bars are shown for various radii.}
	\label{nhprofiles}
\end{center}
\end{figure}

In \autoref{nhprofiles}, we show radial profiles of $\rm \log([NII]/H\alpha )$ for the 62 MWAs in our sample. We calculate these profiles by azimuthally averaging over annuli of width 0.1$R_e$ as done above. We color-code the line for each galaxy according to its classification. The MWA galaxies have a relatively narrow range of $\rm \log([NII]/H\alpha )$ values beyond approximately 0.5$R_e$, compared to smaller radii. The CR galaxies have similar profiles to AGN galaxies, with both types typically producing negative radial gradients in $\rm \log([NII]/H\alpha )$ which flatten beyond roughly 0.5 0.5$R_e$ . SF galaxies typically display profiles that are flat or close to flat, meanwhile with CI galaxies displaying mild negative gradients at low radii.

\subsubsection{Gas-phase metallicity} \label{gas_metallicity}

We estimate the gas-phase metallicity of the MWAs using the ratios $\rm \log ([NII] \lambda 6583/[SII])$ and $\rm \log ([NII] \lambda 6583/H\alpha)$ along with the O3N2 metric of \citet{alloin1979}, which  combines the H$\alpha$, H$\beta$, [OIII]$\lambda$5007, and [NII]$\lambda$6583 fluxes via

\begin{equation}
\rm    O3N2 = \log\left(\frac{\rm [OIII]\lambda 5007}{H\beta}\times \frac{H\alpha}{\rm [NII] \lambda 6583}\right).
\end{equation}

By construction, the O3N2 metric is  insensitive to the effects of dust extinction. The $\rm \log ([NII] \lambda 6583/[SII])$ metric is likewise only weakly dependent on the dust correction, due to the small separation of the respective spectral lines.

A large number of calibrations have been proposed in the literature, aimed of estimating gas-phase metallicities from measured emission line flux ratios, and the choice of calibrator can have a significant impact on derived results \citep[e.g.][]{kewley2008}; we direct the interested reader to Section 3.1 of \citet{sanchez2017} for a comprehensive summary, and to Section 3 of \citet{maiolino2019} for a more detailed review. One choice is to calibrate emission line measures based on results from photoionisation models. This approach has been employed in the derivation of simple gas-phase metallicity calibrators \citep[e.g.][]{Dopita_2016_EmLineDiagnostic}, as well as in more complex fitting methodologies \citep[e.g. IZI;][]{blanc2015}. An objection to this approach \citep[e.g.][and references therein]{sanchez2017} is that such models rely on various strong assumptions on the physical behaviours of ionising populations' atmospheres, which are in practice only loosely understood. Emission line calibrations derived this way are also dependent on assumptions of various abundance ratios, explicitly or otherwise, as well as on the behaviour of the ionisation parameter.

An alternative to calibrating on photoionisation models is to calibrate on gas-metallicity measurements from observational electron temperatures ($T_e$) \citep[e.g.][]{marino2013}. This removes the need for detailed assumptions, but suffers from the inability of associated calibrators to reach metallicities much beyond the solar value. This problem can be accounted for by supplementing data with model results at higher metallicities, as for instance done in \cite{pettini2004} (hereafter PP04). However, such an approach requires the same set of assumptions as those that use ionisation models alone.

Considering the situation, we have chosen to explore multiple gas-phase metallicity calibrators in order to better understand the range of metallicity behaviour allowed by the DAP emission line measurements. Such an approach is similar to that employed in Pipe3D gas metallicity measurements of CALIFA data \citep{sanchez2017}, MaNGA data \citep{sanchez2018} and SAMI data \citep{sanchez2019}. We use the 03N2 calibrator of PP04 along with that of \citet[][hereafter C17]{curti2017}, the latter of which which was developed from $Te$ measurements of local star-forming galaixes in SDSS DR7. We also use the calibration of \citet[][hereafter D16]{Dopita_2016_EmLineDiagnostic}, which was derived through photoionisation models alone. These three calibrators can be written as

\begin{equation}\label{gasmetaleq2pp04}
\rm    [12+\log(O/H)]_{PP04} = 8.73 - 0.32(O3N2),
\end{equation}

\begin{multline}\label{gasmetaleq2d16}
\rm   [12+\log(O/H)]_{D16} = 8.77 + \log ([NII] \lambda 6583/[SII]) \\ 
   \rm + 0.264 \log ([NII] \lambda 6583/H\alpha),
\end{multline}

\begin{equation}\label{gasmetaleq2c17}
\rm    O3N2 = 0.281 - 4.765 x^2 - 2.286 x^2,
\end{equation}

\noindent
where in the final equation, $\rm x$ is the oxygen abundance normalised to the solar value of 8.69 \citep{ap2001}: $\rm x = [12+\log(O/H)]_{C17} - 8.69$.

These expressions all implicitly assume that the ionising gas flux is due to star formation, and are thus inaccurate in cases where the gas ionisation is due to some other source. We minimise the impact of non-star-forming emission on our gas-phase metallicity measurements as follows. We restrict our analysis to reliably ``star-forming'' spaxels by masking any spaxels that do not fall onto the star-forming region of the BPT-NII diagram. We then further mask spaxels for which $EW(H\alpha ) < 14$ \AA\, following the rationale and adopted criterion of \citet{lacerda2018}. We further mask spaxels in which any of the other relevant  emission lines ($\rm [OIII]\lambda 5007$, $H\beta$, $\rm [NII] \lambda 6583$, $\rm [SII]\lambda 6718$, $\rm [SII]\lambda 6733$) have observed equivalent widths lower than 0.5\AA. The spaxels left unmasked from this process are hereafter referred to as star-forming spaxels. 

We then construct one-dimensional gas-phase metallicity profiles in a similar manner as other radial profiles above. We place annuli of width 0.1$R_e$ over our galaxies based on their ellipticity and photometric position angles, and we azimuthally average over all star-forming spaxels for each annulus. We subsequently discount any annuli without any star-forming spaxels within them. We choose the outermost annulus to have a radius lower than that of the largest elliptical aperture containing 85\% of a galaxy's spectra, using the ellipticities and position angles as before. In this case, we perform the calculation using MaNGA datacubes without applying a S/N cut. Approximately two-thirds of our profiles have their maximum radii set by this limit, with the remaining third instead being set by limited availability of star-forming regions. We calculate gradients for these profiles by performing maximum-likelihood linear fits between $0.5-1.5 R_e$ for profiles that extend to at least 1.5$R_e$ and contain at least three star-forming spaxels between 0.5$R_e$ and 1.5$R_e$; this enables gradient calculations for 41 MWAs. As with the stellar populations case, this fitting range in radius is chosen to minimise the impact of the bulge while also being within the FOV of most sample galaxies. Taking means and standard deviations, we find from the PP04 calibration an average sample gradient of $-0.1 \pm 0.1$~dex~$R_e^{-1}$. For the D16 calibration we obtain an average sample gradient of $-0.16 \pm 0.1$~dex~$R_e^{-1}$, while for the C17 calibration we obtain $-0.066 \pm 0.065$~dex~$R_e^{-1}$. In physical units, we find thus find a mean gradient of 0.014~dex/kpc with a robust standard deviation of 0.013~dex/kpc using the PP04 calibration. For the D16 calibration, we obtain a mean gradient of -0.022~dex/kpc with robust standard deviation of -0.012~dex/kpc, while from C17 we obtain a mean of -0.0086~dex/kpc and a robust standard deviation of -0.0082~dex/kpc.

For all three calibrators, we construct mean MWA gas-phase metallicity profiles by taking means of all profiles in bins of 0.1$R_e$, with the radius defined as the mean radius of all points in a given bin. We calculate gradients for these profiles by performing a least-absolute-deviation fit between 0.5$R_e$ and 1.5$R_e$.

\begin{figure*}
\begin{center}
	\includegraphics[trim = 0cm 1.5cm 0cm 14cm,scale=0.9]{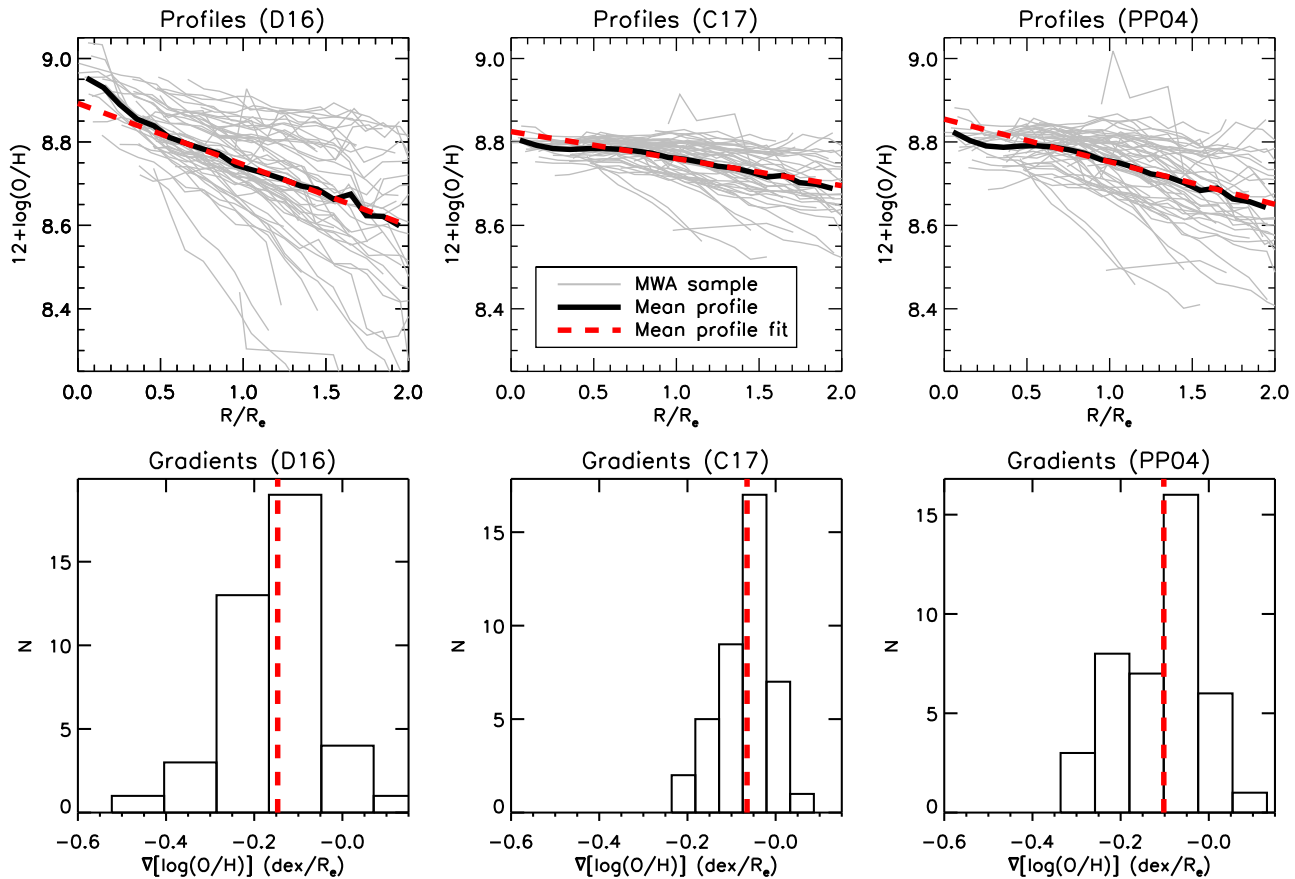}
	\caption{Profiles of $12+\log(O/H)$ for the MWA sample, calculated using our three chosen calibrators. The grey lines reperesent indiviudal galaxies, with the thick black lines showing the mean profiles. The dashed red lines show the results of least-absolute-deviation fits to the black lines, over radii between 0.5 $R_e$ and 1.5$R_e$.}
	\label{gasprofiles}
\end{center}
\end{figure*}

We present our gas-phase metallicity profiles in \autoref{gasprofiles}, wherein we show the profiles for the different diagnostics separately. We also show histograms of calculated gradients between 0.5--1.5$R_e$ for the 41 sample galaxies with profiles extending over the whole of this region.

We now consider how our profiles compare to the literature, as well as how they compare with gas metallicity measurements taken for the Milky Way. In \autoref{litgas} we compare the mean and spread of our gas-phase metallicity gradients with a number of other measurements from straight-line fits in the literature: The \citet{belfiore2017} sample consists of 550 nearby MaNGA galaxies with a stellar mass range of $\log(M/M_\odot ) = 9.5-11.5$. The \citet{sm2016} sample consists of 122 face-on spiral galaxies observed by CALIFA, while the \citet{sanchez2014} comprises 306 CALIFA-obseved galaxies. The \citet{sm2018} sample consists of 102 MUSE-observed spiral galaxies, while the \citet{poetrodjojo2018} sample is made up of 25 face-on star-forming spiral galaxies obtained from SAMI. As shown in \autoref{litgas}, our calculated gradients are in good overall consistency with existing literature. 

For the MW, \citet{esteban2017} obtain a gradient of $-0.04 \pm 0.005$~dex/kpc over a Galactocentric radial range of $5.1-17.0$~kpc from metallicity measurements of individual HII regions. \citet{balser2015} report significant azimuthal variations in the metallicities of HII regions across the MW; between Galactic azimuths of $90^\circ$ and $130^\circ$, they report [O/H] gradients of $-0.082 \pm 0.014$~dex/kpc, while for azimuths between $0^\circ$ and $60^\circ$ they find a gradient of $\approx -0.04$~dex/kpc. \citet{genovali2014}, meanwhile, obtain a slope\footnote{\citet{genovali2014} calculate [Fe/H], so the slope in [O/H] in this case would depend upon any [O/Fe] gradients in their sample.} of $0.06 \pm 0.002$ dex/kpc from fitting to the metallicities of Cepheids, which trace young ($\sim 20-400$~Myr) stars and are expected to mirror the metallicity behaviour of the Milky Way's ISM. As such, we find gradients of the MW gas metallicity profiles to be substantially steeper than the average gradients from the MWA sample, regardless of which calibrator we use and regardless of which specific MW observable we consider.

\begin{figure*}
\begin{center}
	\includegraphics[trim = 1cm 1cm 0cm 16cm,scale=1]{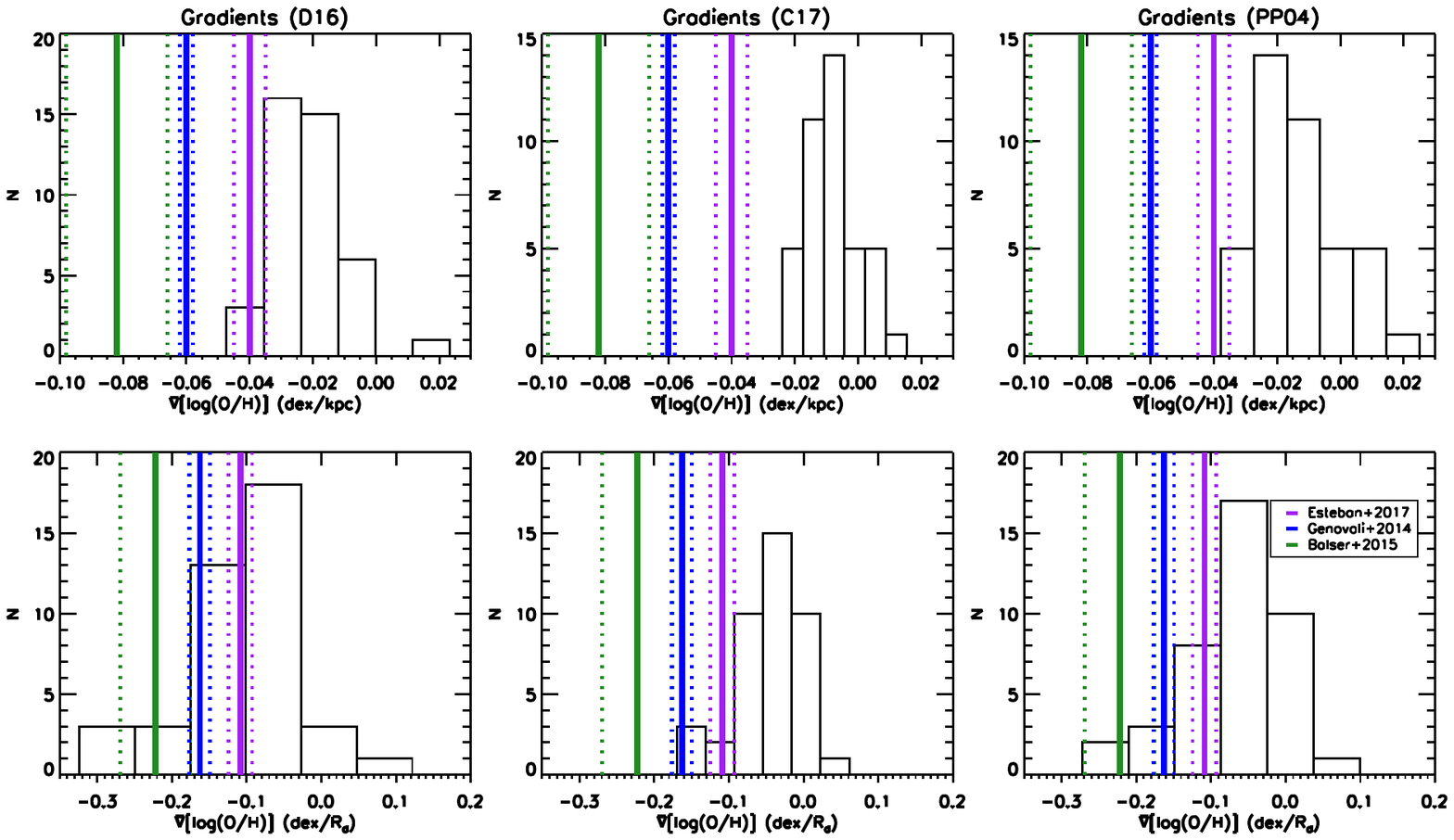}
	\caption{Histograms of gas metallicity gradients calculated for our MWAs using the (left) \citet{Dopita_2016_EmLineDiagnostic}, the (middle) \citet{curti2017} and (right) \citet{pettini2004} calibrations. We also show the \citet{esteban2017} best-fit line slope for the MW gas-phase metallicity (purple), the \citet{genovali2014} slope as inferred from young Cepheids, and the \citet{balser2015} slope for gas-phase metalicity for Galactic azimuths between $90^o$ and $130^o$ (green). We find the MW gradients to be signficantly steeper than is typical for the MWAs (top); however, the MWA sample gradients are more consistent with the MW gradients when gradients are scaled in terms of disk scale radius (bottom), particularly for the D16 and PP04 calibrations.}
	\label{mwgashist}
\end{center}
\end{figure*}

However, we reiterate at this point that the MW appears to be unusually compact for a galaxy of its type, and that our MWAs largely have disk scale radii significantly higher than has been calculated for the MW. In \autoref{mwgashist}, we show histograms of the 41 galaxies for which we calculated individual gradients. We show in the same figure the \citet{esteban2017} gradient and the $-0.08$~dex/kpc gradient reported in \citet{balser2015} for azimuths between $90^o$ and $130^o$. We show the gradients both in units of dex/kpc and in units of dex/$R_d$. We find that scaling gas metallicities by disk scale radius largely removes the discrepancy between the MW and the MWAs. Using the PP04 calibration, we obtain a gradient mean and robust standard deviation of $-0.074 \pm 0.072$~dex~$R_d^{-1}$, for the D16 calibration: $-0.11 \pm 0.076$~dex~$R_d^{-1}$, and for the C17 calibration: $-0.046 \pm 0.045$~dex~$R_d^{-1}$.  The PP04 and D16 values compare well with the scaled \citet{esteban2017} gradient of $-0.11 \pm 0.014$~dex~$R_d^{-1}$, though the C17-derived values remain somewhat flatter. However, the \citet{balser2015} gradient for HII regions at azimuths between $90^o$ and $130^o$, at $-0.22 \pm 0.043$~dex~$R_d^{-1}$ continues to be well beyond what we find for the majority of our studied MWAs. The \citet{genovali2014} gradient, $-0.163 \pm 0.005$ ~dex~$R_d^{-1}$, is likewise steeper than is found for many of the sample MWAs, regardless of which calibration is considered.

\begin{figure*}
\begin{center}
	\includegraphics[trim = 1cm 10cm 0cm 8cm,scale=1]{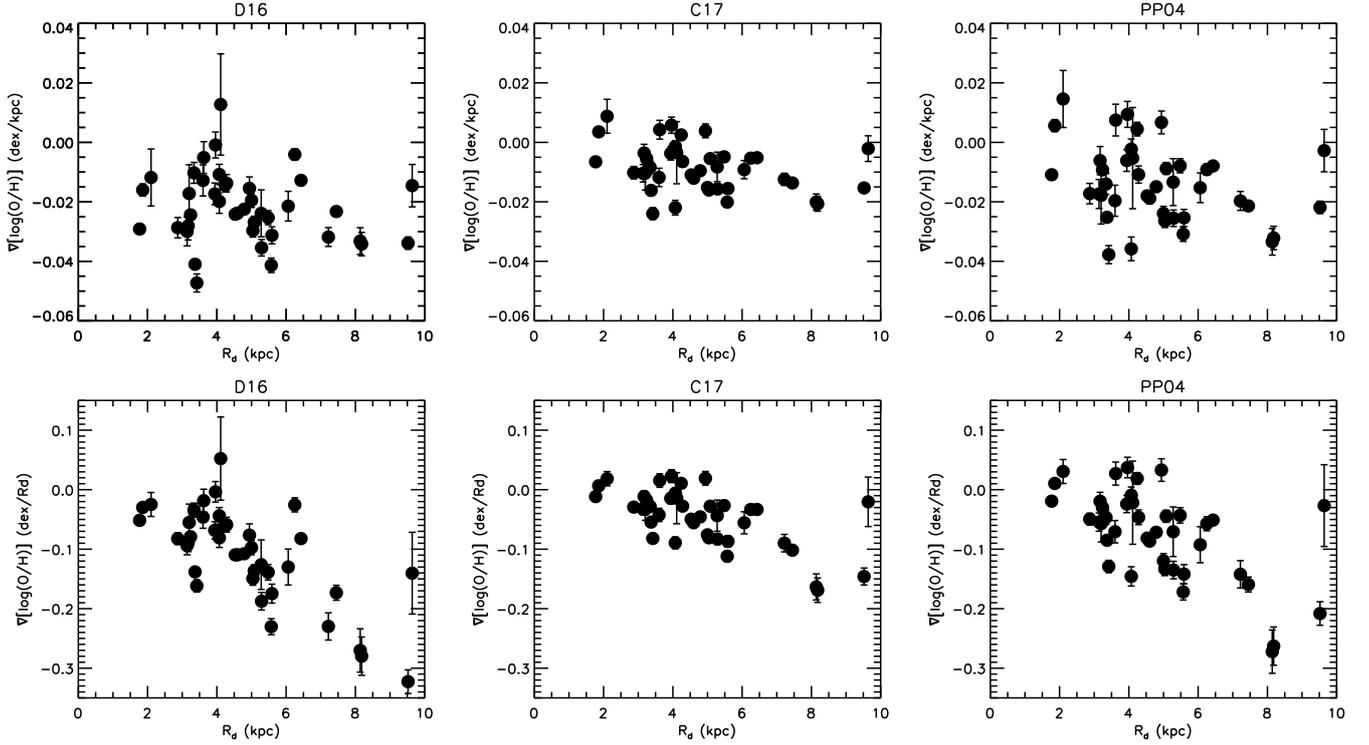}
	\caption{Plot of gas-phase metallicity gradients against disk scale length. We find no trend when gradients are in physical units, and in turn find a mild negative trend when the gradients are scaledd}
	\label{gasmetalgradrd}
\end{center}
\end{figure*}

We note an apparent trend in our particular sample between gas metallicity gradients in scaled units with the disk scale lengths themselves. In \autoref{gasmetalgradrd}, we plot the gas-phase metallicity gradients derived for the galaxies, using all three chosen calibrators, as functions of $R_d$. We plot the gradients both in scaled units, and also in physical units. We find no overall relation between gradients in physical units with $R_d$, and in turn find a mild trend of steeper (more negative) scaled gradients with higher $R_d$. This is different from results obtained from larger samples \cite[e.g.][]{sanchez2014} that support the existence of a characteristic (scaled) gas metallicity gradient. The cause of the trend we find is unclear, but may be a consequence of our specific sample selection.

We find from \autoref{mwgashist} that our MWAs overwhelmingly have flatter gas metallicity gradients in terms of physical units than reported for the MW. However, we find the range of MWA gradients to agree closer with MW measurements when the gradients are parametrised in terms of dex~$R_d^{-1}$. In particular, we find the scaled MWA gradients to be well within the gradient reported by \citet{esteban2017}. As such, we argue that the discrepency in physical units is at least partly due to the relative compactness of the MW disc, with differences in perspective and in observational techniques also likely be to be relevant factors.

\begin{table*}
\begin{center}
{\setlength{\extrarowheight}{1pt}
\begin{tabular}{c|c|c|c|c}

Calibrator & $\nabla[\log(O/H)]$ & Units & Range & Ref.\\[1pt]
\hline
\hline
O3N2 [PP04] & $-0.10 \pm 0.10$ & dex/$R_e$ & $0.5-1.5~R_e$ &  This paper\\[1pt]

[D16] & $-0.16 \pm 0.10$ & dex/$R_e$ & $0.5-1.5~R_e$ & This paper\\[1pt]

O3N2 [C17] & $-0.066 \pm 0.065$ & dex/$R_e$ & $0.5-1.5~R_e$ &  This paper\\[1pt]

\hline
O3N2 [PP04] & $-0.014 \pm 0.013$ & dex/kpc & $0.5-1.5~R_e$ & This paper\\[1pt]

[D16] & $-0.022 \pm 0.012$ & dex/kpc & $0.5-1.5~R_e$ & This paper\\[1pt]

O3N2 [C17] & $-0.0086 \pm 0.0082$ & dex/kpc & $0.5-1.5~R_e$ & This paper\\[1pt]

\hline
O3N2 [PP04] & $-0.1 \pm 0.09$ & dex/$r_e$ & $0.3-2.1~r_e$ & \citet{sanchez2014}\\[1pt]

\hline
O3N2 [PP04] & $-0.11 \pm 0.07$ & dex/$r_e$ & $0.5-2~r_e$ & \citet{sm2016}\\[1pt]

O3N2 [PP04] & $-0.014 \pm 0.012$ & dex/kpc & $0.5-2~r_e$ & \citet{sm2016}\\[1pt]

\hline
O3N2 [PP04] & $-0.08 \pm 0.10$ & dex/$R_e$  & $0.5-2~R_e$ & \citet{belfiore2017}\\[1pt]
\hline
O3N2 [\citet{marino2013}] & $-0.11 \pm 0.07$ & dex/$r_e$  & Various& \citet{sm2018}\\[1pt]
\hline
R23 [\citet{kk2004}] & $-0.12 \pm 0.05$ & dex/$R_e$  & $\sim 0-1.5~R_e$ & \citet{poetrodjojo2018}\\[1pt]
\hline
\end{tabular}}
\end{center}
\caption{Comparison of gas-phase metallicity gradients of comparable galaxy samples from recent studies. All values are given as means and standard deviations. $r_e$ signifies disc effective radius, whereas $R_e$ signifies the effective radius of galaxies as a whole.}
\label{litgas}
\end{table*}

\subsection{Effect of inclination}

Given the known effect of high inclinations on measured
galaxy properties \citep[e.g.][]{belfiore2017,ibarramedel2019}, it is worth considering how important the effect of inclinations is on our reported results. In particular, galaxy inclinations can impact upon measured isophotal radii, due to changes in perspective and also due changes in levels of dust reddening \citep{masters2003}.

\begin{figure}
\begin{center}
	\includegraphics[trim = 3cm 11cm 0cm 0.5cm,scale=0.5]{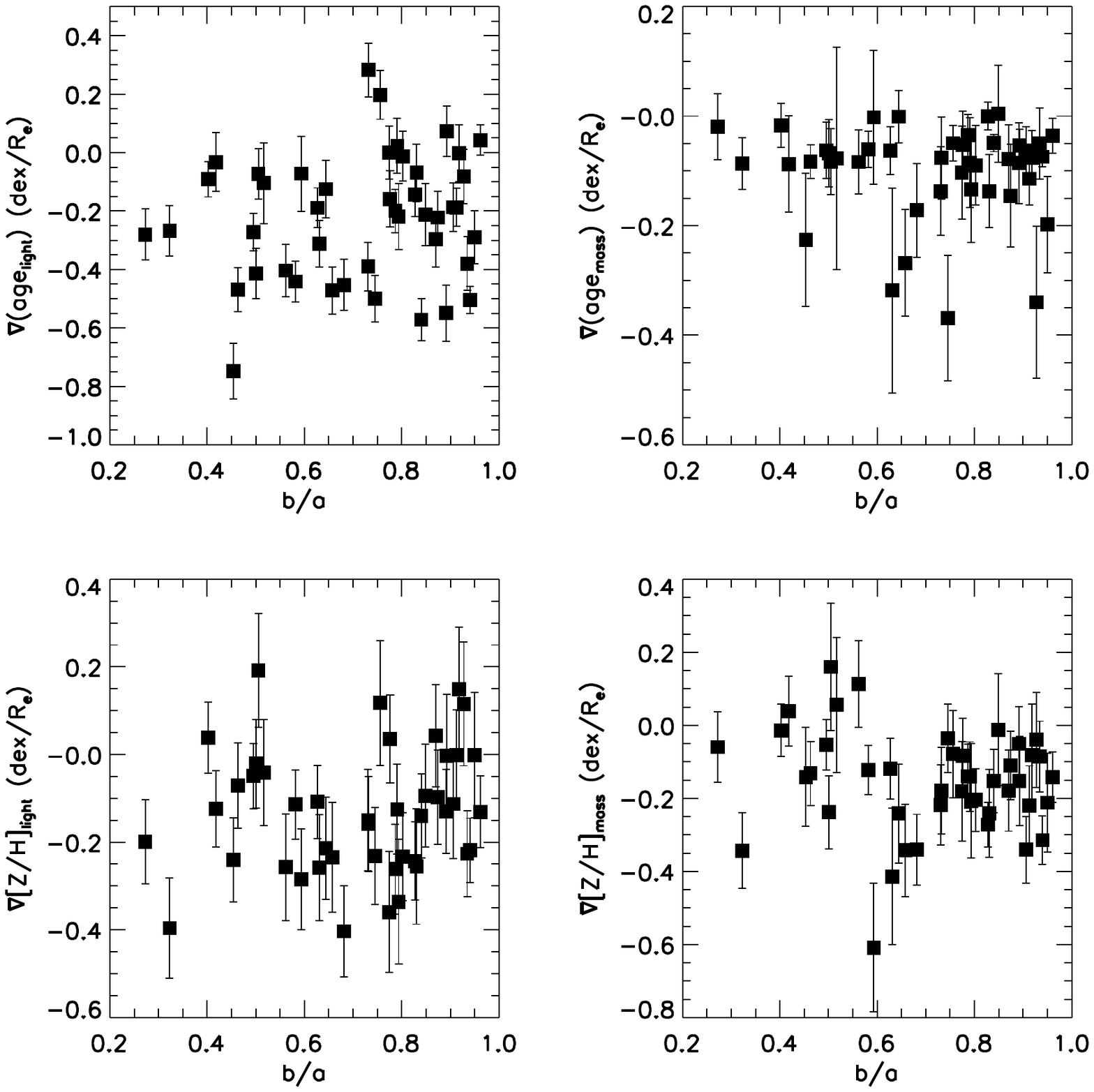}
	\caption{Stellar population gradients plotted as a function of axis ratio $b/a$. The mass-weighted age gradients show a tendency to steeper slopes with increasing axis ratios, but we observe no clear trends in terms of metallicity or light-weighted age.}
	\label{popsba}
\end{center}
\end{figure}

To assess the impact of inclination, in \autoref{popsba} we plot the derived stellar population gradients against axis ratio $b/a$, for all galaxies with sufficient radial coverage. We note a tendency for steeper mass-weighted age gradients at larger $\epsilon$ values, but we find no obvious dependence on axis ratio for the light-weighted age gradients or for the metallicity gradients.

\begin{figure}
\begin{center}
	\includegraphics[trim = 2cm 2cm 0cm 7cm,scale=0.9]{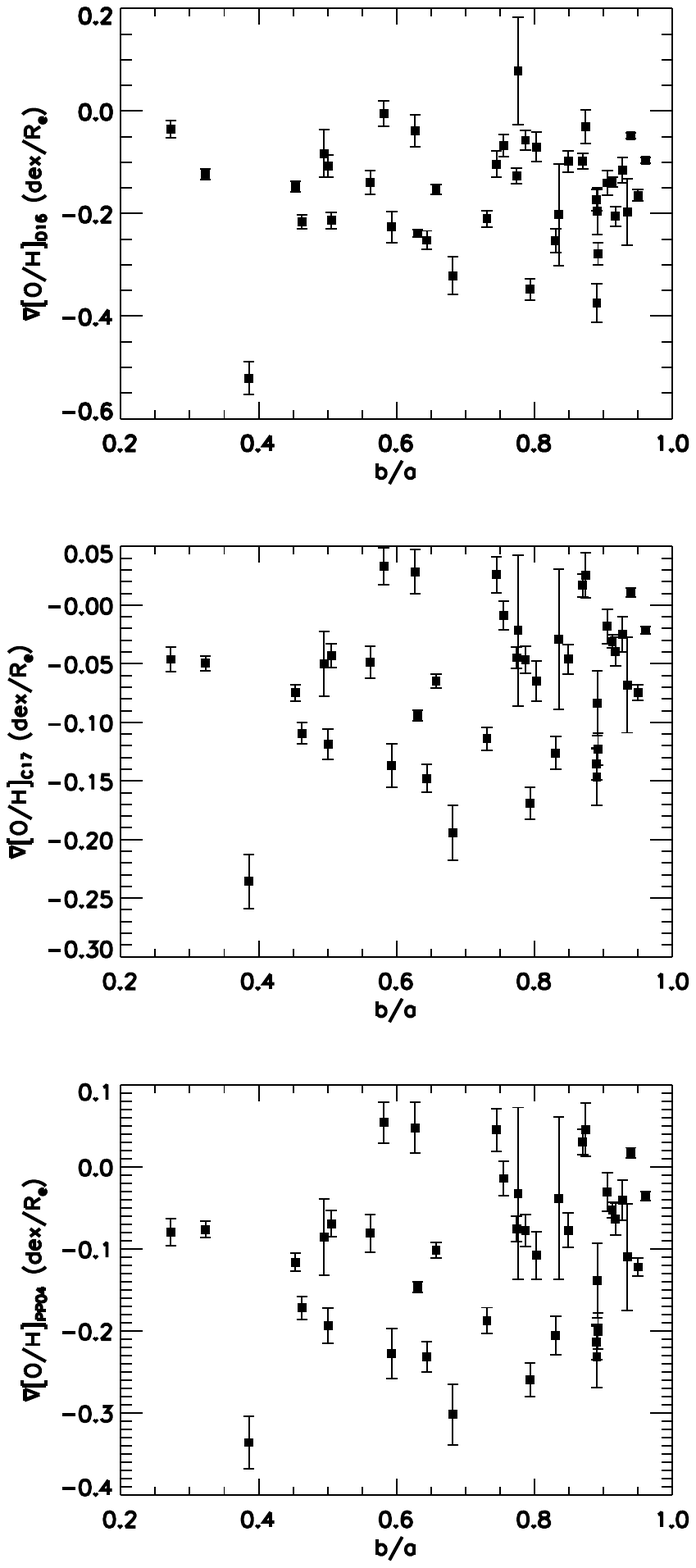}
	\caption{Gas metallicity gradients from all three chosen calibrators plotted as a function of axis ratio $b/a$. The mass-weighted age gradients show a tendency to steeper slopes with increasing axis ratios, but we observe no clear trends in terms of metallicity or light-weighted age.}
	\label{gasba}
\end{center}
\end{figure}

Likewise, our derived gas metallicity gradients show little dependence on galaxies' axis ratios. In \autoref{gasba}, we plot the gradients from all three selected calibrators as a function of $\epsilon$, for all MWAs with sufficient coverage and sufficient numbers of star-forming spaxels. As with the stellar population gradients, we see little dependence of the gas metallicity gradients on galaxies' axis ratios.

In \autoref{gradcompare}, we show the means and robust standard deviations for both the full galaxy sample and for galaxies with axis ratios b/a greater than 0.7, finding that such a cut induces only small changes in the range of gradients we find.

\begin{table*}
\begin{center}
\begin{tabular}{c|c|c|c|c|c}
Cut & $\nabla(Age_{light})$ & $\nabla(Age_{mass})$ & $\nabla([Z/H]_{light}$ & $\nabla([Z/H]_{mass})$ & $\nabla[\log(O/H)]_{PP04}$ \\[1 pt]
\hline
\hline
$0 \leq b/a \leq 1$ (full sample) & $-0.22 \pm 0.22$ & $-0.10 \pm 0.05$ & $-0.13 \pm 0.15$ & $-0.16 \pm 0.13$ & $-0.10 \pm 0.10$\\[1 pt]

$b/a > 0.7$ & $-0.18 \pm 0.23$ & $-0.10 \pm 0.05$ & $-0.12 \pm 0.14$ & $-0.16 \pm 0.09$ & $-0.087 \pm 0.097$\\[1 pt]
\end{tabular}
\end{center}
\caption{Comparison of stellar population and gas metallicity gradients obtained for our sample, with and without cutting for axis ratio b/a. In each case, we show means and robust standard deviations for all galaxies that were included in our final analysis. All gradients are in units of dex/$R_e$.}
\label{gradcompare}
\end{table*}

\subsection{Effect of bar presence}

In this subsection, we consider the effects of bars on our reported gradients, for stellar population parameters and for gas-phase metallicities. Barred galaxies have previously been reported from MaNGA data to contain flatter stellar population gradients within the bar regions \citep{frasermckelvie2019}. Barred galaxies, we note, have also been found to be more likely to contain LINER-like central regions (Krishnarao et al., in prep).

Here, we make use of the MaNGA Galaxy Zoo (GZ) value-added catalog \footnote{\url{https://www.sdss.org/dr15/data\_ access/value-added-catalogs/?vac\_ id=manga-morphologies-from-galaxy-zoo}}, which contains data for 53 out of our 62 galaxies. We exclude galaxies with axis ratios b/a less than 0.4, in order to remove edge-on cases where a bar could be obscured; this leaves us with 50 galaxies. For these galaxies, we obtain the debiased fractions of "bar" and "no bar" answers, which are weighted using the methodology explained in \citet{willett2013} and \citet{hart2016}.

We find a non-negligible proportion of our non-edge-on galaxies to contain bar-like features, with 17/50 found to have debiased GZ "bar" fraction greater than 0.5. We note however that only 4 of these galaxies (8141-12701, 8444-12703, 8615-9102 and 8985-9102) have a debiased "bar" answer fraction of more than 0.7. As such, galaxies with clear bars make up a minority of our sample.

In \autoref{gradcompare2}, we compare the mean measured populations and gas metallicity gradients that we obtain for our sample, along with the robust standard deviations, with and without excluding galaxies with "bar" percentrages greater than 50\%; we perform this calculation for all galaxies with sufficient radial coverage, excluding twelve such galaxies when cutting for bar presense. We also show the results of a more conservative cut, wherein we exclude the four galaxies with bar percentages higher than 70. We show that cutting our galaxy sample in this way produces only a small impact on our findings. The wider effects of bars on MWA sample selection is the subject of further work, and a further exploration is beyond the scope of this present paper.

\begin{table*}
\begin{center}
\begin{tabular}{c|c|c|c|c|c}
Cut & $\nabla(Age_{light})$ & $\nabla(Age_{mass})$ & $\nabla([Z/H]_{light}$ & $\nabla([Z/H]_{mass})$ & $\nabla[\log(O/H)]_{PP04}$ \\[1 pt]
\hline
\hline
$0.4 \leq b/a \leq 1$ & $-0.23 \pm 0.23$ & $-0.11 \pm 0.05$ & $-0.12 \pm 0.15$ & $-0.15 \pm 0.13$ & $-0.10 \pm 0.10$\\[1 pt]

GZ debiased "bar" frac < 0.7, $0.4 \leq b/a \leq 1$  & $-0.21 \pm 0.22$ & $-0.10 \pm 0.04$ & -$0.10 \pm 0.14$ & $-0.15 \pm 0.12$ & $-0.09 \pm 0.09$\\[1 pt]

GZ debiased "bar" frac < 0.5, $0.4 \leq b/a \leq 1$  & $-0.25 \pm 0.20$ & $-0.09 \pm 0.04$ & -$0.11 \pm 0.14$ & $-0.15 \pm 0.12$ & $-0.09 \pm 0.10$\\[1 pt]
\end{tabular}
\end{center}
\caption{Comparison of stellar population and gas metallicity gradients obtained for our sample, with and without cutting out galaxies with bars. For the first row, we show means and robust standard deviations for all galaxies that were included in our final analysis. For the second row, we show galaxies included in the MaNGA Galazy Zoo VAC with debiased "bar" answer fractions no greater than 0.7. All gradients are in units of dex/$R_e$.}
\label{gradcompare2}
\end{table*}

\subsection{Predictors of measured analog properties}
\label{paramcomparison}

Here, we explore how our MWAs appear in other parameter spaces, as well as how our measured parameters compare with those parameters that one would expect to find in a large photometric catalog. In this way, we can begin to consider how ``expensive'' mesureables (such as stellar ages, stellar metallicities, and their gradients) relate to the sorts of ``cheaper'' parameters available from large photometric-based catalogs, which we could expect to be used in designing analog sample selections. At the same time, we can explore if close proximity to the MW in a given parameter space corresponds to greater homogeneity of MWAs in other spaces.

We parametrise the MWAs' closeness to the MW in terms of stellar mass and B/T using the $\Delta_{\rm morph}$ parameter, which was used to set the initial ancillary sample's observing priorities and is defined as

\begin{equation} 
\Delta_{\rm morph} = \sqrt{\Delta_{M}^2 + \Delta_{\rm B/T}^2},
\end{equation}

\noindent
where $\Delta_{M}$ and $\Delta_{\rm B/T}$ are defined respectively using
\begin{equation} 
\Delta_{\rm M} = \frac{|\log_{10}M_* - \log_{10}M_{*,MW}|}{\sigma_{M} }
\end{equation}
and
\begin{equation}
    \Delta_{\rm B/T} = \frac{|\log_{10}(\rm B/T) - \log_{10}(\rm B/T)_{MW}|}{\sigma_{\rm B/T}}.
\end{equation}

\noindent
where the $\sigma$ terms in the above equations denote the 1-$\sigma$ confidence intervals for the MW in those parameters. We also calculate $\Delta_{\rm SFR}$ and $\Delta_{\rm Rd}$, which parametrise the difference between the MWAs and the MW in terms of SFR and $R_d$:

\begin{equation} 
\Delta_{\rm SFR} = \frac{|{\rm SFR}-{\rm SFR}_{MW}|}{\sigma_{\rm SFR}}
\end{equation}

\begin{equation}
    \Delta_{\rm R_d} = \frac{|R_d - R_{d,MW}|}{\sigma_{Rd}}
\end{equation}

\begin{figure*}
\begin{center}
	\includegraphics[trim = 2.0cm 1.52cm 0cm 10cm,scale=0.9]{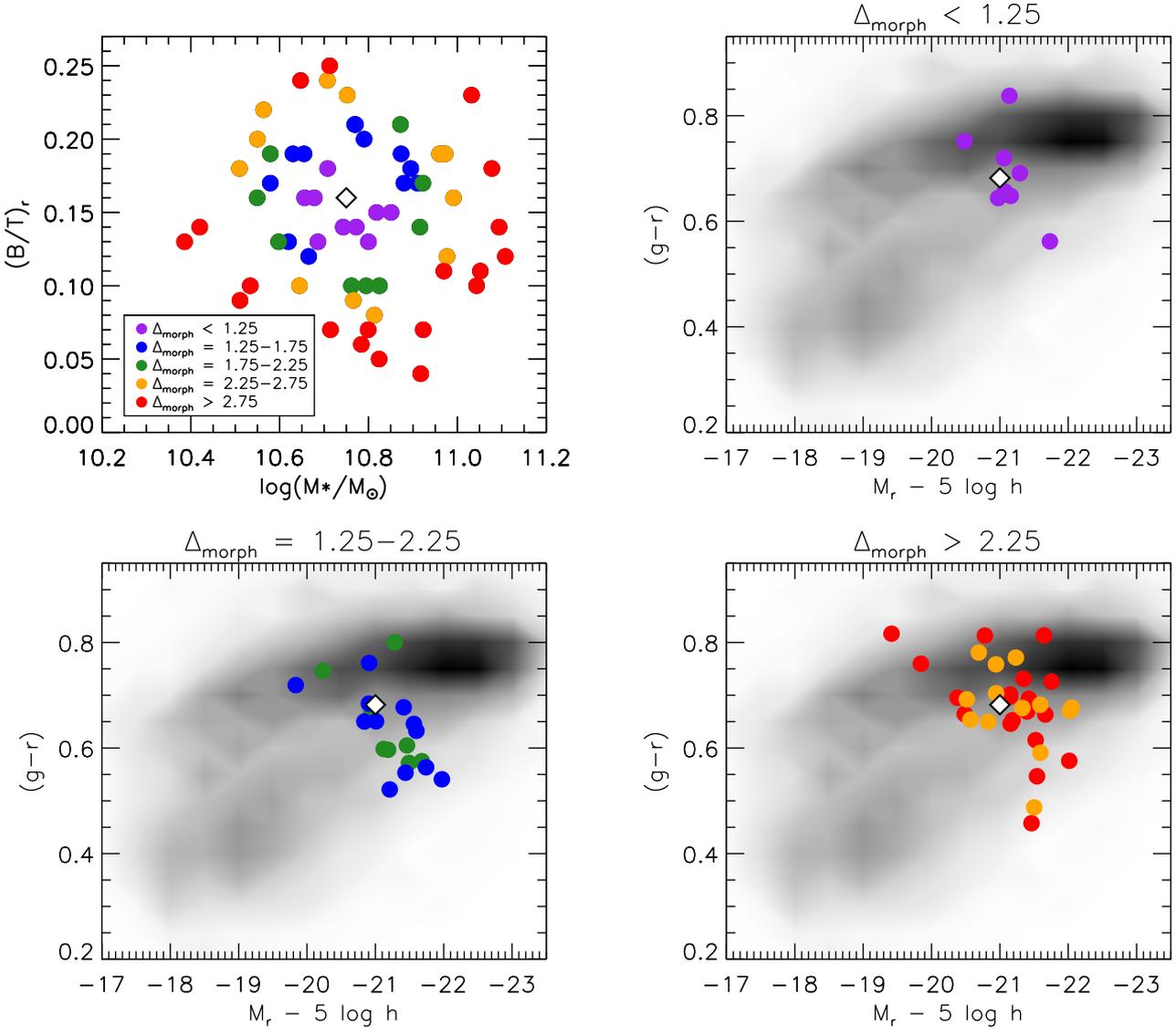}
	\caption{Top left panel: plot of B/T against stellar mass, coloured by $\Delta_{\rm morph}$. Other panels: color-magnitude diagrams showing MWAs grouped by their $\Delta_{\rm morph}$. The MW position on the color-magnitude diagram, as derived by L15, is shown as a white diamond. The color-magnitude positions of the full MPL-8 MaNGA sample are also shown. 
	}
	\label{mbt_priorities}
	\end{center}
\end{figure*}

We use MW values the values for the Milky Way's mass and $R_d$ from \citet{licquia2016}. In all cases, we use the means of the MW probability distribution functions along with the standard deviations, as detailed in \autoref{sigtable}. We show in \autoref{mbt_priorities} how the $\Delta_{\rm morph}$ parameter appears in the $M-B/T$ parameter space, with color-magnitude diagrams also shown for MWAs with various values of $\Delta_{\rm morph}$. 

\begin{table}
\begin{center}
\begin{tabular}{c|c}

$\log(M_*/M_{\odot})$ & $10.75 \pm 0.1$\\[1pt]
\hline
B/T & $0.16 \pm 0.03$\\[1pt]
\hline
SFR & $1.65 \pm 0.19~M_{\odot}$~yr$^{-1}$ \\[1pt]
\hline
$R_d$ & $2.71 \pm 0.21 kpc$\\[1pt]

\end{tabular}
\end{center}
\caption{MW parameters used for assessing MWA proximity to the MW in different parameter spaces. Mass and scale length values are from \citet{licquia2016a}, while B/T and SFR values are from \citet{licquia2015}.}
\label{sigtable}
\end{table}

\begin{figure}
\begin{center}
	\includegraphics[trim = 2.5cm 3cm 0cm 14cm,scale=0.75]{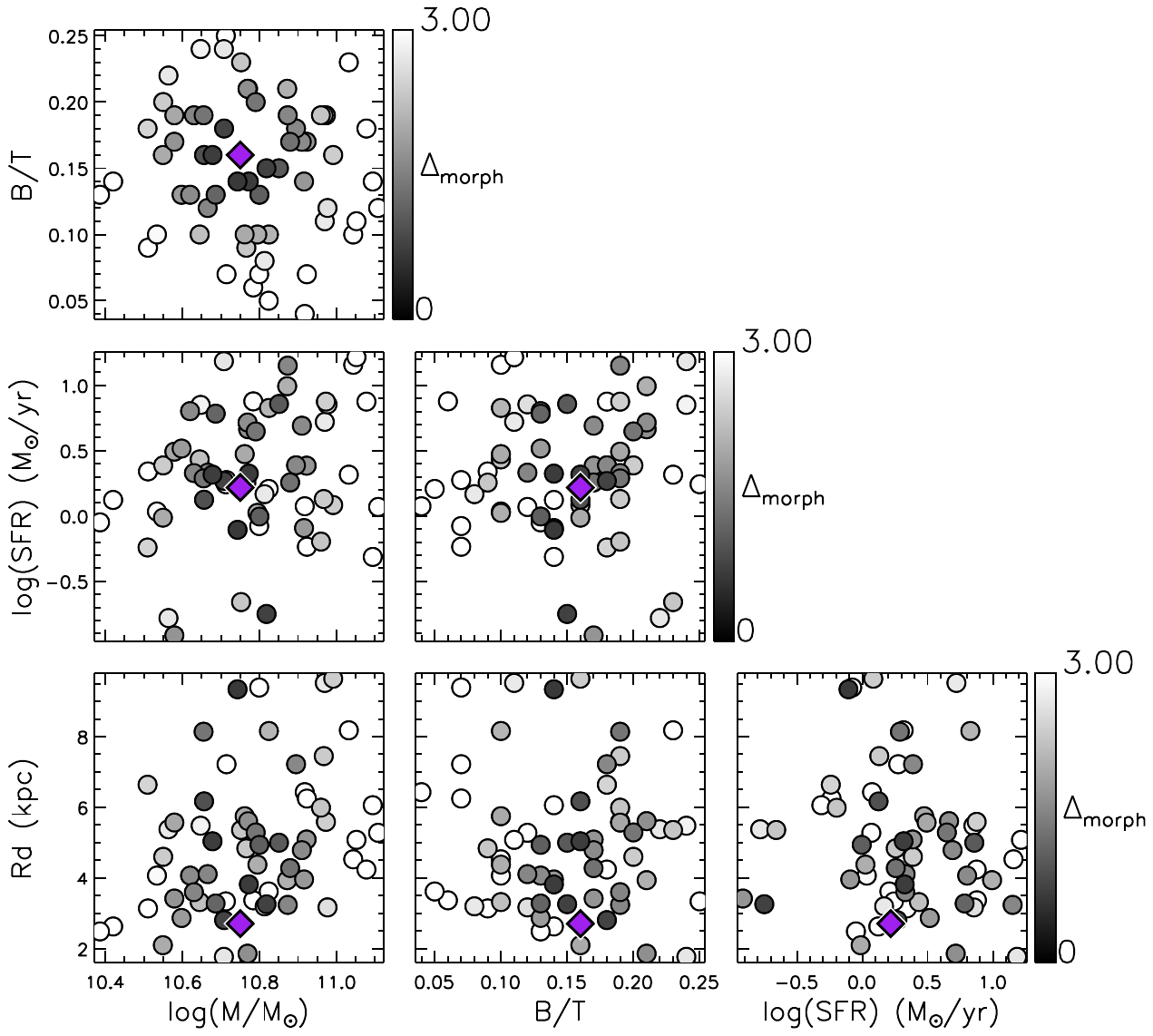}
	\caption{Corner plot of stellar mass, bulge-to-total ratio B/T, star-formation rate and disk scale length. Datapoints are coloured according to $\sigma_{\rm morph}$, which parameterises the galaxies' distance from the MW in $M_*-{\rm B/T}$ space. The purple diamonds indicate the centres of the MW PDF.
	}
	\label{catparams}
	\end{center}
\end{figure}

In \autoref{catparams}, we present a corner plot of the MWAs' $M_*$, B/T, SFR and $R_d$ values, with the points colored according to their value of $\Delta_{\rm morph}$. We immediately see that proximity to the MW in one parameter space does not necessarily imply closeness in other parameter spaces: though $\Delta_{\rm morph}$ depends strongly on $M$ and B/T by design, it shows much less structure in terms of SFR or $R_d$. However, we note a clustering datapoints with low $\Delta_{\rm morph}$ in panels with SFR on one of the axes, suggesting that galaxies with low $\Delta_{\rm morph}$ values are more likely to be similar to the MW in terms of SFR as well. 

\begin{figure*}
\begin{center}
	\includegraphics[trim = 2.cm 2.8cm 0cm 7cm,scale=0.95]{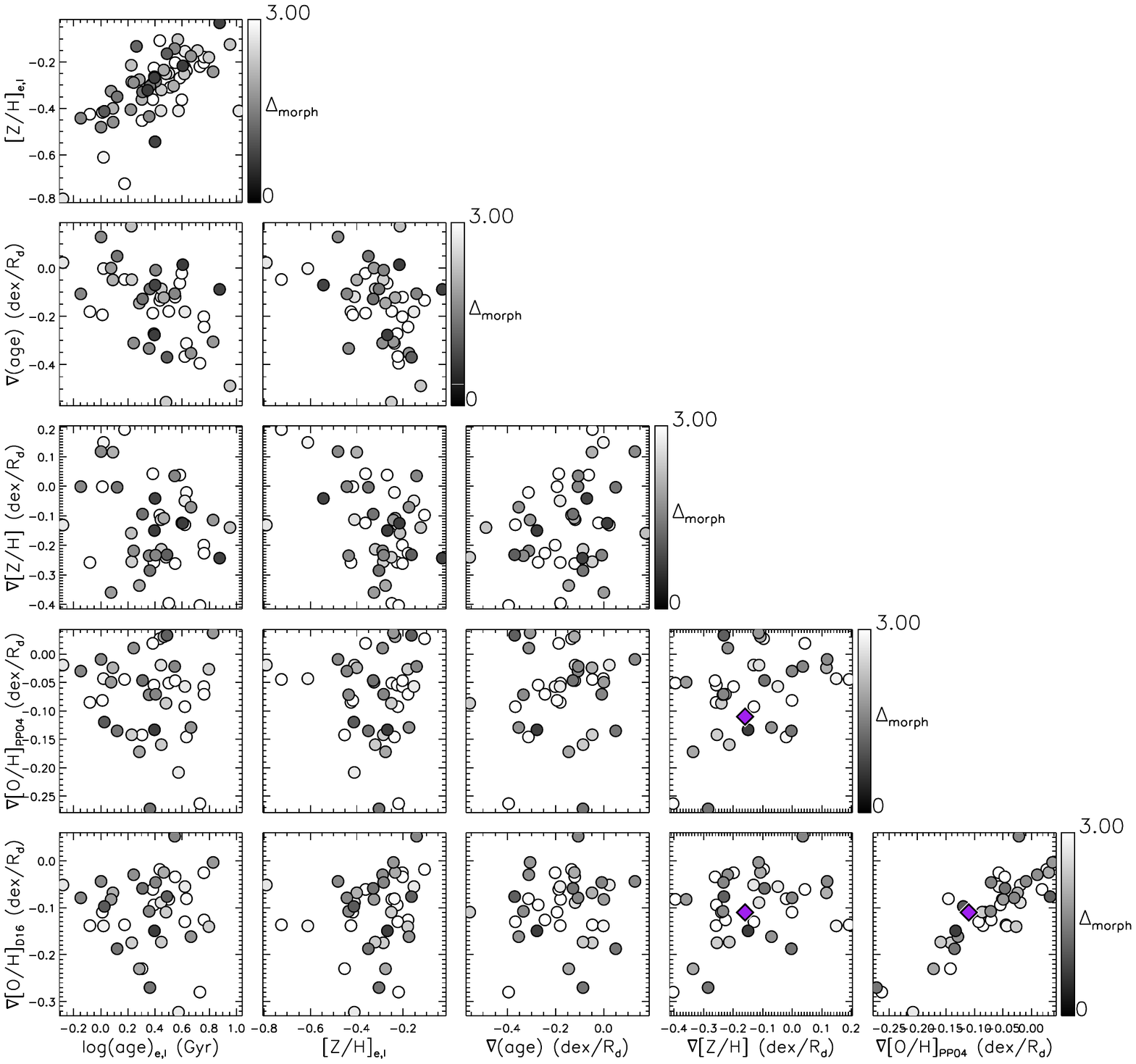}
	\caption{Corner plot showing light-weighed stellar ages and metallicities measured within 1~$R_e$ ellipses, along with stellar age and metallicity gradients and gas metallicity gradients. We find a galaxy's proximity to the MW in $M*-{\rm B/T}$ space, as parametrised by $\Delta_{\rm morph}$, to have only a modest relationship with integrated ages and metallicities. We find no clear relation between $\Delta_{\rm morph}$ and the calculated gradients. The purple diamonds show the \citet{hasselquist2019} metallicity gradient for young MW stars and the \citet{esteban2017} gradients for MW HII regions. 
	}
	\label{mparams}
	\end{center}
\end{figure*}

Next, we consider whether closeness to the MW mass and B/T values corresponds to homogeneity in the stellar population and gas measurements. To this end, we present in \autoref{mparams} a corner plot of light-weighted stellar population parameters and gradients (Section 4.2), along with the PP04- and D16-based gas-phase metallicity gradients (Section 4.3), with gradients scaled according to disk scale length. We also show the MW metallicity gradients for young stars and gas, using values from \citet{hasselquist2019} and \citet{esteban2017} respectively, in the lower right three panels of the figure.  We calculate ages and metallicities within $1~R_e$ ellipses, using the same procedure described for the stellar population profiles in \autoref{results_pops}. As before, we colour datapoints by their value of $\Delta_{\rm morph}$.  We find the properties of MWAs' stellar populations and gas are connected only loosely with their proximity to the MW in $\rm{M_*}-{\rm B/T}$: we note a mild preference for younger ages and greater metallicities, while finding no relationship between $\Delta_{\rm morph}$ and the calculated gradients. We also find no clear relationships between the different gradients considered, save for the expected correlation between the two gas-phase metallicity gradient estimators.

\begin{figure*}
\begin{center}
	\includegraphics[trim = 2.cm 2.8cm 0cm 7cm,scale=0.95]{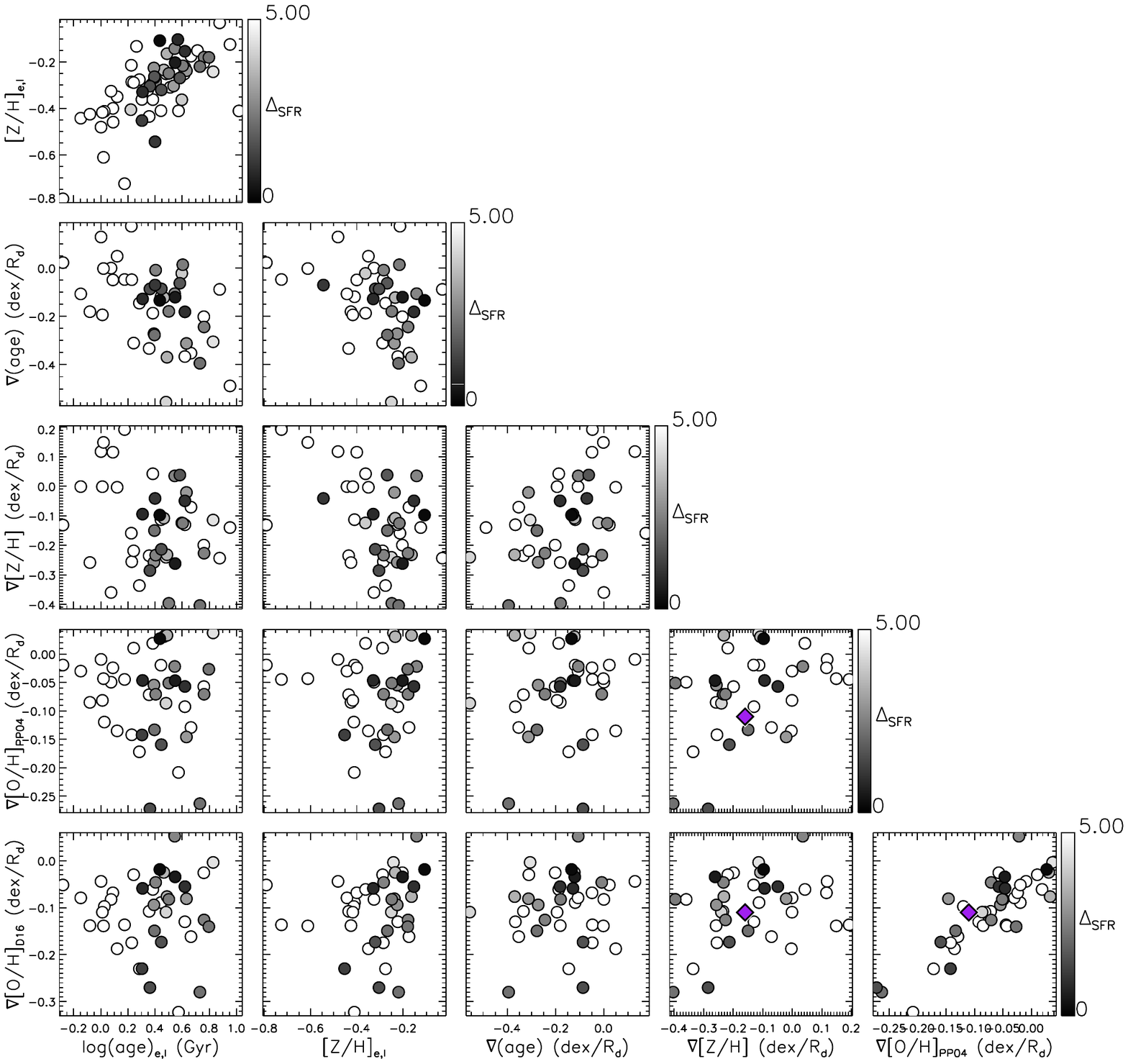}
	\caption{Corner plot showing light-weighed stellar ages and metallicities measured within 1~$R_e$ ellipses, along with stellar age and metallicity gradients and gas metallicity gradients. Datapoints are coloured according to $\Delta_{\rm SFR}$. We find that galaxies with low $\Delta_{\rm SFR}$ occupy a narrower range of integrated light-weighted ages and age gradients. The purple diamonds show the \citet{hasselquist2019} metallicity gradient for young MW stars and the \citet{esteban2017} gradients for MW HII regions.
	}
	\label{mparamssfr}
	\end{center}
\end{figure*}

In \autoref{mparamssfr}, we present the same parameters and gradients as before, with datapoints instead coloured based on $\Delta_{\rm SFR}$. In this case, we note that galaxies with low $\Delta_{\rm SFR}$ seem to cluster within a smaller range of ages, metallicities and age gradients than those with high $\Delta_{\rm SFR}$. This suggests that closeness to the MW in SFR space implies greater homogeneity in certain stellar population properties. This pattern is tentative due to the low numbers of points involved, but is unsurprising in light of the known connection between SFR and color.
Values related to metallicity gradients, by contrast, show no such behaviour with $\Delta_{\rm SFR}$.

We may quantify the behaviour in \autoref{mparamssfr} by considering the means and robust deviations of various derived parameters for the MWA sample. For the integrated light-weighted age, we obtain a value of $log(age/Gyr) = 0.41 \pm 0.27$; however, if we restrict to MWAS with $\Delta_{\rm SFR} < 3$, we instead find $log(age/Gyr) = 0.51 \pm 0.15$, indicating a significant decrease in the spread of measured values. We find for the full MWA sample a metallicity mean and deviation of $[Z/H] = -0.31 \pm 0.13$; restricting to $\Delta_{\rm SFR} < 3$ in this instance, we find $[Z/H] = -0.25 \pm 0.09$, finding the $\Delta_{\rm SFR}$ cut to lead to a modest reduction in the spread of obtained values. Take the age gradient for all galaxies with sufficient coverage in MaNGA, we find $\nabla (age) = -0.16 \pm 0.16$ dex/$R_d$; performing the same $\Delta_{\rm SFR}$ cut as before, we find $\nabla (age) = -0.16 \pm 0.12$ dex/$R_d$, again finding a slight reduction in the spread of measured values. For the metallicity gradient, meanwhile, we find -0.09 $\pm$ 0.11 dex/$R_e$ and -0.12 $\pm$ 0.12 dex/$R_e$ before and after restricting to $\Delta_{\rm SFR} < 3$ respectively, finding  no reduction in the spread of values in this case.

\begin{figure*}
\begin{center}
	\includegraphics[trim = 2.cm 2.8cm 0cm 7cm,scale=0.95]{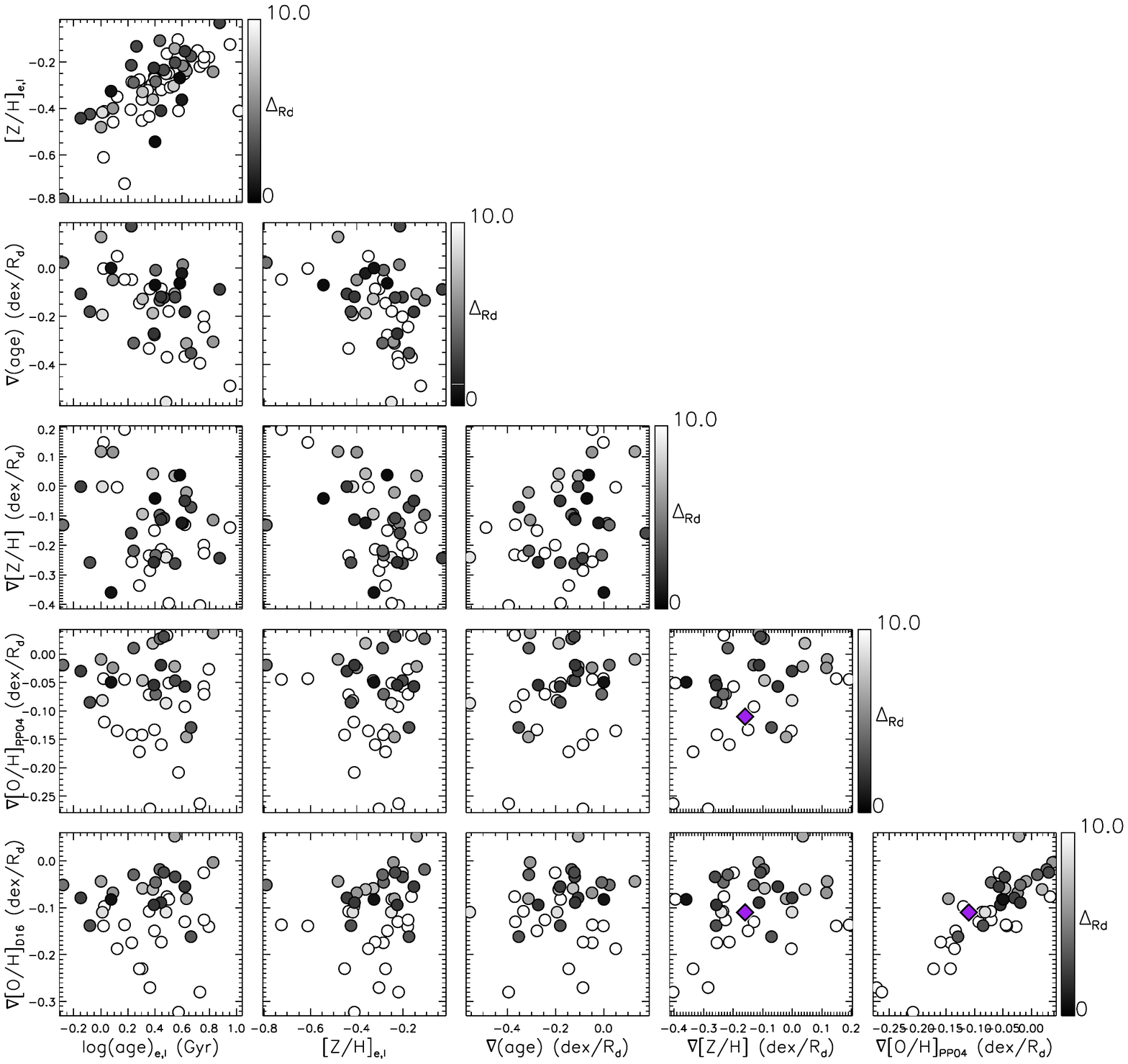}
	\caption{Corner plot showing light-weighed stellar ages and metallicities measured within 1~$R_e$ ellipses, along with stellar age and metallicity gradients and gas metallicity gradients. Datapoints are coloured according to $\Delta_{Rd}$. The majority of our sample galaxies have high $\Delta_{Rd}$, and proximity to the MW in terms of $R_d$ appears to have little relation to the plotted parameters, though we note a preference for flatter gas metallicity gradients. The purple diamonds show the \citet{hasselquist2019} metallicity gradient for young MW stars and the \citet{esteban2017} gradients for MW HII regions.
	}
	\label{mparamsrd}
	\end{center}
\end{figure*}

In \autoref{mparamsrd}, we present the same measurements of parameters and gradients, with datapoints coloured by $\Delta_{\rm Rd}$. We find little dependence on datapoint color with any plotted properties here aside from $\nabla[O/H]$, for which proximity to the MW in terms of $R_d$ favours smaller absolute values; as shown in the figure, and as already implied in \autoref{gasmetalgradrd}, this means that having a MW-like scale length disfavors galaxies from having MW-like gas metallicity gradients. Although the MW has been repeatedly suggested to have an unusually small scale length, deviation from the measured MW scale length does not to be correlated with the measured stallar population parameters we obtain. As an example of this, we take the means and standard deviations the integrated light-weighted ages and restrict to galaxies with $\Delta_{\rm Rd} < 3$. We find $\log (age/Gyr) = 0.41 \pm 0.26$, and thus find that proximity to the MW scale length does not lead to increased age homogeniety in any meaningful way.

Overall, we predict that considering SFR values in analog sample definition could allow such analogs to be more constraining of the stellar age properties that an external observer would obtain for the MW. As such, an analysis of a sample selected simultaneously on stellar mass, B/T, and SFR would be a promising area of future work.

\section{Discussion}\label{disc}

In the preceding sections, we presented measurements for the MWAs' stellar kinematics, stellar populations, and ionised gas properties. We also compared the gaseous and stellar metallicity gradients of our galaxies with equivalent gradients calculated for the Milky Way. We note at this point that our measurements are overall in good qualitivative consistency with both the Pipe3D Value Added Catalog\footnote{available at https://www.sdss.org/dr14/manga/manga-data/manga-pipe3d-value-added-catalog/} (VAC) \citep{sanchez2018} and the DAP, as demonstrated in Appendix B. As such, the findings we will now discuss do not depend significantly on our choices of adopted procedures.

We find our galaxies to have a wide range of central light-weighted stellar ages, along with a wide range of central gas ionisation sources. As argued in the preceding section, both of these points are consistent with the emerging view of the MW as a ``green valley" galaxy in the process of transitioning to quiescence; it is natural to expect analogs of the MW to be at different stages of such a transition, which is reflected in the range of stellar and gaseous properties we find. On the other hand, our results further challenge the idea of the MW being a typical star-forming spiral galaxy, which we would expect MWA properties to reflect if this were the case.

We identified seven MWA galaxies with central AGN-like emission. This is interesting, considering that the MW itself possibly hosted an AGN in its past. However, we caution that none of the seven identified AGN galaxies in this sample contain broad-line spectra; indeed, since these galaxies also all display old central ages, it remains possible that it is stellar populations that are ionising the gas in these galaxies' centres.

We find our galaxies' gas and stellar metallicity gradients to generally be significantly flatter than those calculated for the MW. This is particularly the case for gradients measured in units of dex/kpc, wheras for gradients measured in dex/$R_d$ we obtain significantly better agreement with Galactic values. Such an agreement in scaled units is good consistency with previous work: \citet{belfiore2017} compare their gas gradients to MW metallicity gradients calculated from Cepheid data \citep{genovali2014} and similarly find good consistency between the MW gradients in terms of dex/$R_e$ and the gradient predicted from their data for galaxies of MW-like mass.  As such, we argue the discrepancies in dex/kpc gradients to be largely a reflection of the MW's apparent morphological compactness with respect to the bulk of the MWA sample.

It has previously been suggested that the MW's disk scale length is atypically short. \citet{bovy2013}, for instance, find a mass-weighted scale length of $2.15$ kpc $\pm 0.14$ kpc that is well below the $\approx 4.5$ kpc $\pm 2$ kpc that would be expected for disc galaxies with the MW's mass, based on the \citet{pizagno2005} sample of disk galaxies. \citet{licquia2016a} likewise report their calculated MW $R_d$ to be anomalously small with respect to the MW's luminosity and stellar rotational velocity. The MW disk scale length from \citet{licquia2016} is indeed shorter than the vast majority of our MWA galaxies, further supporting the view of the MW disk being more compact than most galaxies of its mass. 

An interesting analysis in the future, therefore, woiuld be to use $R_d$ as an additional sample selection parameter in a future MWA sample, before performing a similar analysis to that presented in this paper as well as that presented in L15. One could then test if such a selection reduces the scatter of measured properties in other parameter spaces, as well as whether $R_d$ scales with any of these properties. Such an analysis could not be done in detail with this present sample, seeing as most of the galaxies have $R_d$ values well beyond what has been calculated for the MW.

It would also be worthwhile to explore SFR as a Milky Way Analog selection parameter. We found in Section 4.6 that galaxies close to the MW in term of SFR appear to have a tighter range of integrated ages age gradients and those that are not. Such a finding is perhaps not surprising. An exploration to this end could include galaxies selected using M$_*$, SFR and B/T simultaneously; at the same time, one could analyse the stellar population and gas properties of galaxies selected using M$_*$ and SFR alone. Comparing and contrasting the bulk properties of analog samples selected via different means would allow us to further investigate the Milky Way's place within the wider galaxy population, particularly with respect to the question of whether the Milky Way is typical amongst galaxies of its kind. It remains unclear whether structural analogs are more or less constraining of the MW than "star-formation analogs" selected on M$_*$ and SFR, so a such a comparison would be timely to perform.

It is important to remember that a number of observational challenges exist in performing a true ``apples to apples'' comparison between the MW and other galaxies. Crucially, our MWA metallicity gradients are based on the integrated light of galaxies; the equivalent MW values are calculated through combined measurements of individual objects, be they individual HII regions or individual stars. As such, dilution and smoothing effects in integrated light data are a concern when comparing other galaxies to the MW \citep[e.g.][]{mast2014}, particularly for galaxies that are viewed close to edge-on; this is another possible explanation for the relative flatness of the MWA metallicity gradients in physical units when compared to the MW. We also note that we do not directly account for the MaNGA PSF in this study, though given that we do not fit for radial gradients within $0.5R_e$ we do not expect the PSF to have a significant effect. We additionally note that many of the galaxies in our sample were observed using the smaller IFU bundle sizes, and have per-spaxel S/N ratios that necessitated significant binning; these can also serve to flatten measured stellar population ratios, as demonstrated for instance in \citet{ibarramedel2019} using simulated MaNGA data.

A related challenge in the case of stellar population measurements is the fact that total population gradients can appear very different to those of individual mono-age populations. This can be seen for instance in the results of \citet{minchev2013}. \citet{minchev2013} find a strong negative metallicity gradient in their chemo-dynamical model at the solar region, even though the individual mono-age populations in the model displayed metallicity gradients that were close to flat. This means that one should ideally split observed stellar populations up in terms of age or birth radius \citep{minchev2019} before calculating metallicity gradients, which is increasingly feasible in MW studies but far more challenging in extragalactic work. 

It is also important to consider the definition of a ``Milky Way analog''. There is no single definition of what makes a galaxy a ``Milky Way analog'', meaning that the definition in a given study will depend on the goals of that study. In our case, we have chosen to define a MWA as a galaxy that is similar to the MW in terms of mass and bulge-to-total ratio; we do not assert that galaxies selected in this way match the MW in detail, but rather we simply make the assumption that the MW should not be wildly different than galaxies close to it in this parameter space. It is certainly possible that by selecting analogs using a larger number of parameters at once, one could tighten the range of measured properties in MWA galaxies; however, such a selection does not produce a sufficiently large sample in MaNGA. We also reiterate our previous comment that the disk scale radius $R_d$ could be a useful selection parameter for the purpose of selecting MWAs, given the apparent relative compactness of the MW when compared to most other galaxies of its mass. At the same time, it should be remembered that ``disk scale lengths'' measured in external galaxies are perhaps not perfectly equivalent to those reported in the MW. Considering differently-aged stars separately, it is known the MW scale length is larger for younger stellar populations \citep[e.g.][]{bensby2011,anders2014,martig2016} - a fact that cannot be captured by a single disc scale length measure. Such a situation is also expected for external galaxies, given the large amounts of observational evidence for inside-out galaxy formation \citep[e.g.][]{perez2013,lopezfernandez2018, rowlands2018}.

Overall, the Milky Way appears to contain metallicity gradients that are steeper (in dex/kpc) than for a typical MWA from this sample. Part of this discrepancy could be due to difference between the internal perspective offered by the MW and the external perspective available for other galaxies. Another could be that the MW's disc is indeed unusually compact for a galaxy of its type. Further investigation of MWAs using complimentary selection criteria will enable us to further understand the Milky Way's place within the wider galaxy population, and such investigation will be the subject of further work.

\section{Summary and conclusion}\label{conclusion}

We have presented a sample of 62 galaxies selected as Milky Way Analogs, observed as part of the MaNGA IFU survey. We have analysed the galaxies in terms of their stellar kinematics, stellar populations, and ionised gas properties, and we compared the galaxies' gas-phase and stellar metallicity gradients to equivalent measurements made for the MW itself. As indicated in Appendix B, our measurements are in good qualitative consistency with those obtained with other methods.

To summarise, we find that our MWAs have a wide range of stellar population properties, which we interpret as reflecting the MWAs in various stages of a transition to from star-forming to quiescent. We find gas ionisation properties that are in support of this view, finding both MWAs dominated by star-forming gas and MWAs with a significant component of gas ionised by old stars. We also find a significant fraction of the MWAs to contain central ionised gas emission consistent with AGN or composite activity, which is interesting in light of evidence that the MW itself contained an AGN in its past. We found the MWAs to contain radial stellar metallicity gradients and gas-phase metallicity gradients that were flatter than inferred for the MW in physical units, but find greater consistency betetween the MWAs and our own Galaxy when scaling by disk scale length. We also noted the scaled gas metallicity gradients of galaxies to be steeper on average for galaxies with larger scale lengths, possibly as a consequence of our particular sample selection.

The results we have presented in this paper contains much opportunity for follow-up work. In particular, we will explore alternative criteria for selecting MWAs, such as star-formation rates and galaxy morphologioes. We also intend to explore MWAs' neutral gas contents, with the ongoing HI-MaNGA program \citep{masters2019} providing an ideal opportunity in this regard. We also plan in the future to perform a dynamical modelling study on a selected subset of MWAs.

\section*{Acknowledgements}

We thank Mike Merrifield for his useful feedback on this manuscript. The support and resources from the Center for High Performance Computing at the University of Utah are gratefully acknowledged. We thank the anonymous referee for their constructive comments, which served to improve various aspects of this manuscript.

Funding for the Sloan Digital Sky Survey IV has been provided by the Alfred P. Sloan Foundation, the U.S. Department of Energy Office of Science, and the Participating Institutions. SDSS-IV acknowledges support and resources from the Center for High-Performance Computing at the University of Utah. The SDSS web site is \url{www.sdss.org}.

SDSS-IV is managed by the Astrophysical Research Consortium for the Participating Institutions of the SDSS Collaboration including the Brazilian Participation Group, the Carnegie Institution for Science, Carnegie Mellon University, the Chilean Participation Group, the French Participation Group, Harvard-Smithsonian Center for Astrophysics, Instituto de Astrof\'isica de Canarias, The Johns Hopkins University, Kavli Institute for the Physics and Mathematics of the Universe (IPMU) / University of Tokyo, Lawrence Berkeley National Laboratory, Leibniz Institut f\"ur Astrophysik Potsdam (AIP),  Max-Planck-Institut f\"ur Astronomie (MPIA Heidelberg), Max-Planck-Institut f\"ur Astrophysik (MPA Garching), Max-Planck-Institut f\"ur Extraterrestrische Physik (MPE), National Astronomical Observatories of China, New Mexico State University, New York University, University of Notre Dame, Observat\'ario Nacional / MCTI, The Ohio State University, Pennsylvania State University, Shanghai Astronomical Observatory, United Kingdom Participation Group, Universidad Nacional Aut\'onoma de M\'exico, University of Arizona, University of Colorado Boulder, University of Oxford, University of Portsmouth, University of Utah, University of Virginia, University of Washington, University of Wisconsin, Vanderbilt University, and Yale University.

\bibliographystyle{mnras}
\bibliography{bibliography}

\appendix

\section{Summary of measurements}

We give the stellar population gradients, gas-phase metallicity gradients, and gas-ionisation classifications in \autoref{table3}. In supplementary online material, we show maps of the parameters measured for each of the MWAs via pPXF fitting.

\begin{table*}
\begin{center}
\begin{tabular}{c|c|c|c|c|c|c|c}
Plate & IFU Dsg & $\nabla(Age_{light})$ & $\nabla(Age_{mass})$ & $\nabla([Z/H]_{light}$ & $\nabla([Z/H]_{mass})$ & $\nabla[\log(O/H)]_{PP04}$ & Ion. class.\\[0.3 pt]
\hline
\hline
8079 & 12705 & -0.07 $\pm$ 0.087 & -0.08 $\pm$ 0.060 & 0.19 $\pm$ 0.13 & 0.16 $\pm$ 0.17 & -0.04 $\pm$ 0.016 & SF\\[0.3 pt]
8085 & 12704 & -0.41 $\pm$ 0.086 & -0.07 $\pm$ 0.062 & -0.02 $\pm$ 0.10 & -0.24 $\pm$ 0.10 & -0.12 $\pm$ 0.022 & CI\\[0.3 pt]
8137 & 12703 & -0.39 $\pm$ 0.083 & -0.14 $\pm$ 0.081 & -0.15 $\pm$ 0.12 & -0.22 $\pm$ 0.11 & -0.11 $\pm$ 0.016 & CI\\[0.3 pt]
8141 & 12701 & -0.75 $\pm$ 0.094\textsuperscript{a} & -0.23 $\pm$ 0.12\textsuperscript{a} & -0.24 $\pm$ 0.096\textsuperscript{a} & -0.14 $\pm$ 0.13\textsuperscript{a} & -0.07 $\pm$ 0.011\textsuperscript{a} & CI\\[0.3 pt]
8146 & 12702 & -0.13 $\pm$ 0.099 & 0.00 $\pm$ 0.048 & -0.21 $\pm$ 0.12 & -0.24 $\pm$ 0.14 & -0.15 $\pm$ 0.018 & AGN\\[0.3 pt]
8241 & 3704 & -0.29 $\pm$ 0.091 & -0.20 $\pm$ 0.088 & 0.00 $\pm$ 0.14 & -0.21 $\pm$ 0.14 & -0.07 $\pm$ 0.011 & SF\\[0.3 pt]
8244 & 6101 & -0.30 $\pm$ 0.094\textsuperscript{a} & -0.08 $\pm$ 0.064\textsuperscript{a} & 0.042 $\pm$ 0.12\textsuperscript{a} & -0.18 $\pm$ 0.11\textsuperscript{a} & 0.017 $\pm$ 0.016\textsuperscript{a} & CI\\[0.3 pt]
8244 & 9101 & -0.55 $\pm$ 0.096\textsuperscript{a} & -0.09 $\pm$ 0.073\textsuperscript{a} & -0.13 $\pm$ 0.096\textsuperscript{a} & -0.05 $\pm$ 0.10\textsuperscript{a} & -0.08 $\pm$ 0.045\textsuperscript{a} & CI\\[0.3 pt]
8257 & 12705 & -0.45 $\pm$ 0.19\textsuperscript{a} & -0.32 $\pm$ 0.20\textsuperscript{a} & -0.50 $\pm$ 0.21\textsuperscript{a} & -0.59 $\pm$ 0.33\textsuperscript{a} & -0.28 $\pm$ 0.019\textsuperscript{a} & AGN\\[0.3 pt]
8263 & 6104 & -0.27 $\pm$ 0.064 & -0.06 $\pm$ 0.051 & -0.05 $\pm$ 0.075 & -0.05 $\pm$ 0.069 & -0.05 $\pm$ 0.047\textsuperscript{a} & AGN\\[0.3 pt]
8315 & 12705 & 0.00 $\pm$ 0.097\textsuperscript{a} & -0.08 $\pm$ 0.049\textsuperscript{a} & 0.15 $\pm$ 0.14\textsuperscript{a} & -0.08 $\pm$ 0.14\textsuperscript{a} & -0.04 $\pm$ 0.019 & CI\\[0.3 pt]
8444 & 12703 & -0.07 $\pm$ 0.099 & -0.14 $\pm$ 0.067 & -0.26 $\pm$ 0.13 & -0.24 $\pm$ 0.12 & -0.13 $\pm$ 0.023 & SF\\[0.3 pt]
8549 & 3703 & -0.57 $\pm$ 0.072 & -0.05 $\pm$ 0.016 & -0.14 $\pm$ 0.096 & -0.15 $\pm$ 0.087 & -0.01 $\pm$ 0.034 & CR\\[0.3 pt]
8567 & 3701 & -0.19 $\pm$ 0.069\textsuperscript{a} & -0.06 $\pm$ 0.044\textsuperscript{a} & -0.11 $\pm$ 0.084\textsuperscript{a} & -0.12 $\pm$ 0.083\textsuperscript{a} & 0.029 $\pm$ 0.031 & SF\\[0.3 pt]
8568 & 6103 & -0.27 $\pm$ 0.086\textsuperscript{a} & -0.09 $\pm$ 0.047\textsuperscript{a} & -0.40 $\pm$ 0.11\textsuperscript{a} & -0.34 $\pm$ 0.10\textsuperscript{a} & -0.05 $\pm$ 0.010\textsuperscript{a} & SF\\[0.3 pt]
8595 & 3702 & -0.42 $\pm$ 0.11 & -0.02 $\pm$ 0.094 & -0.14 $\pm$ 0.13 & -0.24 $\pm$ 0.14 & -0.15 $\pm$ 0.050\textsuperscript{a} & CR\\[0.3 pt]
8601 & 6103 & -0.09 $\pm$ 0.060 & -0.02 $\pm$ 0.041 & 0.038 $\pm$ 0.081 & -0.01 $\pm$ 0.072 & -0.01 $\pm$ 0.037\textsuperscript{a} & CI\\[0.3 pt]
8613 & 12702 & -0.50 $\pm$ 0.092 & -0.07 $\pm$ 0.050 & -0.22 $\pm$ 0.11 & -0.10 $\pm$ 0.10 & -0.23 $\pm$ 0.021 & CR\\[0.3 pt]
8614 & 6101 & -0.31 $\pm$ 0.079\textsuperscript{a} & -0.32 $\pm$ 0.19\textsuperscript{a} & -0.26 $\pm$ 0.12\textsuperscript{a} & -0.41 $\pm$ 0.19\textsuperscript{a} & -0.09 $\pm$ 0.0068\textsuperscript{a} & SF\\[0.3 pt]
8615 & 9102 & -0.40 $\pm$ 0.090 & -0.08 $\pm$ 0.058 & -0.26 $\pm$ 0.12 & 0.11 $\pm$ 0.12 & -0.05 $\pm$ 0.023\textsuperscript{a} & CI\\[0.3 pt]
8934 & 12701 & -0.22 $\pm$ 0.11 & -0.13 $\pm$ 0.096 & -0.34 $\pm$ 0.14 & -0.21 $\pm$ 0.16 & -0.17 $\pm$ 0.020 & CI\\[0.3 pt]
8947 & 12703 & -0.38 $\pm$ 0.091\textsuperscript{a} & -0.05 $\pm$ 0.065\textsuperscript{a} & -0.23 $\pm$ 0.099\textsuperscript{a} & -0.09 $\pm$ 0.097\textsuperscript{a} & -0.07 $\pm$ 0.065 & CR\\[0.3 pt]
8978 & 9102 & -0.14 $\pm$ 0.075\textsuperscript{a} & 0.00 $\pm$ 0.026\textsuperscript{a} & -0.24 $\pm$ 0.089\textsuperscript{a} & -0.27 $\pm$ 0.062\textsuperscript{a} & N/A & CR\\[0.3 pt]
8979 & 6102 & -0.45 $\pm$ 0.11 & -0.11 $\pm$ 0.083 & -0.13 $\pm$ 0.16 & -0.01 $\pm$ 0.15 & -0.12 $\pm$ 0.026 & AGN\\[0.3 pt]
8979 & 9101 & -0.43 $\pm$ 0.11 & -0.01 $\pm$ 0.065 & -0.23 $\pm$ 0.12 & -0.10 $\pm$ 0.12 & -0.03 $\pm$ 0.099 & CR\\[0.3 pt]
8985 & 9102 & -0.45 $\pm$ 0.088 & -0.17 $\pm$ 0.086 & -0.40 $\pm$ 0.10 & -0.34 $\pm$ 0.097 & -0.19 $\pm$ 0.037 & CR\\[0.3 pt]
9028 & 9101 & -0.20 $\pm$ 0.10 & -0.43 $\pm$ 0.17 & -0.30 $\pm$ 0.16 & -0.31 $\pm$ 0.18 & -0.14 $\pm$ 0.022 & SF\\[0.3 pt]
9031 & 12703 & -0.36 $\pm$ 0.21\textsuperscript{a} & -0.17 $\pm$ 0.15\textsuperscript{a} & -0.52 $\pm$ 0.21\textsuperscript{a} & -0.55 $\pm$ 0.22\textsuperscript{a} & N/A & CI\\[0.3 pt]
9040 & 6102 & -0.22 $\pm$ 0.089\textsuperscript{a} & -0.15 $\pm$ 0.094\textsuperscript{a} & -0.10 $\pm$ 0.11\textsuperscript{a} & -0.11 $\pm$ 0.094\textsuperscript{a} & 0.026 $\pm$ 0.032\textsuperscript{a} & CR\\[0.3 pt]
9040 & 9102 & -0.44 $\pm$ 0.069 & -0.06 $\pm$ 0.033 & -0.11 $\pm$ 0.079 & -0.12 $\pm$ 0.067 & 0.033 $\pm$ 0.025 & AGN\\[0.3 pt]
9095 & 9102 & 0.28 $\pm$ 0.091 & -0.08 $\pm$ 0.075 & -0.16 $\pm$ 0.11 & -0.18 $\pm$ 0.12 & -0.02 $\pm$ 0.025 & SF\\[0.3 pt]
9189 & 9101 & -0.01 $\pm$ 0.085 & -0.09 $\pm$ 0.073 & -0.23 $\pm$ 0.100 & -0.20 $\pm$ 0.086 & -0.06 $\pm$ 0.029 & CI\\[0.3 pt]
9192 & 12704 & 0.073 $\pm$ 0.087 & -0.05 $\pm$ 0.029 & 0.00 $\pm$ 0.14 & -0.15 $\pm$ 0.12 & -0.12 $\pm$ 0.022 & SF\\[0.3 pt]
9196 & 6104 & -0.08 $\pm$ 0.093 & -0.34 $\pm$ 0.14 & 0.12 $\pm$ 0.14 & -0.04 $\pm$ 0.13 & -0.02 $\pm$ 0.024 & SF\\[0.3 pt]
9196 & 9101 & -0.39 $\pm$ 0.16 & -0.10 $\pm$ 0.12 & -0.28 $\pm$ 0.18 & 0.044 $\pm$ 0.20 & -0.13 $\pm$ 0.035 & CR\\[0.3 pt]
9485 & 1901 & 0.043 $\pm$ 0.052 & -0.04 $\pm$ 0.032 & -0.13 $\pm$ 0.082 & -0.14 $\pm$ 0.069 & -0.02 $\pm$ 0.0054 & SF\\[0.3 pt]
9486 & 12702 & -0.17 $\pm$ 0.15 & -0.11 $\pm$ 0.14 & 0.12 $\pm$ 0.19 & -0.19 $\pm$ 0.21 & -0.11 $\pm$ 0.063 & CI\\[0.3 pt]
9487 & 12701 & -0.60 $\pm$ 0.20 & -0.16 $\pm$ 0.29 & -0.26 $\pm$ 0.23 & -0.65 $\pm$ 0.27 & -0.21 $\pm$ 0.051 & SF\\[0.3 pt]
9491 & 12704 & 0.20 $\pm$ 0.084\textsuperscript{a} & -0.05 $\pm$ 0.032\textsuperscript{a} & 0.12 $\pm$ 0.14\textsuperscript{a} & -0.08 $\pm$ 0.12\textsuperscript{a} & -0.01 $\pm$ 0.021\textsuperscript{a} & SF\\[0.3 pt]
9496 & 9102 & -0.47 $\pm$ 0.075 & -0.08 $\pm$ 0.030 & -0.07 $\pm$ 0.097 & -0.13 $\pm$ 0.088 & -0.11 $\pm$ 0.014\textsuperscript{a} & AGN\\[0.3 pt]
9499 & 12703 & -0.50 $\pm$ 0.080 & -0.37 $\pm$ 0.11 & -0.23 $\pm$ 0.11 & -0.04 $\pm$ 0.094 & 0.026 $\pm$ 0.026 & AGN\\[0.3 pt]
9506 & 3701 & -0.19 $\pm$ 0.065 & -0.11 $\pm$ 0.048 & 0.00 $\pm$ 0.10 & -0.22 $\pm$ 0.11 & -0.03 $\pm$ 0.0093\textsuperscript{a} & SF\\[0.3 pt]
9514 & 9101 & -0.20 $\pm$ 0.074\textsuperscript{a} & -0.04 $\pm$ 0.039\textsuperscript{a} & -0.26 $\pm$ 0.10\textsuperscript{a} & -0.14 $\pm$ 0.088\textsuperscript{a} & -0.05 $\pm$ 0.020\textsuperscript{a} & CR\\[0.3 pt]
9866 & 12701 & -0.21 $\pm$ 0.11 & 0.0041 $\pm$ 0.088 & -0.09 $\pm$ 0.12 & -0.01 $\pm$ 0.15 & -0.05 $\pm$ 0.021 & CR\\[0.3 pt]
9868 & 12705 & -0.47 $\pm$ 0.081 & -0.27 $\pm$ 0.097 & -0.23 $\pm$ 0.13 & -0.34 $\pm$ 0.13 & -0.06 $\pm$ 0.0096 & CI\\[0.3 pt]
9894 & 6104 & 0.00031 $\pm$ 0.091 & -0.10 $\pm$ 0.085 & -0.36 $\pm$ 0.14 & -0.18 $\pm$ 0.14 & -0.04 $\pm$ 0.015 & SF\\[0.3 pt]
10213 & 1902 & -0.69 $\pm$ 0.39\textsuperscript{a} & -0.33 $\pm$ 0.51\textsuperscript{a} & -0.29 $\pm$ 0.49\textsuperscript{a} & 0.21 $\pm$ 0.58\textsuperscript{a} & -0.05 $\pm$ 0.058\textsuperscript{a} & SF\\[0.3 pt]
10215 & 9102 & -0.03 $\pm$ 0.10 & -0.09 $\pm$ 0.088 & -0.12 $\pm$ 0.087 & 0.039 $\pm$ 0.096 & N/A & CI\\[0.3 pt]
10215 & 12705 & -0.33 $\pm$ 0.12\textsuperscript{a} & -0.52 $\pm$ 0.25\textsuperscript{a} & -0.18 $\pm$ 0.18\textsuperscript{a} & -0.14 $\pm$ 0.17\textsuperscript{a} & -0.15 $\pm$ 0.037\textsuperscript{a} & CI\\[0.3 pt]
10216 & 6102 & -0.09 $\pm$ 0.088\textsuperscript{a} & -0.22 $\pm$ 0.10\textsuperscript{a} & -0.09 $\pm$ 0.091\textsuperscript{a} & -0.14 $\pm$ 0.095\textsuperscript{a} & N/A & CI\\[0.3 pt]
10216 & 6104 & -0.14 $\pm$ 0.16 & -0.15 $\pm$ 0.17 & -0.16 $\pm$ 0.17 & -0.15 $\pm$ 0.19 & -0.02 $\pm$ 0.023 & SF\\[0.3 pt]
10217 & 3703 & -0.50 $\pm$ 0.046 & -0.07 $\pm$ 0.018 & -0.22 $\pm$ 0.073 & -0.31 $\pm$ 0.067 & 0.011 $\pm$ 0.0061 & CI\\[0.3 pt]
10218 & 12701 & -0.16 $\pm$ 0.096 & -0.05 $\pm$ 0.062 & 0.035 $\pm$ 0.10 & -0.08 $\pm$ 0.10 & -0.02 $\pm$ 0.10\textsuperscript{a} & SF\\[0.3 pt]
10220 & 6104 & -0.11 $\pm$ 0.14 & -0.08 $\pm$ 0.20 & -0.04 $\pm$ 0.12 & 0.056 $\pm$ 0.18 & -0.01 $\pm$ 0.033 & SF\\[0.3 pt]
10220 & 12702 & -0.28 $\pm$ 0.088 & -0.02 $\pm$ 0.061 & -0.20 $\pm$ 0.096 & -0.06 $\pm$ 0.096 & -0.05 $\pm$ 0.016 & CI\\[0.3 pt]
10220 & 12705 & -0.66 $\pm$ 0.11 & -0.18 $\pm$ 0.11 & -0.17 $\pm$ 0.13 & -0.29 $\pm$ 0.16 & -0.12 $\pm$ 0.035 & CI\\[0.3 pt]
10493 & 3701 & -0.12 $\pm$ 0.11 & -0.03 $\pm$ 0.069 & 0.00098 $\pm$ 0.14 & -0.18 $\pm$ 0.18 & -0.08 $\pm$ 0.029 & SF\\[0.3 pt]
10494 & 6101 & -0.07 $\pm$ 0.13 & 0.00 $\pm$ 0.12 & -0.29 $\pm$ 0.11 & -0.61 $\pm$ 0.18 & -0.14 $\pm$ 0.031 & SF\\[0.3 pt]
10497 & 6101 & -0.19 $\pm$ 0.083 & -0.06 $\pm$ 0.043 & -0.11 $\pm$ 0.13 & -0.34 $\pm$ 0.091 & -0.02 $\pm$ 0.024 & CI\\[0.3 pt]
10498 & 3701 & 0.023 $\pm$ 0.094 & -0.08 $\pm$ 0.082 & -0.13 $\pm$ 0.10 & -0.14 $\pm$ 0.095 & -0.20 $\pm$ 0.10 & CR\\[0.3 pt]
10509 & 3704 & 0.59 $\pm$ 0.071\textsuperscript{a} & -0.01 $\pm$ 0.039\textsuperscript{a} & 0.0086 $\pm$ 0.088\textsuperscript{a} & -0.20 $\pm$ 0.075\textsuperscript{a} & -0.07 $\pm$ 0.079\textsuperscript{a} & SF\\[0.3 pt]
10517 & 12703 & -0.39 $\pm$ 0.27 & -0.13 $\pm$ 0.29 & -0.09 $\pm$ 0.28 & -0.84 $\pm$ 0.37 & -0.24 $\pm$ 0.032 & CI\\[0.3 pt]
\end{tabular}
\end{center}
\caption{Stellar population and gas-phase metalicity gradients and (formal) errors measured for the MWA sample, along with gas-ionisation classifications. Gradients are in units of dex/$R_e$. Gradients marked with \textsuperscript{a} superscripts were excluded from our analysis. Galaxies with no gas metallicity gradient are those with less than three star-forming spaxels between 0.5 $R_e$ and 1.5 $R_e$.}
\label{table3}
\end{table*}

\section{Comparison with previous analyses}

A number of analyses have been performed on the MaNGA sample to date, several of which have generated publicly-available VACs. It is worthwhile, then, to compare our findings with those that could be derived from the VACs alone.

Stellar population parameters can be obtained from the Pipe3D VAC, as well as the \citet{goddard2017} VAC produced from calculations with the Firefly code \citep{wilkinson2017}. Both of these VACs, we note, were generated from studies of large statistical MaNGA samples, and so are not necessarily ideal for analyzing smaller and more specialised samples such as that which is presented in this paper.

\begin{figure}
\begin{center}
	\includegraphics[trim = 3cm 2cm 0cm 11cm,scale=0.5]{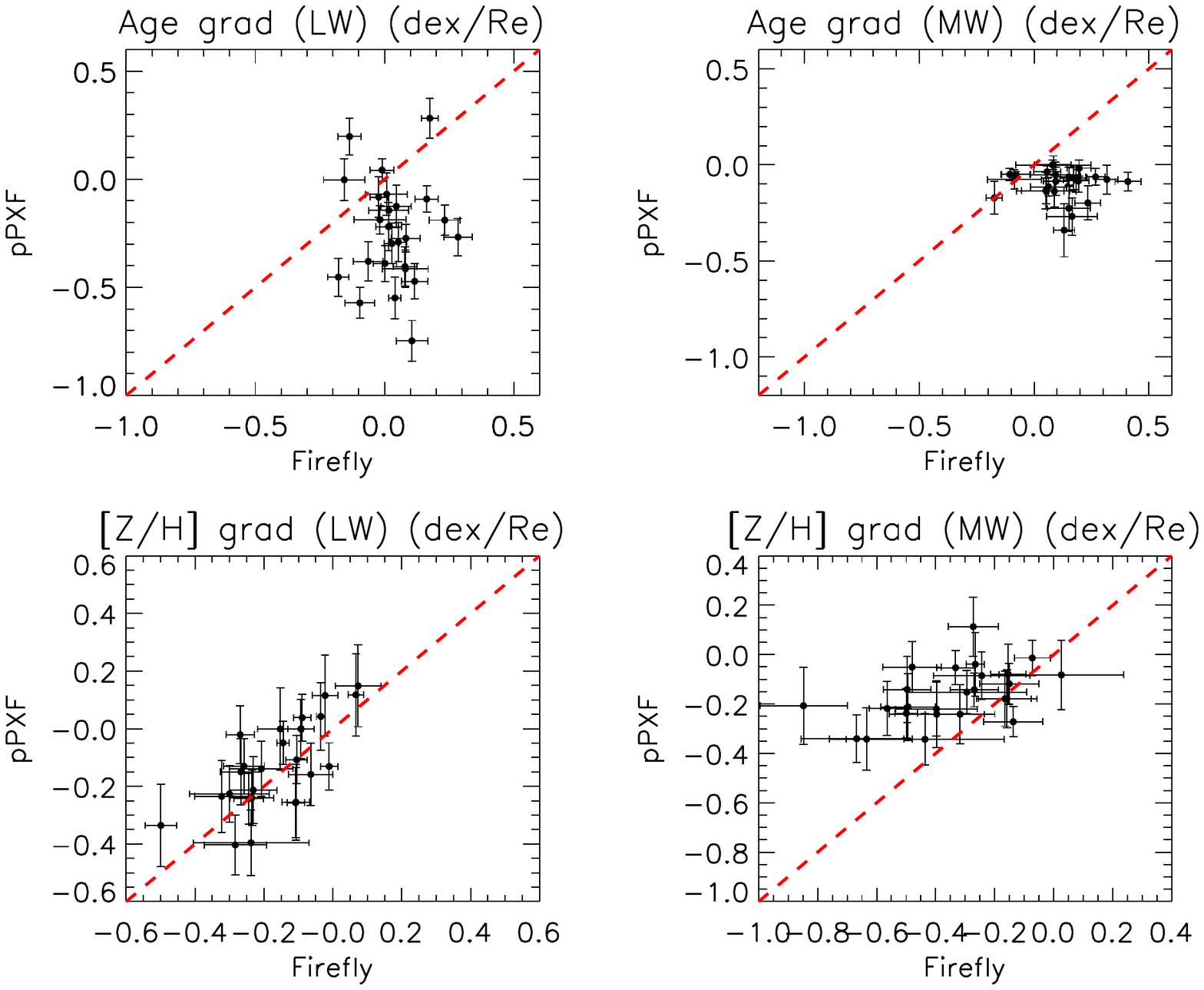}
	\caption{Comparison of population gradients derived from our pPXF implementation with those reported on the MaNGA firefly VAC for MWAs avilable in MPL-7. We find the two sets of gradients to not be consistent with one another. Notably, the Firefly VAC typically reports light-weighted age gradients that are flat or slightly positive, whereas we find negative gradients for the light-weighted ages}
	\label{ppxffirefly}
	\end{center}
\end{figure}

In \autoref{ppxffirefly}, we compare our calculated stellar population gradients (age and metallicity, light-weighted and mass-weighted) with those obtained from the Firefly VAC, with the latter scaled to account for different $R_e$ values (Sersic fit $R_e$ instead of eliptical Petrosian aperture values). Since the Firefly VAC is not available in MPL-8 as of writing, we restrict this comparison to MWAs available in MPL-7 or earlier. This is not a true apples-to-apples comparison, since the Firefly VAC is calculated over a difference radial range (0-1.5 $R_e$). However, we may immediately see that this VAC is not even in qualitative consistency with our own results for this galaxy sample: while our metallicity gradients (particularly light-weighted values) agree reasonably well, the Firefly measurements indicate flat-to-positive age gradients, contrary to our own findings. We note that firefly uses linear stellar ages and metallcities when calculating light-weighted and mass-weighted values \footnote{As implemented in StellarPopulationModel.py and firefly\_library.py, available at https://github.com/FireflySpectra/firefly\_release}, rather than logarithmic ages and metallicities as done in our own analysis, which can potentially lead to higher weighting of old metal-rich stellar populations \citep[e.g.][]{zheng2017}

\begin{figure}
\begin{center}
	\includegraphics[trim = 3cm 2cm 0cm 11cm,scale=0.5]{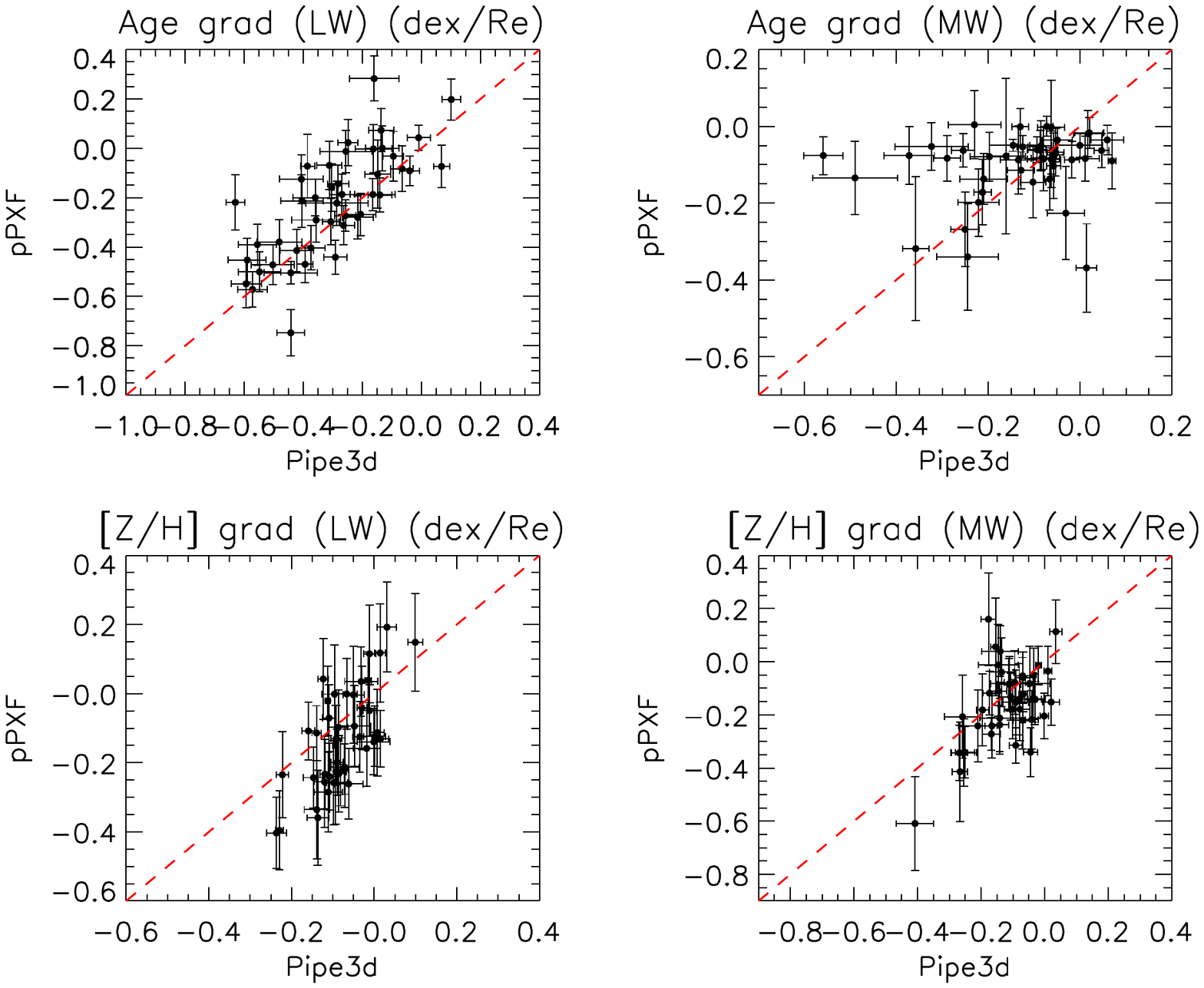}
	\caption{Comparison of population gradients derived from our pPXF implementation with those reported on the MaNGA Pipe3D VAC. We find reasonable consistency between the two sets of gradients overall - particularly for light-weighted age - though with significant levels of scatter.}
	\label{ppxfpipe3d}
	\end{center}
\end{figure}

Next, we consider how our population gradients compare with Pipe3D in \autoref{ppxfpipe3d}. Here, we find much better consistency in the light-weighted age gradients, while also finding reasonable consistency in the metallicity gradients. The comparison with mass-weighted age gradients, meanwhile, is heavily scattered. We note that Pipe3D uses logarithmic ages and metallicities, in a similar manner to our own analysis, when calculating light-weighted and mass-weighted values \citep{sanchez2016}.

\begin{figure}
\begin{center}
	\includegraphics[trim = 3.5cm 2cm 0cm 11cm,scale=0.5]{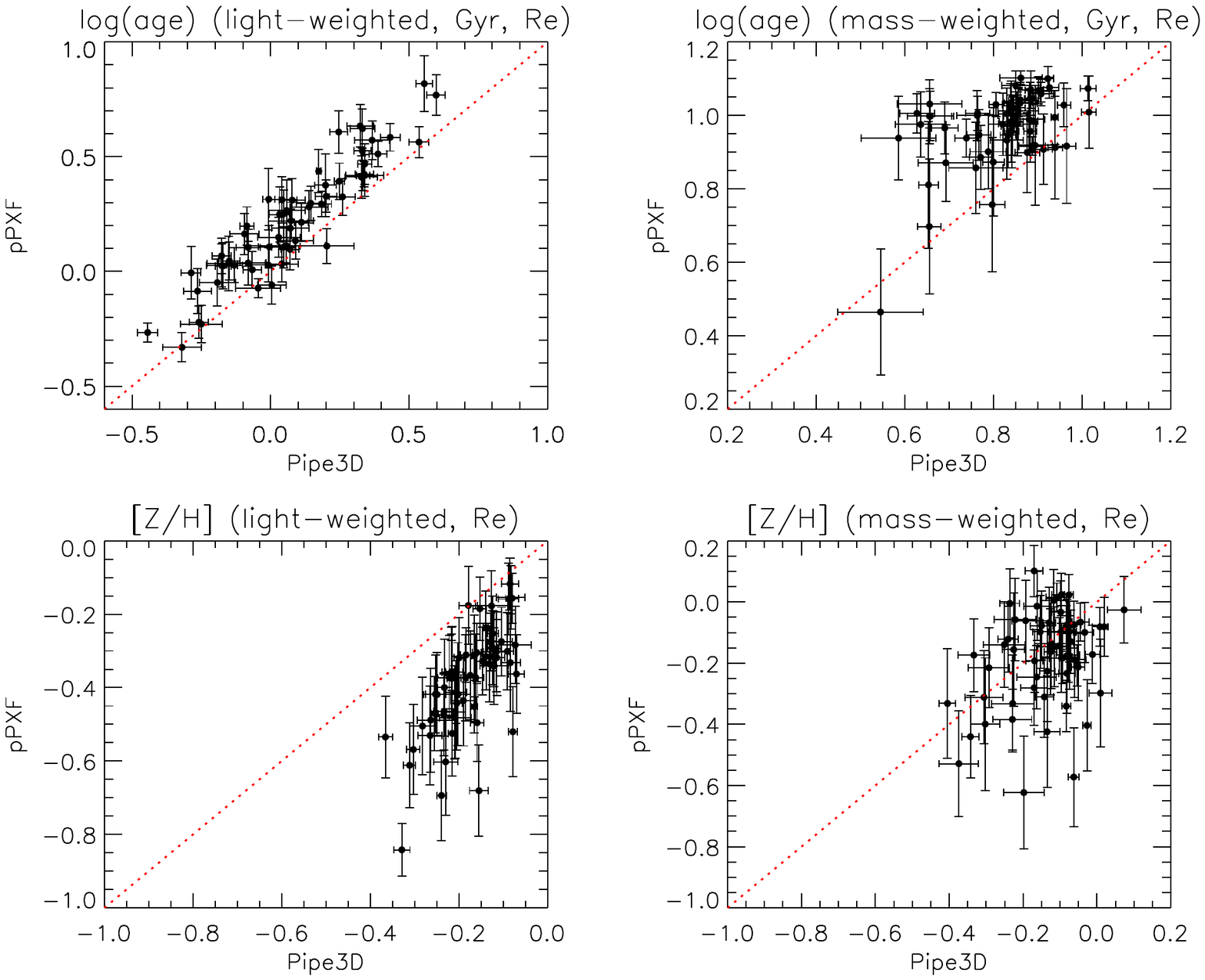}
	\caption{Comparison of ppulation parameters derived at 1 $R_e$ from our pPXF implementation with those reported on the MaNGA Pipe3D VAC. We find good consistency in the light-weighted ages and reasonable consistency in mass-weighted metallicities, though our light-weighted metallicities and mass-weighted ages are notably offset.}
	\label{ppxfpipe3dre}
	\end{center}
\end{figure}

We next consider in \autoref{ppxfpipe3dre} how our absolute measurements of the stellar populations compare to those on the pipe3D VAC. We extract the pipe3D measurements reported for 1 $R_e$ annuli, and we compare these with 1$R_e$ values from our own measurements obtained from interpolating the data and variance profiles. We find the light-weighted ages and metallicities of the two datasets to correlate tightly; however, our ages are systematically higher than pipe3D, while our metallicities are systematically lower. The comparisons for the mass-weighted population values are heavily scattered, meanwhile, with the mass-weighed ages from our pPXF measurements systematically higher than the pipe3d values.

\begin{figure}
\begin{center}
	\includegraphics[trim = 2cm 10cm 0cm 9cm,scale=0.95]{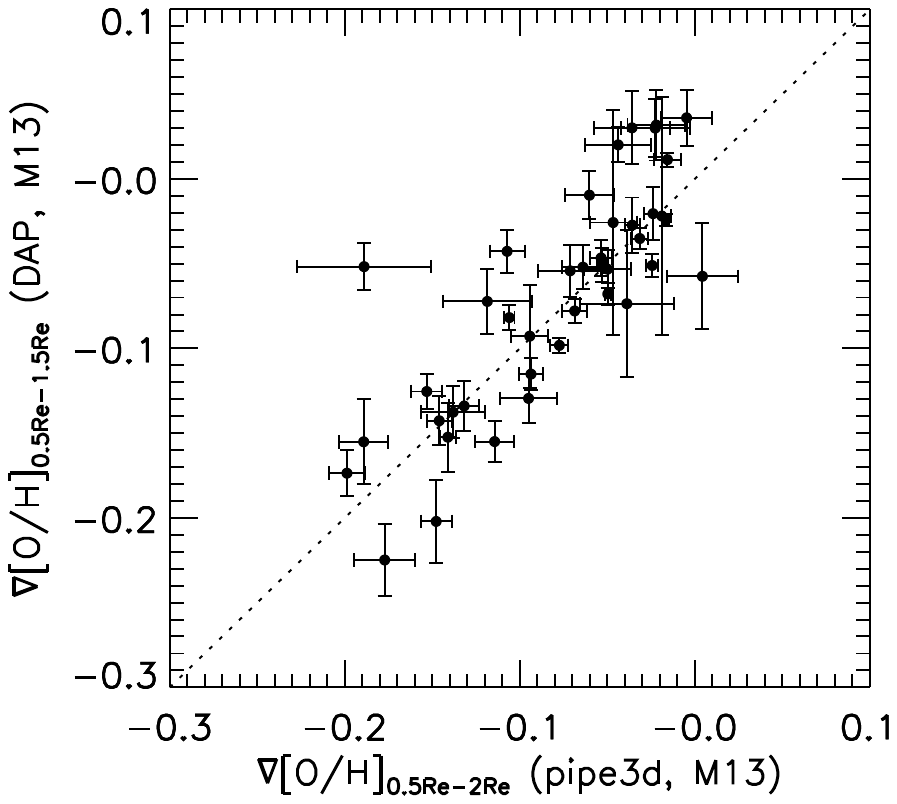}
	\caption{Comparison of gas-phase metallicity gradients calculated from the DAP with those calculated in Pipe3D, with the \citet{marino2013} O3N2 used in both cases. We find good agreement with Pipe3d overall, though with a degree of scatter.}
	\label{ppxfpipe3dgas}
\end{center}
\end{figure}

Next we consider how our calculated gas-phase metallicity gradients compare to Pipe3D. For this, we obtain gas-phase metallicity gradients for our galaxies using the \citet{marino2013} O3N2 calibrator, using the same method as described previously for other calibrators. We compare to the Pipe3D gradients from the same calibrator in \autoref{ppxfpipe3dgas}, for all galaxies used in our presented gas metallicity analysis. We find our gradients to correlated tightly with those from Pipe3D.

\begin{figure}
\begin{center}
	\includegraphics[trim = 1cm 1cm 0cm 12cm,scale=0.55]{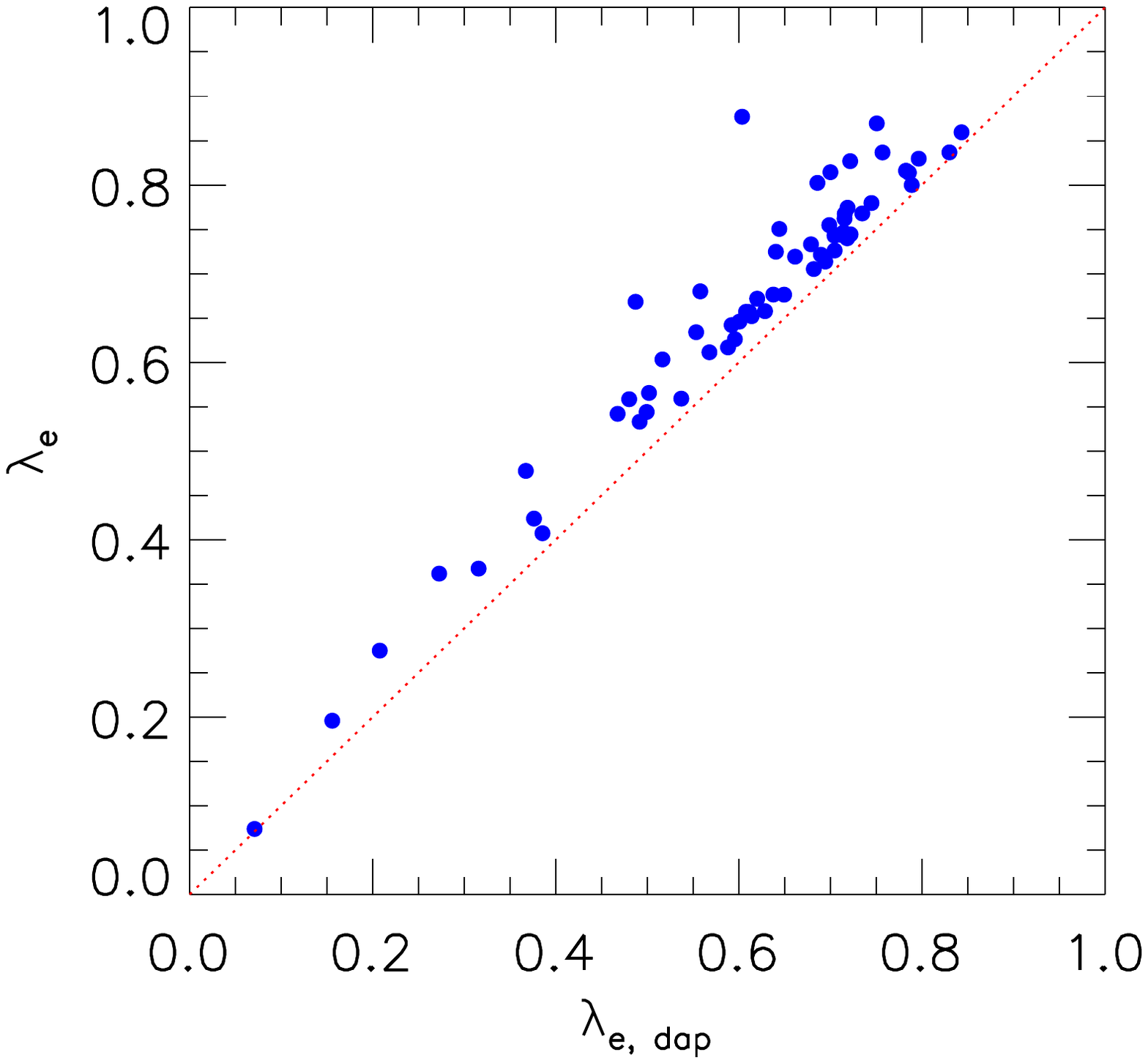}
	\caption{$\lambda_R$ values at $1R_e$ ($\lambda_e$) as calculated from pPXF fits with ELODIE, compared to equivalent values calculated from the MaNGA DAP ($\lambda_{e,dap}$). The dotted red line shows the 1--1 relation. Overall, we find a slight offset with respect to the DAP, with a mean difference of 0.05.}
	\label{lamrdap}
\end{center}
\end{figure}

In \autoref{lamrdap}, we compare our reported $\lambda_e$ values with those that can be extracted from the MaNGA DAP. We extracted these values, hereafter termed $\lambda_{e,dap}$, using MaNGA DAP velocities and kinematics in the same manner as for our derived ELODIE kinematics. Overall, we find that we obtain slightly higher values of $\lambda_e$ overall, with a mean offset of 0.05. However, this offset does not affect any of our qualitative conclusions. Possible reasons for this discrepancy include the use of different methods for correcting data-model resolution offsets, as well as the use of slightly different wavelength ranges for performing the fitting.

\bsp	
\label{lastpage}
\end{document}